\documentclass[prl,a4paper,twocolumn,preprintnumbers,nofootinbib]{revtex4-1}
\usepackage[a4paper, hdivide={1.91cm,,1.165cm}, vdivide={1.83cm,,3.8cm}]{geometry}
\usepackage{color}
\usepackage[hyperfootnotes=false,colorlinks,citecolor=magenta]{hyperref}
\usepackage{ulem}
\usepackage{youngtab}
\usepackage[english]{babel} 
\usepackage{amsmath}
\usepackage{float}
\usepackage{amssymb}
\usepackage{braket}
\usepackage{amsfonts,bbm}
\usepackage{dsfont,tensor}
\usepackage{enumerate}
\makeatletter\AtBeginDocument{\let\@elt\relax}\makeatother
\usepackage{graphicx}
\usepackage{caption}
\usepackage{adjustbox}
\usepackage{bbm}
\usepackage{array}
\usepackage{booktabs}
\usepackage{units}
\usepackage{textcomp}
\usepackage{mathtools}
\usepackage{cancel}
\usepackage{slashed}
\usepackage{multirow}
\usepackage{slashed}
\usepackage{comment}
\usepackage{relsize}
\usepackage{tcolorbox}
\usepackage{csquotes}
\usepackage{simpler-wick}
\usepackage{subcaption}
\usepackage{lipsum}

\allowdisplaybreaks

\newcommand{\dd}{\text{d}}

\newcommand{\kp}{k^{\prime}}
\newcommand{\eps}{\epsilon_-}

\newcommand{\pts}[1]{\phantom{.}\hfill(\textit{#1~point}\ifthenelse{\equal{#1}{1}}{}{\textit{s}})}

\newcommand{\psibar}{\bar{\psi}}

\newcommand{\dtild}{\tilde{\Delta}}
\newcommand{\ktild}{\tilde{\kappa}}

\newcommand{\Msun}{{\ifmmode{{\rm{M_{\odot}}}}\else{${\rm{M_{\odot}}}$}\fi}}

\newcommand{\beq}{\begin{equation}}
\newcommand{\eeq}{\end{equation}}
\newcommand{\bea}{\begin{eqnarray}}
\newcommand{\ena}{\end{eqnarray}}

\newcommand{\lsim}{\mathrel{\mathop{\kern 0pt \rlap
{\raise.2ex\hbox{$<$}}}
\lower.9ex\hbox{\kern-.190em $\sim$}}}
\newcommand{\gsim}{\mathrel{\mathop{\kern 0pt \rlap
{\raise.2ex\hbox{$>$}}}
\lower.9ex\hbox{\kern-.190em $\sim$}}}

\begin{document}

\title{Condensed dark matter with a Yukawa interaction}
\author{Raghuveer Garani}
\email{garani@fi.infn.it}
\affiliation{INFN Sezione di Firenze, Via G. Sansone 1, I-50019 Sesto Fiorentino, Italy}

\author{Michel~H.G.~Tytgat}
\email{michel.tytgat@ulb.be}
\affiliation{Service de Physique Th\'eorique, Universit\'e Libre de Bruxelles, Boulevard du Triomphe, CP225, 1050 Brussels, Belgium}

\author{J\'{e}r\^{o}me~Vandecasteele}
\email{jerome.vandecasteele@tum.de}
\affiliation{Physik-Department, Technische Universität München,
James-Franck-Straße, 85748 Garching, Germany}

\begin{abstract}
 We explore the possible phases of a condensed dark matter (DM) candidate taken to be in the form of a fermion with a Yukawa coupling to a scalar particle, at zero temperature but at finite density. This theory essentially depends on only four parameters, the Yukawa coupling, the fermion mass, the scalar mediator mass, and  the DM density. At low fermion densities we delimit the Bardeen-Cooper-Schrieffer (BCS), Bose-Einstein Condensate (BEC) and crossover phases as a function of model parameters using the notion of scattering length. We further study the BCS phase by consistently including emergent effects such as the scalar density condensate and superfluid gaps. 
 Within the mean field approximation, we derive the consistent set of gap equations, retaining their momentum dependence, and valid in both the non-relativistic and relativistic regimes. We present numerical solutions to the set of gap equations, in particular when the mediator mass is smaller and larger than the DM mass. Finally, we discuss the equation of state (EoS) and possible astrophysical implications for asymmetric DM. 
 \end{abstract}

\maketitle
{
  \hypersetup{linkcolor=black}
}
\section{{Introduction}}
\label{sec:introduction}
The nature of Dark Matter (DM) remains to a large extent a mystery. Like ordinary matter, it is possible that it consists of an asymmetric population of particles~\cite{Kaplan:2009ag}. Additionally, DM may be in a cold and dense state in several regions of the Universe, for instance at the core of galaxies and dwarf galaxies~\cite{Destri:2013pt,Domcke:2014kla,Randall:2016bqw,Bar:2021jff} or, if captured, inside of dense and cold stars~\cite{Goldman:1989nd,Garani:2018kkd,Graham:2018efk} or in the form of nuggets of DM~\cite{Gresham:2017cvl,Bai:2018dxf}. It is even considered that DM particles can condense and make dark stars~\cite{Narain:2006kx,Hippert:2021fch}. Much like ordinary condensed matter, it is then conceivable that DM particles manifest complex, emergent behavior at low densities, like superfluidity. This may have various implications, from the formation of proto-DM halos to the dissipationless transport of heat, to vortex formation during DM halo mergers, or simply for the equation of state of DM, which is  essential for the study of compact DM object formation and evolution. Incidentally, superfluid DM is a possibility that is much studied, usually in the form of fundamental scalar DM, see e.g.~\cite{Berezhiani:2015bqa,Fan:2016rda, Ferreira:2020fam}. Here we focus on the less studied possibility that condensed fermionic DM particles can be superfluid, see e.g.~\cite{Alexander:2016glq,Alexander:2020wpm}.

More precisely, we explore the possible superfluid phases of a very simple and yet quite rich asymmetric DM model. It consists of a DM Dirac fermion $\psi$ with a Yukawa coupling to a spin 0 ($\phi$, a real scalar) mediator. This theory rests essentially on 3 parameters on top of the DM density: the DM mass, the mass of the mediator and the Yukawa coupling. For simplicity, we neglect a possible self-coupling of the mediator. Since spin 0 boson exchange is attractive, at low densities and depending on the parameters, this fermionic DM may either form di-molecules that could be in a Bose-Einstein condensate (BEC phase), or Bardeen-Cooper-Schrieffer pairs (BCS phase). The thermodynamic description of the BEC and BCS phases are qualitatively different, since in the BEC phase there are true di-fermionic bound states, see e.g.~\cite{Grimm}. But, in either cases, such DM should be superfluid at low enough temperatures. The problem turns out to be quite complex, so our aim here will be to set the ground for more phenomenological studies. In particular, we will focus on the thermodynamic properties of this simple DM model at zero temperature and will delineate which phases can be formed as a function of model parameters and DM density. This will allow us to address a question regarding the gravitational collapse of such DM, if it is accumulated at the core of neutron stars \cite{Kouvaris:2018wnh,Gresham:2018rqo}. This requires the determination of the overall normalization of the gaps and hence necessitates a more careful treatment that goes beyond the standard, textbook BCS approximations. 

Our plan is as follow. In section~\ref{sec:model} we setup the model and establish which phases (BCS or BEC) DM may form at low temperatures, depending on the model parameters. To do so, we characterize the strength of Yukawa interaction at finite density using the concept of scattering length (see also appendix~\ref{app:a}). Next, in section~\ref{sec:gaps}, we focus on the case where DM particles are in a BCS phase. We derive the consistent set of gap equations, taking into account the change of DM mass at finite density due to the formation of a scalar condensate. The gaps equations are solved numerically in section~\ref{sec:sols} and, where possible, comparison is made with analytical solutions. We show that in the large DM density limit the system is generically in the BCS phase. We next discuss some possible implications for DM phenomenology, ranging from DM in neutron stars to dwarf galaxies in section~\ref{sec:astro}. Finally, we draw conclusions in section~\ref{sec:con}. 
Several technical results which may be of interest to a broad audience can be found in the appendices~\ref{app:walecka}\,--\,\ref{sec:phase}.

\section{Low density phases of a Yukawa theory}
\label{sec:model}
Our starting point is a degenerate gas of Dirac fermions $\psi$ (the asymmetric DM) with a Yukawa coupling $g$ to a real scalar $\phi$ (the mediator)
\beq\label{eq:lagr1}
\mathcal{L} = i \bar{\psi}\slashed{\partial} \psi - m \bar \psi\psi +  \mu \bar\psi\gamma^0 \psi + \frac{1}{2} \partial_\mu \phi \partial^\mu \phi - {1\over 2} m^2_\phi \phi^2 - g \,\bar{\psi}  \psi \phi~.
\eeq
 The fermion $\psi$ and the mediator $\phi$ are singlets of the Standard Model (SM) but $\psi$ is charged under a global dark $U(1)$ symmetry. Therefore $\mu$ is the chemical potential conjugate to DM fermion number $N$ in a volume $V$, corresponding to a DM density $n= N/V \equiv \langle\bar \psi \gamma^0 \psi\rangle$. The expectation value $\langle \ldots \rangle$ is on the ground state of the system, here taken to be at finite fermion density but at zero temperature. For a degenerate gas of free fermions $g\rightarrow 0$, the chemical potential is equal to the Fermi energy $E_F$, $\mu = E_F \equiv \sqrt{m^2 + k_F^2}$ and 
$
n = {k_F^3/3 \pi^2}
$. 

\begin{figure}[H]
	\centering
\includegraphics[width=0.25\textwidth]{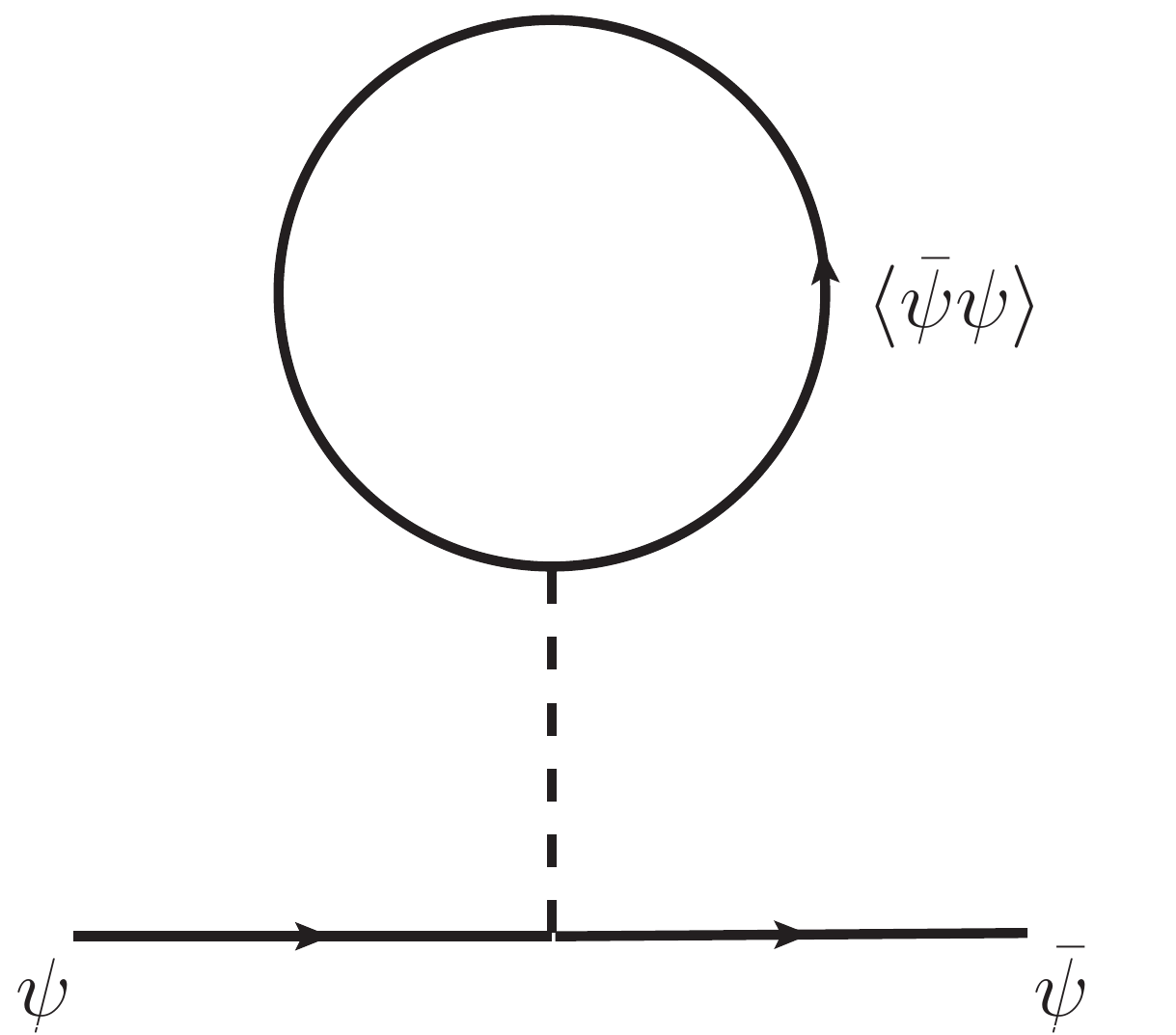}
	\caption{ The tadpole is non-vanishing at finite density. This corresponds to a scalar density condensate $\langle \bar \psi \psi\rangle$ which modifies the effective mass of the fermions in the medium.}
	\label{fig:feynman_cond}
\end{figure}

In eq.~\eqref{eq:lagr1}, $m$ and $m_\phi$ denote the bare fermion and boson masses at zero density. Both are modified in a medium and, in particular, at finite density. The most dramatic effect is the change of the fermion mass. In this case, physically, the scalar operator $\bar \psi \psi$ has a non-zero mean, $n_s = \langle \bar \psi \psi \rangle >0$ \cite{Walecka:1974qa}. In turn, $n_s$ sources the scalar field due to its Yukawa interactions with the fermions
$$
{\delta {\cal L}\over \delta \phi} = 0 \rightarrow m_\phi^2 \langle \phi \rangle + g \langle \bar \psi\psi\rangle = 0~,
$$
which consequently changes the mass of the DM 
\beq
m_\ast= m + g \langle \phi \rangle \rightarrow m_\ast = m - {g^2\over m_\phi^2} n_s(m_\ast),
\label{eq:ns}
\eeq
see fig.~\ref{fig:feynman_cond} for a diagrammatic representation. We have expressed the fact that $n_s$ is itself a function of the effective fermion mass.\footnote{See also ref.~\cite{Esteban:2021ozz, Smirnov:2022sfo} for an application of this effect to SM neutrino clustering.} 
This effect implies that the effective fermion mass decreases with increasing fermion density. For a given value of Yukawa coupling and the mediator mass,
at low enough densities, $n_s$ is close to the DM fermion number and $m_\ast \approx m$. However, as the density increases , the in-medium mass $m_\ast$ decreases and
 eventually tends to zero ($m_\ast \rightarrow 0$), see appendix~\ref{app:walecka}.\footnote{{Inclusion of self coupling term of the mediator in the scalar potential ($\lambda \phi^4$) qualitatively changes the picture in two ways. The equilibrium values of in-medium fermion mass gets bounded by below, i.e. $m_\ast \rightarrow m \sqrt{\lambda}$ in the limit of large $\lambda$. The second effect is related to the effective mass of the mediator, which would scale as $m^{\rm eff}_\phi = m_\phi \sqrt{1+ 2 V(\langle \phi \rangle)/m_\phi^2\langle \phi \rangle^2}$~\cite{Gresham:2017zqi}. }}

The change of the fermion mass is of course not the only effect of Yukawa interactions. If the attraction due to spin 0 particle exchange is strong enough, fermions can be bound into di-molecules at low densities. These bound states behaves as bosons and can condense into a Bose-Einstein condensate (BEC) at low enough $T$. Nevertheless, even if attraction is weak such that no true bound state can be formed, Cooper instability can lead to the formation of Bardeen-Cooper-Schrieffer (BCS) pairs. The distinction between the two possibilities is not sharp and which of these situation is realized at low densities depends on the parameters of the theory. The transition between the BEC and BCS states as these parameters vary is continuous or a crossover~\cite{Nozieres:1985zz}.

In the non-relativistic limit, the nature of the low density phase can be qualitatively understood by examining the scattering length ($a$) of the fermions. While this is part of the standard toolbox of condensed matter physics, it is less common in the high energy physics literature. A notable and interesting exception is ref.~\cite{Chu:2019awd}, in which the scattering length is put forward as a convenient tool to discuss the properties of interacting DM. The most relevant feature of the scattering length is that, while in the Born approximation $a_{\rm Born}\lesssim 0$ for an attractive potential, $a >0$ signals the possibility of forming a bound state~\cite{Landau:1991wop} and so a BEC. For convenience, we summarize the basic and relevant properties of the scattering length in appendix~\ref{app:a}.

For a Yukawa interaction, the s-wave scattering length for the singlet, spin 0 channel reads
\beq
\lim\limits_{k \to 0} k \cot{\delta_0(k)} = -\frac{1}{a},
\eeq
where $\delta_0(k)$ is the s-wave phase shift. Obtaining this quantity
requires solving the Schr\"odinger equation for the scattering problem. A much used approach, valid in the limit of large particle separation compared to the effective range of the interaction, is to approximate the attractive potential by a contact interaction, a delta function in other words, and to re-express the scattering problem in terms of the scattering length~\cite{Thorn:1978kf}. Applied to a degenerate system of fermions at finite density with Fermi momentum $k_F$, the phases can then be characterized in terms of the dimensionless parameter $k_F\, a$, with a dilute system corresponding to $\vert k_F\, a\vert \rightarrow 0$, while the sign indicates the phase of the system~\cite{Pethick:2008,Schmitt:2014eka,PhysRevB.55.15153}. 
Consequently, the BEC phase corresponds to small positive values of $k_F\, a$, physically corresponding to a scattering length smaller than the particle separation, while the BCS regime is characterized by small negative values of $k_F\, a$. Finally, the crossover regime is realized for large absolute values of $k_F\, a$. This corresponds formally to a diverging scattering length $\vert a\vert$, and is called a unitary Fermi gas in the literature~\cite{Heiselberg:2000bm}. 

\begin{figure}[H]
	\centering
\includegraphics[width=0.48\textwidth]{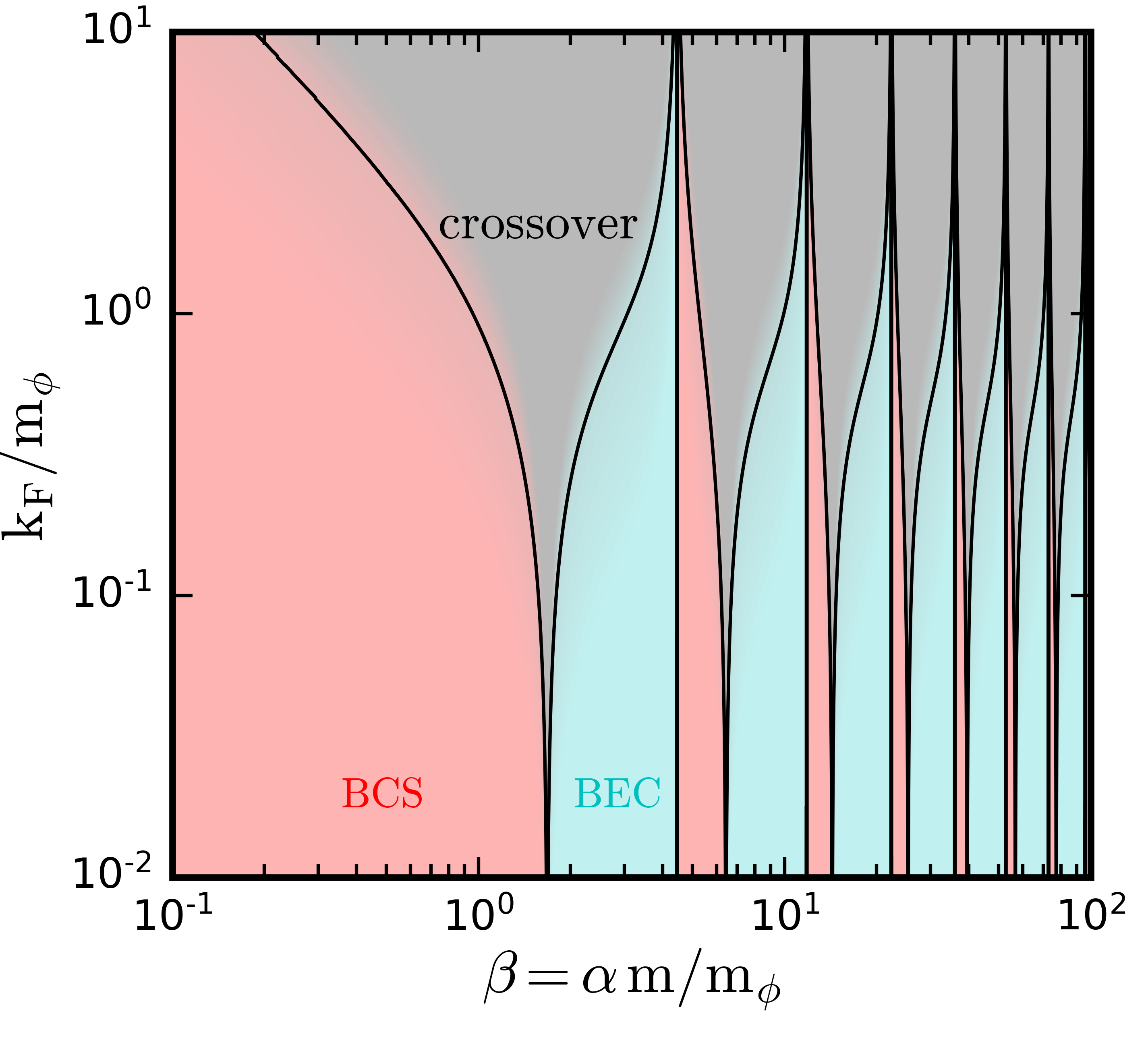}
	\caption{Contours of $(k_F a)^{-1}$. Red shaded regions are characterized by $(k_F a)^{-1} < -1 $ indicating BCS regime. In the cyan shaded regions $(k_F a)^{-1} > 1 $ indicating BEC regime, and the gray regions correspond to possible BEC-BCS crossover with $-1 < (k_F a)^{-1} < 1 $. }
	\label{fig:scattering_length_regions}
\end{figure}

 In this work, so as to  map the model parameters of the theory~(\ref{eq:lagr1}) to the possible phases of condensed DM, we go beyond the above contact interaction approximation and compute directly the scattering length by solving the Schr\"odinger equation. We solve the scattering problem using the numerical method proposed in~\cite{Chu:2019awd} which is also summarized in appendix~\ref{app:a} for reference. In doing so, we can determine parameters of the Yukawa theory for which  DM particles are clearly in the BCS (large negative $k_F a$) or in the BEC (large positive $k_F a$) phases. This result is depicted in fig.~\ref{fig:scattering_length_regions} as function of the dimensionless parameters $\beta = \alpha m/m_\phi$ with $\alpha = g^2/4 \pi$ and the ratio of length scales $k_F/m_\phi$. The Born approximation corresponds to $\beta \ll 1$ (also denoted $b$ in the DM literature \cite{Tulin:2013teo}), so we can expect the onset of bound state formation (and thus BEC phases) to be around $\beta \gtrsim 1$. However, the sign of $a$ changes each time a new bound state channel opens, so the relation between the possible phases and the parameters is complex. The other parameter ($k_F/m_\phi$) is simply a measure of the mean particle separation over the range of the Yukawa potential, large $k_F/m_\phi$ corresponding to large densities. In fig.~\ref{fig:scattering_length_regions}, the red shaded regions indicate the BCS phase, which we define to correspond to $(k_F a)^{-1} < -1$, a value motivated by the results obtained based on the contact interaction approximation, see e.g. \cite{Schmitt:2014eka}. The cyan shaded regions are characterized by $(k_F a)^{-1} > 1$ and are delimiting the BEC phase. The gray shaded area show the intermediate crossover phase, $-1 < (k_F a)^{-1} < 1$. The unitarity limit is reached when $k_F |a|\rightarrow \infty$, indicating crossover regime at all densities, which is seen as a feature in fig.~\ref{fig:scattering_length_regions}. Further, for finite densities, the cases of anti-resonance $k_F a\rightarrow 0$ are captured by the peaks delimiting BEC to BCS transitions. 
 
 We conclude that, for fixed $k_F/m_\phi$, the BCS, BEC, and crossover regimes alternate as we change the value of $\beta$. However, for large values of $\beta$, i.e. large coupling or small mediator mass limit, the system is more likely to be in a BEC or, at large densities, in crossover regimes. Whereas, in the opposite limit, the interaction is not strong enough to form bound states and the DM fermions are expected to be in a BCS phase at low temperatures. 

The above considerations are based on a very simple criteria, essentially the sign of the scattering length, a low momentum/short range parameter, indicative of the possible outcomes of a system of fermions at relatively low DM densities $k_F \ll m$. Notice that there only two possibilities for fixed parameters, either BCS or BEC at low densities with a crossover at increasing densities. This behaviour is manifest in the approximation of a contact interaction, see \cite{Schmitt:2014eka}. These considerations however neglect the fact that the effective mass of the fermion $m_\ast$ changes, an effect which is particularly dramatic at larger densities. To take that effect in a self-consistent way, in the next section we derive the gap equations for the BCS phase in complete generality by retaining the mediator mass and momentum dependence of the anomalous propagators. On that basis, we shall argue that, regardless of the phase at low densities, the fermions are in a pure BCS phase at large densities, corresponding to $k_F \gtrsim m_\ast$. That only a BCS phase is possible in the relativistic regime is probably due to the fact that it is not possible to form bound state of relativistic particles with a Yukawa interaction.

\section{BCS phases}
\label{sec:gaps}

\begin{figure}[H]
	\centering
 \includegraphics[width=0.3\textwidth]{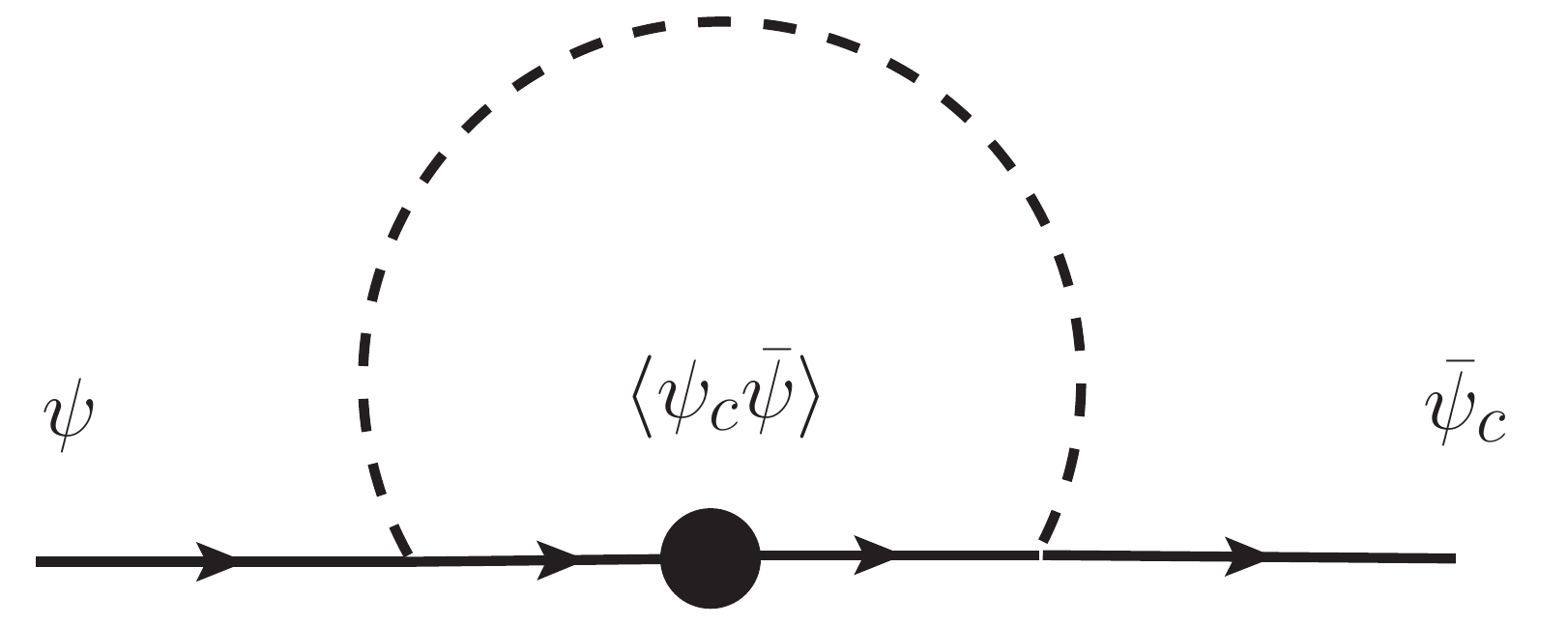} 
	\caption{Diagrammatic representation of fermion BCS condensate.}
	\label{fig:feynman_gap}
\end{figure}

We have so far described the low density phases of the Yukawa theory with no reference to the details of superfluidity, such as the bound-state wave-functions \cite{Nozieres:1985zz}. In this section, we study in detail the BCS gap matrix $\Delta\equiv\langle \psi_c\bar{\psi}\rangle$, with $\psi_c$ being the charge conjugate of $\psi$, see fig.~\ref{fig:feynman_gap}. We analyze its Dirac structure so as to decompose them into simple gap functions, derive the corresponding gap equations and finally solve them, taking into account their interplay with other in-medium density effects, in particular the scalar condensate and its impact on the DM mass. We aim at showing the evolution of the gaps with varying density, from the non-relativistic to the ultra-relativistic regimes, an aspect that has not yet been studied as far as we know. This implies that we will focus our study on the red regions shown in fig.~\ref{fig:scattering_length_regions}, i.e. for small values of $\beta$, in the non-relativistic regime. However, we will argue that, for fixed $\beta$, a BEC phase becomes a true BCS phase at large densities. 

The techniques to derive the gap equations for a potentially relativistic fermionic system at finite density are well-established. In this section, we first analyze the Dirac structure of the gap matrix, $\Delta$, which is a $4\times 4$ matrix (for one flavor of DM) which can be decomposed using the Dirac matrices. We follow the approach set up in~\cite{Bailin:1983bm} and in particular \cite{Pisarski:1999av}, in which the Yukawa theory is studied in detail in the massless limit, $m = 0,\, m_\phi = 0$. Then to derive the gap equations, we follow the energy functional approach of \cite{Alford:2003fq,Alford:2004hz}, (see also \cite{Alford:2017ale} for the Yukawa theory), which is based on the so-called Hubbard–Stratonovich transform. In particular, we were inspired by \cite{Kleinert:2011rb} to take into account the variation of the DM mass through the scalar condensate. 

\subsection{Gap Dirac structure}
\label{sec:form}

We will consider a general ansatz for the $4\times 4$ gap matrix $\Delta = \langle \psi_c\bar{\psi}\rangle$. We work in the rest frame of the fermion gas, which is assumed to be infinite and homogeneous. In this case, one can be easily convinced that the $\Delta$ matrix can be written as a sum of up to 8 translation invariant terms that can be expressed using the Clifford basis of matrices built upon the Dirac $\gamma^\mu$ and $\gamma_5$ ~\cite{Bailin:1983bm,Pisarski:1999av}. As the Yukawa theory preserves parity, we expect that the gap matrix is also parity symmetric. Also, the ground state is expected to be rotationally invariant. This implies that pairing of fermions should be in $J^P=0^+$ channel. This allows us to express the gap matrix $\Delta$ in terms of only 3 gap functions (the gaps in the sequel): 
\beq\label{eq:gap_ansatz}
\Delta \equiv \langle \psi_c \bar{\psi}\rangle = \Delta_1\, \gamma_5 + \Delta_2 \,{\vec\gamma\cdot\hat{k}}\,\gamma_0 \gamma_5 + \Delta_3\, \gamma_0 \gamma_5 ~.
\eeq
The task is then to determine self-consistently these gaps, together with the scalar condensate 
\beq
n_s = \langle \bar{\psi}\psi\rangle~.
\eeq
While this is not a new problem, to our knowledge it has not been worked out in the framework of the Yukawa theory.
We will show that the $\Delta_i$'s strongly depend on $n_s$ though the effective fermion mass $m_\ast$, while the dependence of the scalar condensate on the gaps is mild, see below. 

\subsection{Quasi-particle dispersion relations}
\label{sec:disp}

The next step is to examine the spectrum of fermionic excitations (quasi-particles) near the Fermi surface, see appendix \ref{app:subsec:dr} and \cite{Schmitt:2014eka}. For a free degenerate gas, the dispersion relations are simply 
\beq\label{eq:disp_free}
 \epsilon_\pm(k)= \vert \omega_k \pm \mu \vert~,
 \eeq
with $\omega_k= \sqrt{k^2 + m^2}$ and the subscript $-(+)$ corresponding to fermion (anti-fermion) excitations. These dispersion relations can be read off directly from Lagrangian~\eqref{eq:lagr1}. Simply put, this means that it costs little energy to create a fermionic excitation near the Fermi surface, 
\beq 
\epsilon_-(k)\approx v_F \vert k - k_F\vert~,
\eeq
with $v_F = d\epsilon/dk\vert_{k_F} = k_F/\mu$ and $k> k_F$ for a particle-like excitation while $k< k_F$ for a hole-like excitation. The linearity of the dispersion relation for particle/hole excitations near the Fermi surface is valid both in the non-relativistic and relativistic regimes. For anti-particles, however, the cost is at least $\epsilon_+ \approx 2 \mu$.  

If the gaps are non-vanishing, it is tedious to derive the dispersion relations but the final result can be approximated by the following fairly simple expression
 \beq\label{eq:disp_approx}
 \epsilon_\pm^2\approx \left(\omega\pm \mu \right)^2+ \left(\Delta_1\pm\left(\frac{k}{\omega}\Delta_2+\frac{m_\ast}{\omega}\Delta_3\right)\right) ^2~,
 \eeq
where again $-(+)$ corresponds to particle (anti-particle) excitations, see appendix~\ref{app:subsec:dr} for details. We have assumed that the gaps are smaller than the chemical potential, which we expect to be the case in the BCS phase. As far as we could judge, our ansatz is consistent with results derived in~\cite{Bailin:1983bm}, albeit with a distinct approach. It generalizes the results presented in~\cite{Pisarski:1999av}, where the focus was only on the ultra-relativistic regime, $m=0$ in which case only $\Delta_{1,2}\gg \Delta_3$ are relevant. As the gap functions are momentum dependent, in principle the dispersion relation eq.~\eqref{eq:disp_approx} interpolates between the non-relativistic and relativistic regimes. This also goes beyond the simple ansatz put forward in~\cite{Schmitt:2014eka}, where only $\Delta_1$ pairing channel is retained in the non-relativistic regime. This is consistent with setting $\Delta_2 =0$, whose contribution to the dispersion relation is suppressed if $k \ll m$. This leaves open the possible relevance of $\Delta_3$ in this regime. Notice however that its contribution is also related to a possible distinction between particle and antiparticle excitations. 

This motivates the introduction of the new gap functions 
\beq\label{eq:dtilde}
\dtild_\pm = \Delta_1\pm\left(\frac{k}{\omega}\Delta_2+\frac{m_\ast}{\omega}\Delta_3\right)~,
\eeq
and
\beq
\ktild = \frac{m_\ast}{\omega}\Delta_2-\frac{k}{\omega}\Delta_3~.
\eeq
We expect and will verify that $\ktild$, which is orthogonal to $\tilde \Delta_\pm - \Delta_1$, is relatively small compared to $\Delta_\pm$ in the non-relativistic and ultra-relativistic regimes. 
Indeed, in the relativistic limit, eq.~(\ref{eq:dtilde}) suggests that $\tilde{\Delta}_\pm \approx \Delta_1 \pm \Delta_2$ and $|\tilde{\kappa}| \approx - \Delta_3$. Whereas in the non-relativistic limit, $\tilde{\Delta}_\pm \approx \Delta_1 \pm \Delta_3$ and $|\tilde{\kappa}| \approx \Delta_2 $. In either case, $\tilde{\kappa}$ should be relatively small. This intuition is supported by the numerical results, see appendix \ref{app:subsec:resgap}. So in the sequel, we simply set $\ktild$ to zero. 
\begin{figure*}[t!]
	\centering
	\captionsetup{singlelinecheck=false,justification=centerlast}
\includegraphics[width=0.48\textwidth]{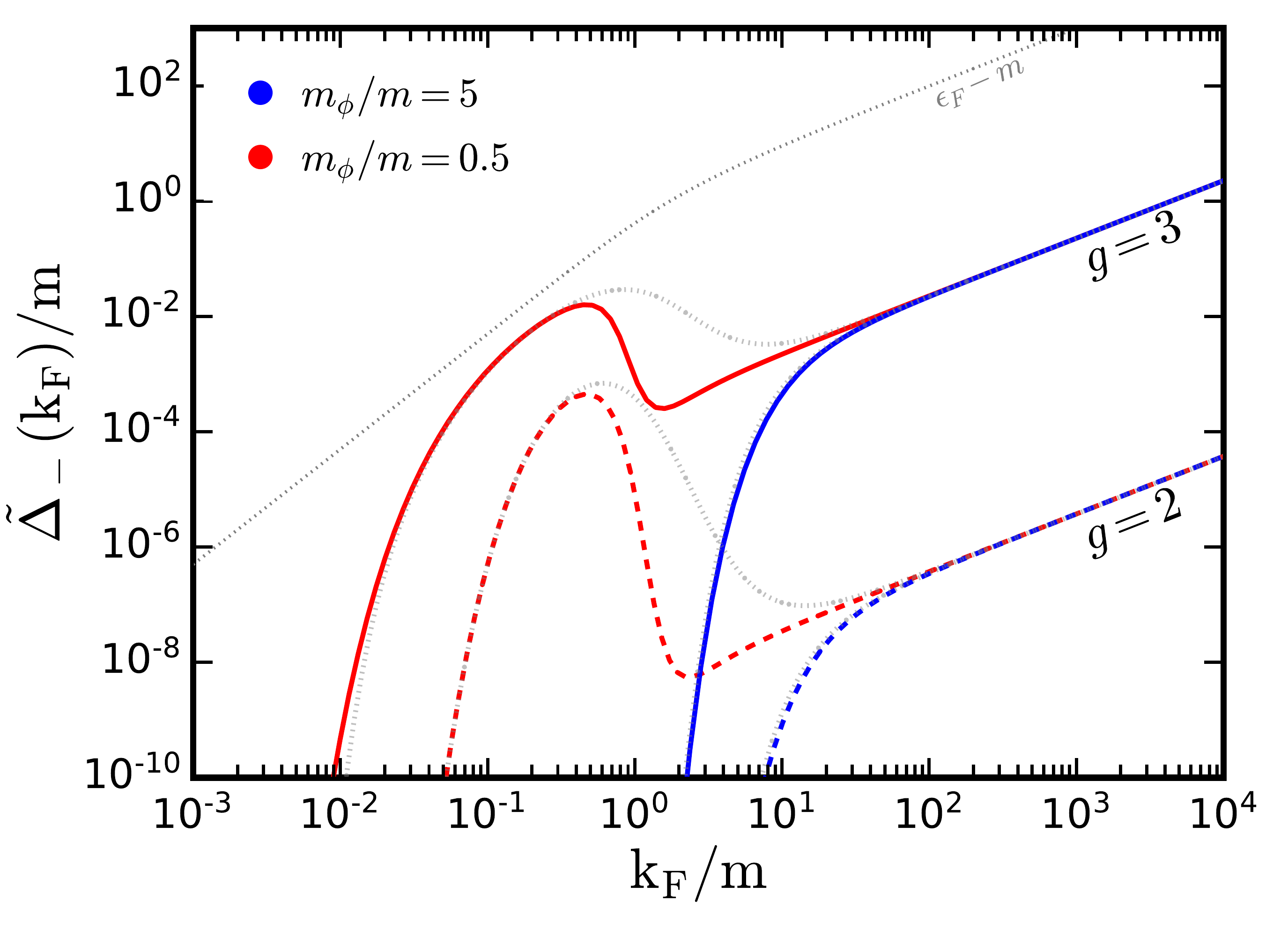}
\includegraphics[width=0.48\textwidth]{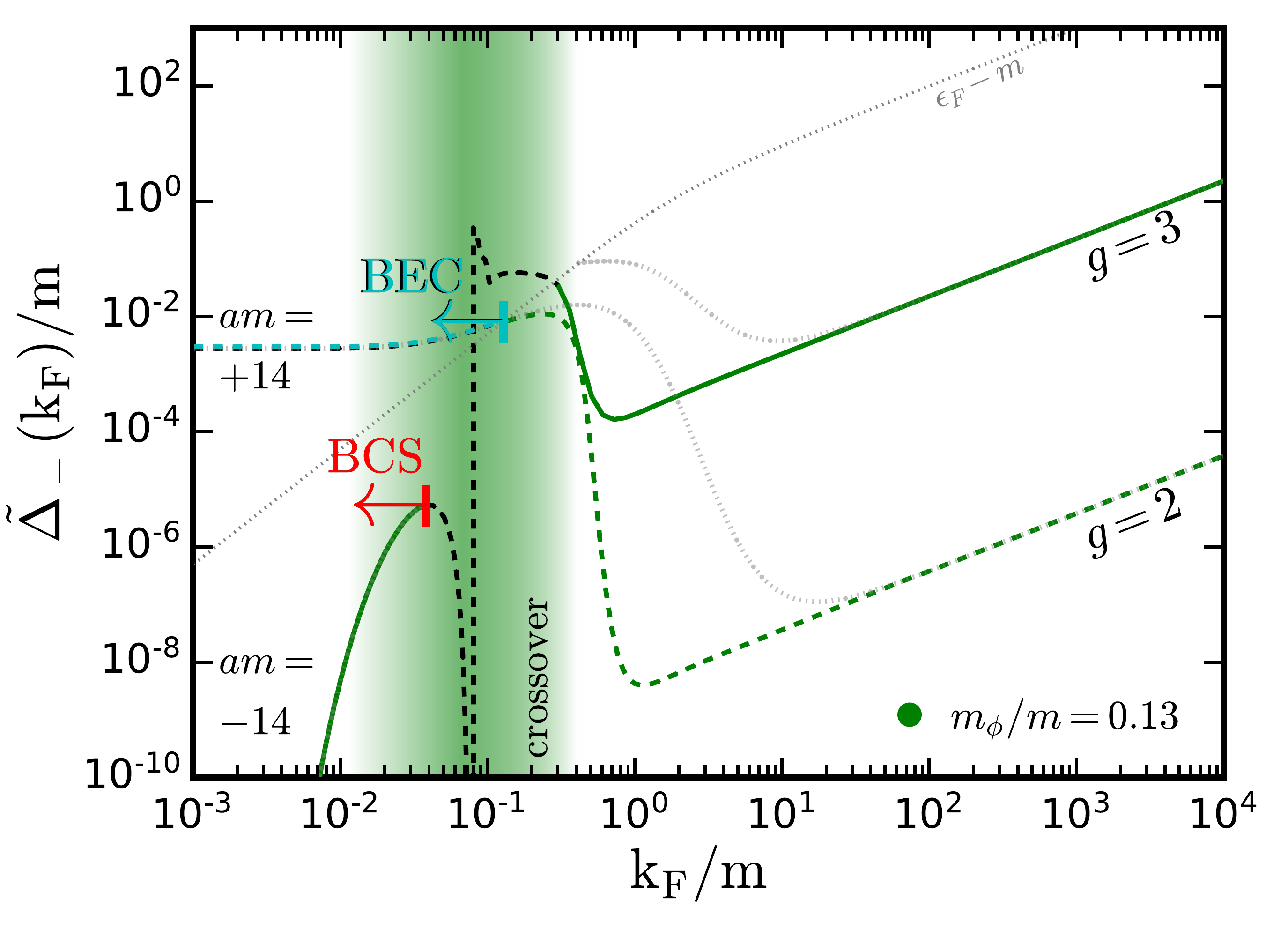}
	\caption{The solution to the gap equation including the effect of scalar density condensate is shown as function of dimensionless variable $k_F/m$. In the left panel we show the results for values of $g=3$ (solid curves) and $g=2$ (dashed curves). Blue (red) colored curves correspond to mediator mass $m_\phi= 5 \,m $ ($m_\phi= 0.5\, m$). Dotted gray curves represent the solution to the gap equation with $m_\ast=m$. The right panel follows the same scheme as in the left panel but correspond to $m_\phi = 0.13\, m$. The green shaded region represents crossover regime. See text for details.}
	\label{fig:results}
\end{figure*}

\subsection{Gap and scalar condensate equations }
\label{sec:gapsns}

With the  dispersion relation eq.~\eqref{eq:disp_approx} at hand, we now derive the gap equations in the mean-field approximation. To do so, we follow a variational approach based on the so-called Hubbard–Stratonovich transformation \cite{1957SPhD,1959PhRvL}. In this framework, a potential gap function, say $\Delta_i$, is introduced as an auxiliary field in the expression  for the free energy, $\Omega = - T \ln Z/V= - p(\mu,T)$ at finite fermion density (and temperature), i.e. $\Omega \rightarrow \Omega[\Delta_i]$. Minimizing the free energy with respect to the gap function correspond to the true ground state of the system at finite density and, by the same token, a gap equation for the corresponding gap function. An extra complication is the presence of the scalar condensate $n_s$. To take this into account, we follow the approach put forward in \cite{Kleinert:2011rb}. Alternatively, one can derive the gap equations in the mean-field approximation using the the Wick theorem, see e.g. \cite{Fetter}, p.441. Finally, recall that $n= - (\partial \Omega/\partial\mu)_T$ (here assumed to be at $T=0$).

Altogether, the gap functions and scalar condensate are thus determined imposing
\beq\label{eq:minimize}
\frac{\partial \Omega}{\partial \Delta_1} =0\,, \quad \frac{\partial \Omega}{\partial \Delta_2} =0\,,\quad \frac{\partial \Omega}{\partial \Delta_3} =0\,, \quad \frac{\partial \Omega}{\partial n_s}=0~.
\eeq
Thus we obtain the following equations, which we have expressed using the $\tilde{\Delta}_\pm$  
combination of gap functions, 
\begin{widetext}
\bea\label{eq:gap_penultimate}
n_s&=&\frac{-g^2}{m_\phi^2}\sum\limits_{\eta=\pm}\int_0^\infty\frac{dk\, k^2}{2\pi^2}\left\{\frac{m_\ast}{\omega_k}\left(\frac{\omega_k + \eta \mu}{\epsilon_\eta\left(k\right)} -1 \right)-\eta\frac{k}{\omega_k}\frac{\ktild(k)}{\omega_k}\frac{\dtild_\eta(k)}{\epsilon_\eta(k)}  \right\}\,,\\ 
\label{eq:gap_penultimate2}
\dtild_\pm(p)&=& \frac{g^2}{32\pi^2} \sum_{\eta=\pm}\int^\infty_0 d k \frac{k}{p} \left\{\log\frac{m_\phi^2+(p+k)^2}{m_\phi^2+(p-k)^2} \mp \eta \frac{k p}{\omega_p \omega_k}\left(-2 + \frac{m^2_\phi +k^2 +p^2}{2 k p} \log\frac{m_\phi^2+(p+k)^2}{m_\phi^2+(p-k)^2} \right)\right. \\ \nonumber
& & \left. \quad \quad \quad \quad \quad \quad \quad \quad\pm \eta \frac{m^2_\ast }{\omega_p \omega_k} \log\frac{m_\phi^2+(p+k)^2}{m_\phi^2+(p-k)^2}\right\} \frac{\dtild_\eta(k)}{\epsilon_\eta(k)}~. 
\ena
\end{widetext}
Here, the gap equation for $\tilde{\kappa}$ is omitted as it is expected to be subdominant in all regimes, see appendix~\ref{app:sub:gesc} for the complete set of gap equations. While having a relatively simple structure, these expressions call for some explanations. Consider first the equation for $n_s$. It is easy to verify that, in the limit of zero gap, it reduces to eq.~\eqref{eq:scden_gap0} with an integral over the volume of the Fermi sphere of radius $k_F$. Note that we have renormalized the expression for $n_s$ by subtracting the vacuum part; hence the $-1$ in the first term of eq.~\eqref{eq:gap_penultimate}. We further find that $n_s$ is momentum independent, and the presence of non-zero gaps manifests as the second term in the equation. This term is proportional to $\tilde{\kappa}$ and represents a parametrically minor correction to the gap-less expression.     

Now consider the equation for the gap $\tilde{\Delta}_\pm$, eq.~\eqref{eq:gap_penultimate2}. The limit $m_\phi \gg k_F \gg m$,  corresponds to the case of contact interactions and relativistic fermions. In this case  the above system reduces to the well known simple BCS gap equation~\cite{Schmitt:2014eka}
\begin{align}\label{eq:bcs_rel}
\Delta_1 = \frac{g^2}{2\,m_\phi^2}\int \frac{d^3k}{\left(2\pi\right)^3}\frac{\Delta_1}{2\,\epsilon_-\left(k\right)},
\end{align}
see also appendix~\ref{app:sub:BCSlimit}. Note however that in general the gap functions are momentum dependent. The standard BCS approximation assumes that the gap is momentum independent and hence constant. This requires the introduction of a cut-off on the integral over momentum. The cut-off may be dictated by physical considerations, for instance it may be set by the Debye screening length in the case of ordinary superconductivity. Such cut-off also sets the overall normalization of the gap. In the case of Yukawa interactions, it is not clear a priori which cut-off scale can be introduced. Furthermore, we want to have a handle on the overall scale of the gap functions.   
In such a situation, the best way forward is to keep the full momentum dependence of the gap functions, which by construction should vanish in the limit of large momentum. So, the advantage of our consistent set of gap equations is that we can in principle integrate the above equations all the way to infinity without having the need to introduce spurious cut-off dependence. 
Finally note that we have not considered possible in-medium corrections to the mediator mass $m_\phi$. These corrections are $\mathcal{O}(g \mu)$, and contributes positively to its mass. We neglect them for simplification, but one should keep in mind that for large coupling or massless mediator, they may play an important role~\cite{Pisarski:1999av}.

\subsection{Solutions to the gap equations}
\label{sec:sols}

In this section, we present the numerical solutions to the set of gap equations, together with the scalar condensate. We use a method called matrix inversion and developed in ref.~\cite{KHODEL1996390} (see appendix~\ref{app:numerics}). We present our findings in fig.~\ref{fig:results}, for various representative values of parameters of the theory, i.e. the Yukawa coupling and mediator mass. In the left panel we show the results for both heavy and moderately heavy mediator masses of $m_\phi = 5\, m$ (in blue) and $0.5\,m$ (in red), for $g=3\, (2)$ in solid (dashed), respectively, including the effects of scalar density condensate. For the case of heavy mediator we recover the familiar BCS solution~\cite{Schmitt:2014eka}, see appendix~\ref{app:subsec:resgap}. In the case of the moderately heavy mediator, we obtain solutions which are parametrically different from the BCS case. We interpret this to be due to the fact that interactions are not point-like (or contact interaction) contrary to the standard BCS approximation. In particular, we show that the gaps can be substantially larger at relatively small densities. Nevertheless, the exponential fall off, typical of BCS gap solutions, is recovered at moderate densities. At very large densities or, in other words, in the relativistic limit, we find that the solutions to the gap equations do not depend on the mediator mass. For the whole range of DM densities, we find that the gap is always substantially smaller than the kinetic energy at the Fermi surface, see the dotted line labelled $\epsilon_F - m$. Here, we also clearly see the impact of the scalar density condensate at intermediate densities, where the solution behaves similarly to a relativistic system. For comparison, the  gray dotted lines show the evolution of the gaps neglecting the change of the effective mass $m_\ast$. In other words, the effect of the shift of the mass of the DM is to drive more rapidly the system into the relativistic regime.  

\begin{figure*}[htbp]
	\centering
	\captionsetup{singlelinecheck=false,justification=centerlast}
\includegraphics[width=0.48\textwidth]{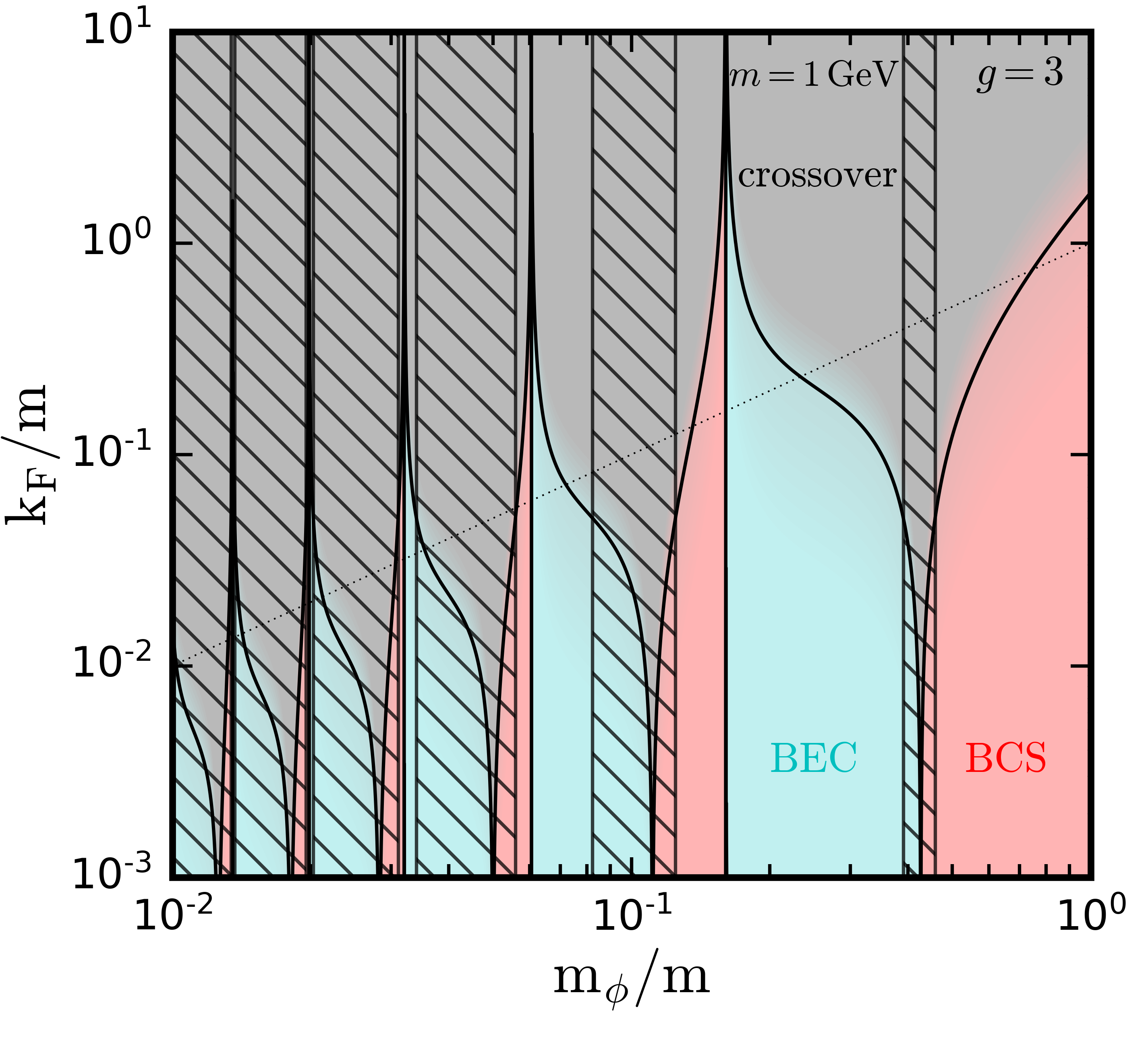}
\includegraphics[width=0.48\textwidth]{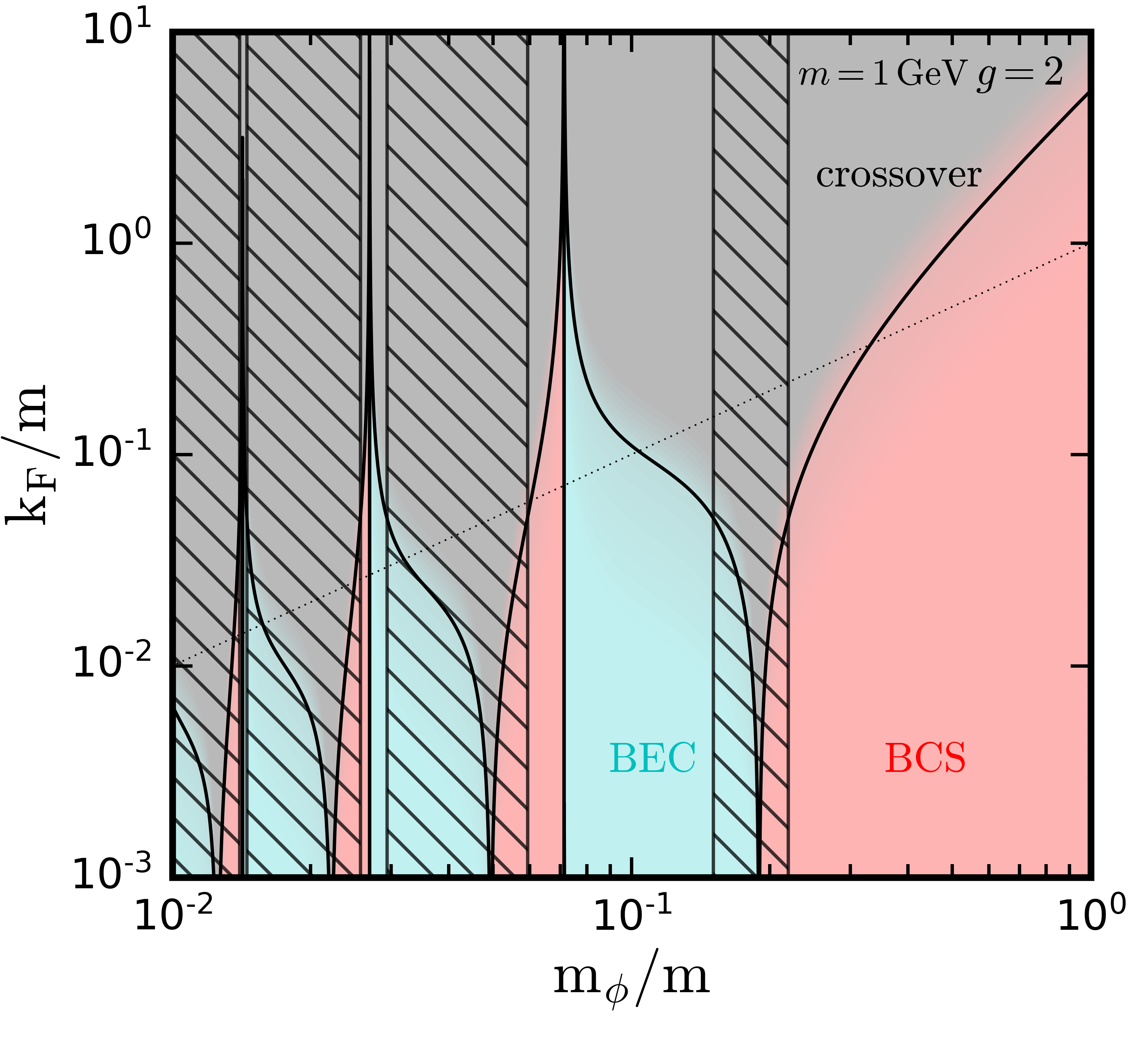}
	\caption{Contours of $(k_F a)^{-1}$ as a function of Fermi momentum and mediator mass. We have set $m =1$ GeV. Color code is the same as fig.~\ref{fig:scattering_length_regions}. The hatched shaded regions are excluded by bullet cluster limits on the self-interaction cross section of DM, $\sigma/m > 1\,{\rm cm^2/g}$.}
	\label{fig:scattering_length_bullet}
\end{figure*}

In the right panel we present the results for light mediator with $m_\phi = 0.13\, m$ for the same values of couplings as in the left panel. As expected, the solution in the high density regime behaves parametrically as in the left panel. However, at low densities the situation could be drastically different from the left panel depending on the value of $g$. This is best understood through the low density phase diagram of the Yukawa theory put forward in fig.~\ref{fig:scattering_length_regions}. For the case $g=3\,(2)$, corresponding to $\beta=5.5\,(2.4)$, we see that the system is in BCS (BEC) phase at low densities. As we increase the density, both the cases are in a crossover regime at $k_F \approx |a^{-1}| = 0.07\,m$. Interestingly, for this choice parameters the scattering length turns out to be the same in magnitude but opposite in sign. In light of these observations, we can now understand the low density regions shown in the right panel of fig.~\ref{fig:results}. For $g=2$, at low densities, we argue that the system is in the BEC phase. The solution yields the gap to be constant and much larger than $k_F^2/2m$. Although we get a solution for the gap, it does not represent a small perturbation to the Fermi surface, i.e. the chemical potential is no longer given by $\sqrt{k^2_F + m^2}$ but should evolve towards the binding energy of the would-be DM molecules~\cite{Nozieres:1985zz}. To obtain the correct solutions in this regime, we would need to simultaneously solve for both the gaps (and the scalar condensate) and for the chemical potential. This would require a more sophisticated analytical and numerical approach than the one we have considered, so we leave this particular situation for possible future works. 
Whereas, for $g=3$, at low densities the gap is exponentially suppressed, indicative of the non-relativistic BCS phase. As we approach densities close to $k_F \approx |a^{-1}| = 0.07\,m$, the system goes to the crossover regime; for which we do not present any solution and it is shown as the shaded region. Regardless, at very large densities, the system becomes relativistic (this is further enhanced by the decrease of the effective mass) and, as the formation of true bound state becomes impossible, the system makes a transition to the relativistic BCS phase, a feature which seems to be novel.

 \begin{figure*}[t]
	\centering
		\captionsetup{singlelinecheck=false,justification=centerlast}
\includegraphics[width=0.48\textwidth]{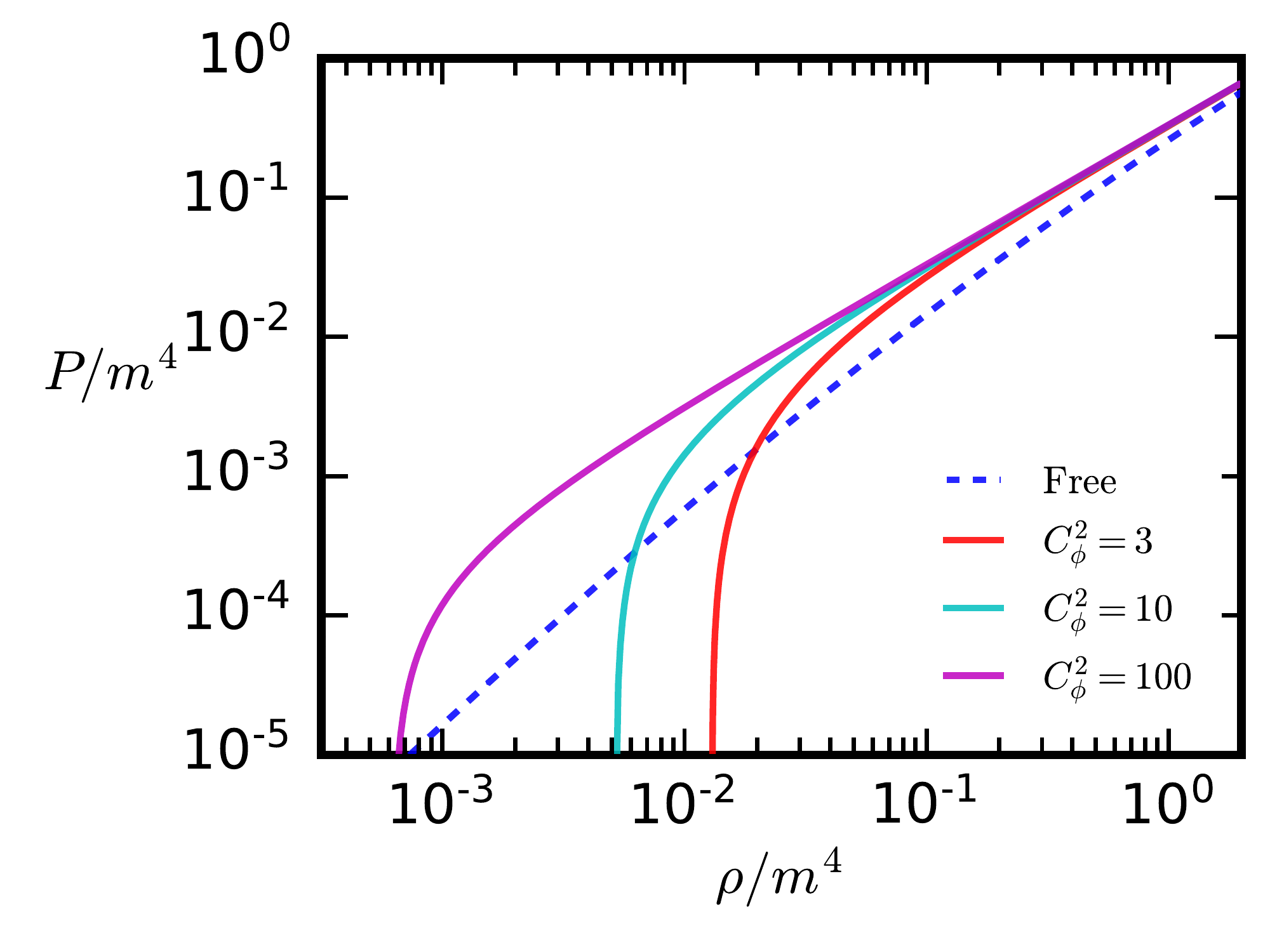}
\includegraphics[width=0.48\textwidth]{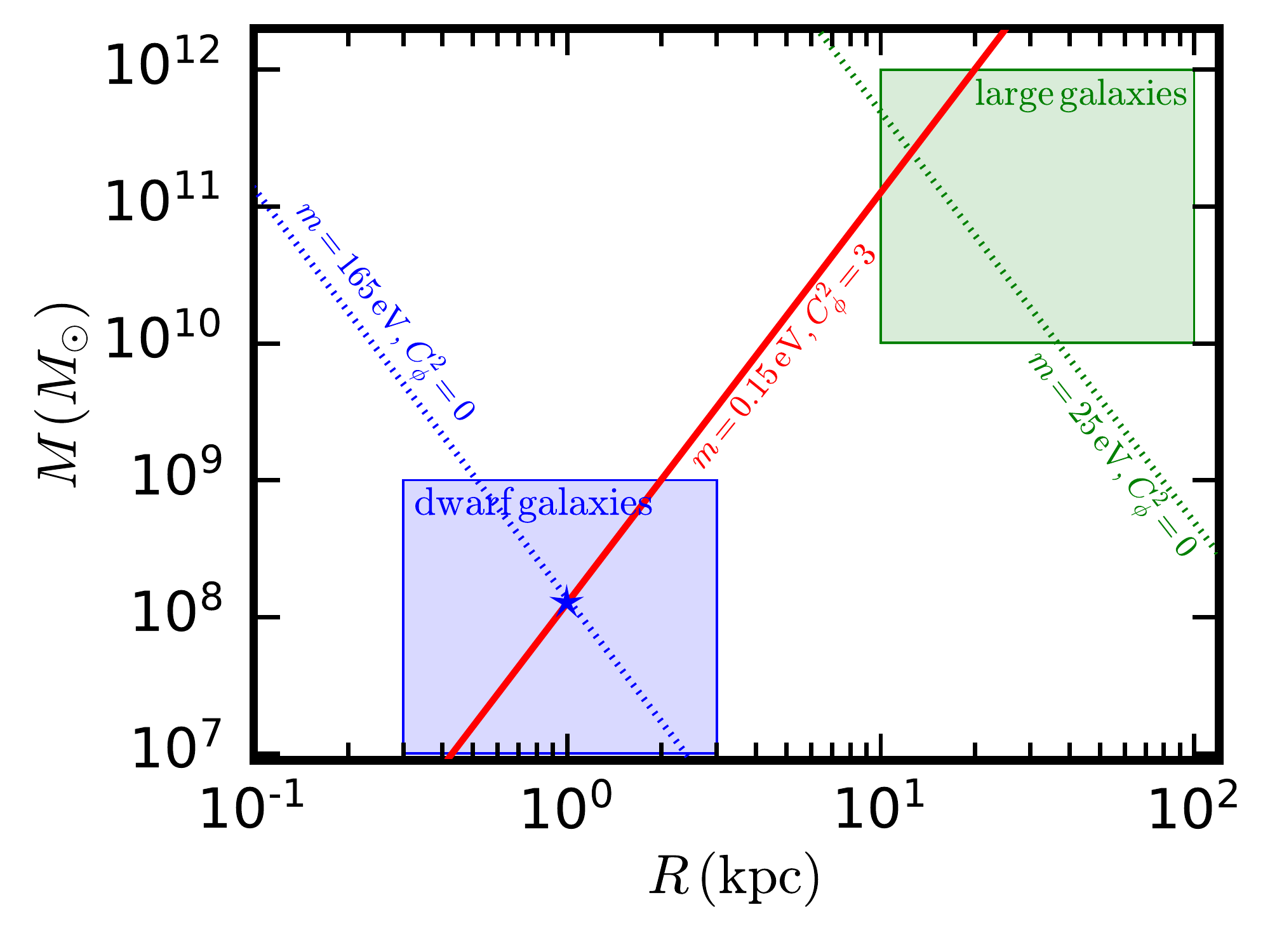}
	\caption{{\it Left panel:} Equation of state of a condensate-dominated system for different values of the effective coupling $C_\phi^2$ (solid). The dashed curve corresponds to the non-interacting case. {\it Right panel:} Mass-Radius relationship of self-gravitating configuration of interacting, condensate-dominated (solid) and non-interacting (dashed) systems for different DM bare masses. The blue (green) shaded regions correspond to the typical size and mass of the DM halo of dwarf spheroidal (large) galaxies \cite{Domcke:2014kla}. The blue star indicate the benchmark point $M= 10^8 M_\odot$ and $R=1 \text{ kpc}$, representative of a typical dwarf spheroidal galaxy.}
	\label{fig:EoSMR}
\end{figure*}

\section{Astrophysical constraints and implications}
\label{sec:astro}

{\it DM self interaction constraints.} -- In the context of DM self interactions, 
 in fig.~\ref{fig:scattering_length_bullet}, we show the phase diagram of the Yukawa theory in the $k_F/m - m_\phi/m$ plane for a dark matter candidate of mass $m=1$ GeV for $g=3$ (left) $g=2$ (right). The gray dotted line corresponds to $k_F=m_\phi$. We overlay the constraints on the dark matter self-interaction cross section at the scale of the Bullet cluster, requiring $\sigma/m \lesssim 1 \text{ cm}^2/\text{g}$ at a velocity of $v=2000 \text{ km}/\text{s}$. For such DM candidate, we find that $a \,m > 20$ is excluded. The corresponding excluded mediator mass range $m_\phi$ is shaded in gray. As could be expected, the unitarity regime are excluded, as is most of the very light mediator regime. The very fine viable intervals of $m_\phi$ correspond to vanishing self-interactions, $i.e.$ $a\rightarrow 0$. This nicely illustrates the possibility for a dark sector to manifest emergent phenomenon like superfluidity, while
 being not entirely excluded by self interaction constraints on DM.

{\it DM in neutron stars.} -- We now turn to discuss the impact of superfluid gaps on the fate of captured asymmetric DM in the core of neutron stars (NS). More precisely, DM particles in the galactic halo can be efficiently captured by NS~\cite{Goldman:1989nd,Kouvaris:2011fi}. For any given neutron star of mass $M$ and radius $R$, the maximal rate of DM accretion is given by the so-called geometric rate corresponding to scattering cross section $ \pi R^2/N \sim 10^{-45}$ cm$^2$, where $N$ is the total number of target neutrons in the NS. The maximum number of accumulated DM in the NS over its life time of 10 Gyrs is given by 
 \beq
 N_{acc}\approx 10^{42}\frac{\rho}{0.4\, {\rm GeV/cm^3}}\frac{1\, {\rm GeV}}{m}. 
\eeq
After DM thermalization with the medium~\cite{Bertoni:2013bsa,Garani:2018kkd,Garani:2020wge,Garani:2021feo}, the DM thermal sphere can collapse to a black hole. For non-interacting fermion DM particles, this happens as soon as the number of accumulated particles exceeds the Chandrasekhar limit given by $N_{\rm ch}\approx M^3_{\rm pl}/m^3$. 

In ref.~\cite{Kouvaris:2018wnh}, it was suggested that in the presence of attractive self-interactions, such as in the Yukawa theory, the above Chandrasekhar limit
could be parametrically reduced to $\sim (m_\phi/m\sqrt{\alpha})^3 N_{\rm ch}$. Soon after, ref.~\cite{Gresham:2018rqo} pointed out that as the DM thermal sphere shrinks the DM density increases therefore the emergence of scalar density condensate should be accounted for consistently in the calculation. Therefore, DM particles become relativistic at smaller densities than in the free case and the Chandrasekhar limit is not parametrically modified.

In this section we comment on the impact of superfluid gaps on the above scenario. To this end, the order parameter $\Delta$ should be evaluated. As an example, we choose the model parameters $m =200$ GeV, $m_\phi=1$ MeV, and $\alpha = 10^{-3}$, as exemplified in ref.~\cite{Kouvaris:2018wnh}. We find that at low densities, for these parameters, the system finds itself in the BEC phase, and consequently the gaps are much larger than the chemical potential. However, at higher densities, fermions become relativistic and the system transitions to the relativistic BCS phase. At this stage we can estimate the values of the gap using our relativistic BCS analytical approximation given by $\Delta\approx \frac{4}{3}k_F\exp\left(-\frac{8\pi^2}{g^2}\right)\exp\left(-\frac{4\pi^2 m_\phi^2}{g^2 k_F^2}\right) $ (see also appendix~\ref{sec:solgaps}, eq.~\eqref{geq:bcs_rel_rp}). This yields the numerical value of the gap to be exponentially suppressed and thus negligible. While the BEC phase may change the equation of state at low DM densities~\cite{Nozieres:1985zz}, we conclude that at large densities, in the BCS phase this has little impact on the equation of state of the DM with a Yukawa interaction and thus the conclusions of~\cite{Gresham:2018rqo} are essentially unchanged.

{\it Condensed DM Halos.} -- The possibility that DM may be in a degenerate fermionic gas  phase and form a core at the center of galaxies and dwarf galaxies has been put forward in the literature, see e.g.~\cite{Domcke:2014kla,Randall:2016bqw,Bar:2021jff}, albeit for non-interacting particles. More recently, possible superfluid phase for DM in the context of dwarf galaxies was studied in ref.~\cite{Alexander:2020wpm}, in a model of ultra-light dark-QCD matter parameterized by rescaling the known QCD CFL phase~\cite{Alford:2007xm}. It is well known that the Yukawa theory shares qualitative features at finite densities with quark matter~\cite{Pisarski:1999av}. We now consider the possibility that DM particles that make up the halos of galaxies (e.g. dwarf spheroidal galaxies) could manifest emergent phenomena described in our work, in a simpler model such as the Yukawa theory. We first present the equation of state (EoS) for the Yukawa theory\footnote{See the description of the so-called \enquote{liquid phase} in appendix~\ref{sec:phase}} and then showcase a possible realization of condensed DM halos at galactic scales.

In the left panel of fig.~\ref{fig:EoSMR} we present the EoS in dimensionless units, for the case of free degenerate fermions (dashed), and for attractive self-interactions (solid), see also appendix~\ref{sec:phase} for further details. The strength of interaction in this case is most conveniently characterized by the parameter $C^2_\phi = 4/3 \,\alpha m^2/(3 \pi^2 m^2_\phi)$. The curves shown in red, cyan, and magenta, correspond to values of $C^2_\phi = 3, 10$ and $100$, respectively.

At relativistic densities ($\rho \gg m^4$), each of the EoS obtained in the presence of attractive interactions reproduce that of the relativistic Fermi gas, $P=\rho/3$. However, in the non-relativistic regime ($\rho \ll m^4$) significant departures from the free case appear.
Due to the emergence of the scalar density condensate, at large values of $C^2_\phi$, $m_\ast \ll m$, and hence the relation $P\sim\rho$ is valid down to densities $\rho\sim m^4_\ast$. Additionally, the presence of attractive self-interactions lead to the appearance of a critical saturation density analogous to the nuclear saturation density, at which $P$ vanishes for a finite $\rho$. Close to the saturation density, the EoS is very stiff and the system becomes incompressible. Interestingly, for such densities, the pressure of the interacting gas is much smaller than that of the free case. This would lead to more compact self-gravitating configurations of the DM cloud, when compared to the free case. It is also seen that, as the system contracts and $\rho$ becomes larger (for example during its evolution towards gravitational collapse) the pressure quickly rises above the values of the free case, hence hindering collapse, as described above.

In the right panel of fig.~\ref{fig:EoSMR}, we present the mass-radius relationship for a few representative cases depicted in the left panel. We have obtained these configurations by solving the Tolman-Oppenheimer-Volkoff equations with the EoS presented in the left panel. Particularly, we have considered the free degenerate EoS ($C_\phi^2=0$) and the mildly interacting EoS ($C_\phi^2=3$). As is well known, for free degenerate fermions $M\propto R^{-3}$~\cite{Domcke:2014kla,Randall:2016bqw,Alvey:2020xsk}, and such DM candidates cannot simultaneously reproduce typical DM halos of both dwarf  spheroidal and large galaxies. However, if DM has sufficient attractive self-interactions, as exemplified in the Yukawa theory, stable solutions exist for incompressible configurations  which behave as $M\propto R^3$. An estimate of the $M-R$ relationship is given by
\beq
\left(\frac{M}{10^8 \, \rm M_\odot} \right) \left(\frac{\rm kpc}{R} \right)^3 \approx 4500 \left(\frac{1}{C^2_\phi}\right) \left(\frac{m}{\rm eV} \right)^4.
\eeq
As shown in the figure, such configurations could lead to halo masses and sizes that are both typical of dwarfs (blue shaded regions) and large galaxies (green shaded regions). Due to their incompressible nature, both halos would exhibit cored profiles. It is interesting to note that perhaps only the inner cores of halos are in such configurations, while the outer regions could be at temperatures high enough for the scalar condensate to not play any role. 
A comprehensive analysis of such phenomena could, for example, be captured by a piece-wise EoS correctly interpolating between the cold, compact core and a hot, less dense outer halo.  

Light fermionic dark matter candidates are clearly challenging to reconcile with the formation of cosmic structures. For $m<2\text{ keV}$, the Fermi pressure is itself sufficient to modify substantially the matter power spectrum \cite{Carena:2021bqm,Bar:2021jff}. Nonetheless, if the DM particles are bound in di-fermionic molecule, for example in the BEC phase, at early time, such bounds could  be alleviated.

\section{Conclusion and perspectives}
\label{sec:con}

We have developed a formalism to study possible phases in the Yukawa theory.
To this end, we have brought together concepts originating from the theory of non-relativistic scattering, the physics of high density nuclear matter, and phenomena that are realized in condensed matter systems, and have presented our findings in the context of particle dark matter at finite density. By computing the scattering length in the Yukawa theory we have found that DM could be in BCS, BEC or crossover phase, depending on the parameters of the theory at low densities. We have further explored the BCS phase. To do so we have computed superfluid energy gaps, including the effects of the scalar density condensate; by retaining the mediator mass dependence and deriving the consistent set of UV finite gap equations for the three most dominant pairing channels. We have found that at large densities (or ultra-relativistic limit) DM finds itself in the BCS phase, irrespective of the phase realized at low densities.

As an application of our formalism we have examined whether captured asymmetric fermion DM in the core of neutron stars can collapse to a black hole in models with attractive DM self-interactions, for DM mass $\sim$ GeV. Even in the presence of large attractive interactions, it was pointed in ref.~\cite{Gresham:2018rqo} that collapse to a black hole is impeded due to the emergence of scalar density condensate, and that the Chandrasekhar limit remains the same parametrically as that for free degenerate fermions. In this work, we have studied the interplay between the scalar density condensate and superfluid gaps consistently. We have found that further accounting for the superfluid gaps does not qualitatively change the conclusions drawn in ref.~\cite{Gresham:2018rqo}, since the gap contribution to the pressure is small and positive, with scaling $\propto \mu^2 \Delta^2$. 

By consistently computing the EoS in the Yukawa theory, we have also explored the possibility that halo DM could realize cored density profiles due to emergent density effects, at both small and large scales, corresponding to dwarf galaxies and other large galaxies. We have found that the impact of the scalar density condensate is important and dominates over that of superfluid gaps.

There are various directions for further phenomenological enquiry. In this work we have focused on identifying the possible phases of condensed DM with a Yukawa interaction, on self-consistently determining the gaps and the scalar condensate and the impact of the latter on the energetics of the system. The next natural step would be to examine the transport properties of such systems. On astrophysical scales high density regions of DM are found, e.g. in dwarf spheroidal galaxies, or at the center of more massive galaxies. 
Examining velocity dispersion data of several Milky-Way dwarf galaxies would in principle provide a window to determine the phase of DM. Moreover, since superfluidity is realized through breaking of the global $U(1)$ we expect formation of strings and vortices. Observations of galaxy-galaxy mergers and/or dwarf galaxy-host galaxy mergers might provide a way to probe DM phases. Finally, depending on how dark sector couples to the visible sector dark stars or hybrid stars could exist in the Universe. Gravitational wave observations of mergers of such objects with black holes, neutron stars, or white dwarfs, would pave the way in testing the scenario through waveform template fitting.

\section*{Acknowledgments}
 We thank Chris Kouvaris and Peter Tinyakov for participation at the early stages of this work. 
 We thank Diego Aristizabal , Camilo Cely, Xioyang Chu, Malcolm Fairbairn, Nathan Goldman, Thomas Hambye, Josef Pradler, Michele Redi, and Andrea Tesi for discussions.
 R.G. is supported by MIUR grant PRIN 2017FMJFMW and acknowledges the Galileo Galilei Institute for hospitality during part of this work. The work of M.T. is supported by the F.R.S./FNRS under the Excellence of Science (EoS) project No. 30820817 - be.h “The H boson gateway to physics beyond the Standard Model” and by the IISN convention No. 4.4503.15. The work of J.V. is supported by the Collaborative Research Center SFB1258 and by the Deutsche Forschungsgemeinschaft (DFG, German Research Foundation) under Germany’s Excellence Strategy - EXC-2094 - 390783311. J.V. acknowledges INFN Florence for hospitality during completion of this work.

\normalem
\bibliographystyle{apsrev4-1}
\bibliography{biblio}

\noindent
\appendix
\maketitle
\onecolumngrid 

\section{Scalar condensate}\label{app:walecka}

The scalar condensate at finite density $n_s = \langle \bar \psi \psi\rangle$ has peculiar properties \cite{Walecka:1974qa}. In particular, it leads to the vanishing of the effective fermion mass at large densities. 

 A simple application of the decomposition of $\psi$ in terms of creation/destruction operators gives
\begin{align}\label{eq:scden_gap0}
n_s = {m\over 2 \pi^2}\left[ k_F E_F - m^2 \log\left((k_F + E_F)/m\right)\right]~, 
\end{align}
for $k_F \ll m$, $n_s \approx n =\langle \bar \psi \gamma^0 \psi\rangle = {k_F^3/3 \pi^2} $ while for $k_F \gtrsim m$, $n_s  \approx m {k_F^2/2 \pi^2}$. This scalar condensate leads to a modification of the fermion mass through $m_\ast = m - g^2/m_\phi^2 n_s(m)$. Replacing $m$ by $m_\ast$ in the argument of $n_s$ gives a self-consistent, mean-field approximation for $m_\ast$. 
Alternatively, one can start from the grand partition function of the Yukawa theory eq.~\eqref{eq:lagr1} 
\beq
Z =  e^{-\beta V \Omega}~,
\eeq
where $\Omega$ is the (Landau) thermodynamical potential $\Omega = -p(T,\mu)$. At non-zero temperature $\beta = 1/T$, integrating over the fermions and replacing the field $\phi$ by a constant mean-field value $\phi_0$ gives
\beq
\Omega = -\int {d^3p\over (2 \pi)^3} \left\{  (\vert \omega+\mu\vert + \vert \omega - \mu\vert) + {2\over \beta} \ln (1 + e^{-\beta\vert\omega- \mu\vert}) +  {2\over \beta} \ln( 1+ e^{-\beta \vert\omega+\mu\vert})\right\}+ {1\over 2} m_\phi^2 \phi_0^2~.
\eeq
One recognizes contributions from fermions and antifermions. Note also, $\omega = \omega_k(m_\ast)$ with $m_\ast = m - g \phi_0$. In the limit $T\rightarrow 0$, the integral reduces to the first term which can be written as
\beq
\Omega = -{2 \over 2 \pi^2}
\left\{ \int_0^\infty dp\, p^2\, \omega + \int_0^{p_F} dp\, p^2\, (\mu - \omega)\right\} + {1\over 2} m_\phi^2 \phi_0^2~.
\eeq
There is an infinite contribution from vacuum which can be renormalized away, see below. Then, minimizing $\Omega$ with respect to $\phi$ gives
\beq
\phi_0 = {g\over m_\phi^2} {1\over \pi^2}  \int_0^{k_F} 
dk k^2 {m_\ast\over \omega} = {g \over m_\phi^2} n_s(m_\ast) ~.
\eeq
Within the same mean-field approximation, the energy density of the Yukawa gas is written as 
\beq
\epsilon = \epsilon^\ast_{FG} + {1\over 2} m_\phi^2 \phi_0^2~,
\eeq
with the free gas contribution 
\beq
\label{eq:energydensity}
\epsilon_{FG}^\ast  = 2 \int_{k < k_F} {d^3 k \over (2 \pi)^3} \sqrt{k^2 + m^{2}_\ast}
 = {1\over 8 \pi^2} \left[ k_F E_F^\ast (E_F^{\ast 2} + k_F^2) - m^{ 4}_\ast \ln\left((k_F + E_F^\ast)/m_\ast\right)\right]~,
\eeq
and the Fermi energy $E_F^\ast = \sqrt{k_F^2 + m_\ast^2}$. Now we examine the energy density above. In the non-relativistic limit, $k_F \ll m $, this gives
\beq
\epsilon \approx m_\ast  {k_F^3\over 3 \pi^2} +  {k_F^5\over 10 m \pi^2} + {1\over 2} m_\phi^2 \phi_0^2~,
\eeq
with $m_\ast \approx m - g^2 k_F^3/(3\pi^2m_\phi^2)$ and $\phi_0 \approx g k_F^3/(3\pi^2m_\phi^2) $, so
\beq
\epsilon \approx m  n - {g^2 \over 2 m_\phi^2} n^2 +  {k_F^5\over 10 m \pi^2}~. 
\eeq
Notice that a factor $1/2$ comes from the $1/2 m_\phi^2 \phi_0^2$ term. 
This reproduces the leading term derived in~\cite{Kouvaris:2018wnh} from the potential energy of a sphere of $N$ particles with a short range attractive interaction, 
\beq
\epsilon = \epsilon_{FG} - {9 \alpha\over 8 m_\phi^2} \left({N\over R^3}\right)^2 ~,
\eeq
here $R$ is the radius of the sphere, $\epsilon_{FG}$ is the free Fermi gas energy density, and $\alpha = g^2/4\pi$. At very small $k_F$, the energy density $\epsilon \propto n$. If $g m/m_\phi$ is large enough, it becomes negative. However (at large densities) when
\beq
n \sim {m_\phi^2 m /g^2}~,
\eeq
$m_\ast$ approaches zero. The energy density is then as a relativistic gas, 
\beq
\epsilon \approx {k_F^4\over 4 \pi^2}~.
\eeq

We conclude this appendix with a brief note about the vacuum contribution to $n_s$. The above discussion is for a renormalized $n_s$ so that $n_s = 0$ for $T$ and $\mu = 0$. The vacuum contribution is potentially infinite being given by 
\beq
\langle \bar \psi \psi\rangle_0 = - 2 m \int {d^3 k\over (2 \pi)^3}\, {1\over \omega(k)} \sim - m \Lambda^2 ~,
\eeq
for some cut-off scale $\Lambda^2$. This leads to a negative $n_s$ and so a positive contribution to the fermion mass. Following the logic of above leads to the NJL mechanism of chiral symmetry breaking, in which case the effective dressed mass is 
\beq
m_\ast = m - G \langle \bar \psi \psi \rangle_0  \geq 0~,
\eeq
where $G$ is some effective coupling (dimension $1/M^2$). The chiral limit is $m=0$ and a scalar condensate contributes dynamically to yield a non-zero fermion mass, $m_\ast \propto \langle \bar \psi \psi\rangle_0$. 
At finite density, the contribution to $n_s$ is positive and non-zero if and only if $m\neq 0$, i.e., the opposite to the vacuum case so to speak.

\section{Scattering in Yukawa theory}\label{app:a}

The scattering length is a basic quantity used to study the onset of BEC and, more generally, low energy scattering processes. It is less commonly encountered in DM literature, see however \cite{Chu:2019awd} for applications to DM self-scattering in galaxies and dwarf galaxies. It is particularly useful to characterize in simple terms complicated, possibly strongly interacting scattering systems in terms of a physical observable such as the scattering length. 

The wave-function in the CM frame of two non-relativistic particles is of form
\beq
\psi = e^{i kz} + f(\theta) {e^{ikr} \over r}~,
\eeq
with scattering amplitude $f(\theta)$ and differential cross section $d\sigma/d\Omega = \vert f(\theta)\vert^2$. In the case of s-wave scattering, which is dominant for many systems (e.g. a pair of identical fermions in a singlet spin configuration), the scattering amplitude is constant $f(\theta) = f_0$. In terms of the s-wave phase shift $\delta_0$, 
\beq
f_0 = {1\over 2 i k}(e^{2 i \delta_0} - 1)~.
\eeq
Then, for low energy scattering $k \rightarrow 0$, the wave function can be written as
\beq
\psi \approx 1 - {a\over r}~,
\eeq
where the scattering length ($a$) is given by, 
\beq
\lim_{k\rightarrow 0} k \cot \delta_0(k) = - {1\over a}~.
\label{eq:sl}
\eeq
In this limit $\sigma = 4 \pi a^2$.
 For $a>0$, the vanishing of the wave-function signals the possibility of bound state formation \cite{Landau:1991wop}. For instance, for a neutron-proton (n-p) pair in a spin $S=1$ triplet state, the scattering length is $a \approx +7$ fm corresponding to the formation of a deuteron bound state. There is no n-p bound state with in the spin singlet state $S=0$, but the s-wave scattering is very large ($\sim$ resonant) at low energies, corresponding to $a\approx - 20$ fm. A resonance corresponds to a phase shift $\delta_0(k_{\rm res})= \pi/2$ and through eq.~\eqref{eq:sl}, to a change of sign of the scattering length. Formally, the scattering length is infinite if $\delta_0\rightarrow \pi/2$ as $k\rightarrow 0$. In the condensed matter literature such systems are called unitary Fermi liquids.
Similarly, for a neutron-neutron (n-n) interactions in the s-wave, $a \approx - 17$ fm; the nuclear interaction between two neutron is attractive, but not large enough to form a real 2-body bound state.

The relevance of all of this is that an attractive interaction, for instance through a Yukawa coupling may, depending on the parameters of the theory,
lead or not to formation of bound state at low energies. Thus corresponding to a BEC or a BCS state, respectively, see e.g.~\cite{Schmitt:2014eka}. In the following we will reproduce the results of scattering length in Yukawa theory by solving the Schr\"odinger equation (following ref.~\cite{Chu:2019awd}) and use these findings to study phases of Yukawa theory.

 The interaction Yukawa potential in co-ordinate space takes the form
 \beq
 V(r) = \pm \alpha \frac{e^{-m_\phi r}}{r},
 \eeq
 where $\alpha =g^2/(4\pi)$ and $m_\phi$ is the mass of the force mediator. The phase shifts due to DM self scattering are obtained by solving the Schr\"odinger equation for the radial wavefunction ($R_{l,k}$) through the equation
 \beq
 \frac{1}{r^2} \frac{d}{d r} \left(r^2 \frac{d R_{l,k}}{d r} \right) + \left(k^2 - \frac{l(l+1)}{r^2} - m\,V(r) \right) R_{l,k} =0~,
 \eeq
with boundary condition $r R_{l,k}= 0$ at $r=0$. As discussed in detail in ref.~\cite{Chu:2019awd}, with change of variables, the above equation could be re-expressed in terms of spherical Hankel function of the first kind $h^{(1)}_l$. This results in the following equation for the phase shift
 \beq
 \frac{d \delta_{l,k}(r)}{dr} = - k\, m \,r^2 \,V(r) \,{\rm Re}\left[e^{i \delta_{l,k}(r)} h^{(1)}_l (k r) \right]^2~,
 \eeq
 which is to be solved with boundary conditions $\delta_{l,k} (0) =0$ and $\delta_{l,k} \rightarrow \delta_l$ at $r \rightarrow \infty$. Having determined the phase shift, the s-wave scattering length is obtained by
 \beq\label{eq:scatl_yuk}
 a = - \lim_{k\rightarrow0} \frac{\tan \delta_l}{k}.
 \eeq

 For comparison we also compute the scattering length for Huelthen potential. The interaction potential in co-ordinate space in this case reads
 \beq
 V(r) = \pm \alpha \frac{e^{-m_\phi r}}{r}\sim \pm \alpha\,\delta \frac{e^{-\delta r}}{1 - e^{-\delta r} }~,
 \eeq
 with $\delta = \sqrt{2\zeta(3)} \,m_\phi$ ~\cite{Cassel:2009wt}. The s-wave scattering length for this potential can be derived by solving the Schr\"odinger equation as above. However, for the Huelthen potential analytical solution exists for the scattering length and the effective range of the interaction~\cite{Cassel:2009wt,Chu:2019awd,Mahbubani:2020knq}, which are given by
 \bea\label{eq:scatl_huethen}
 a&=&  \frac{1}{\delta} \left(\psi^{(0)}(1+\eta) + \psi^{(0)}(1-\eta) +2\gamma  \right) ~,\\
 r_e &=& \frac{2\, a}{3} -\frac{1}{3\,\delta^3 \,\eta\, a^2}\left(3\left[\psi^{(1)}(1+\eta) - \psi^{(1)}(1-\eta) \right]    + \eta \left[ \psi^{(2)}(1+\eta) + \psi^{(2)}(1-\eta)+ 16 \zeta(3) \right] \right)  ~.
 \ena
 Here, $\eta = \sqrt{\alpha \,m/\delta}$ and $\psi^{(n)}(z)$ are polygamma functions of order $n$ and $\gamma$ is Euler-Mascheroni constant. 

\begin{figure}[H]
	\centering
\includegraphics[width=0.48\textwidth]{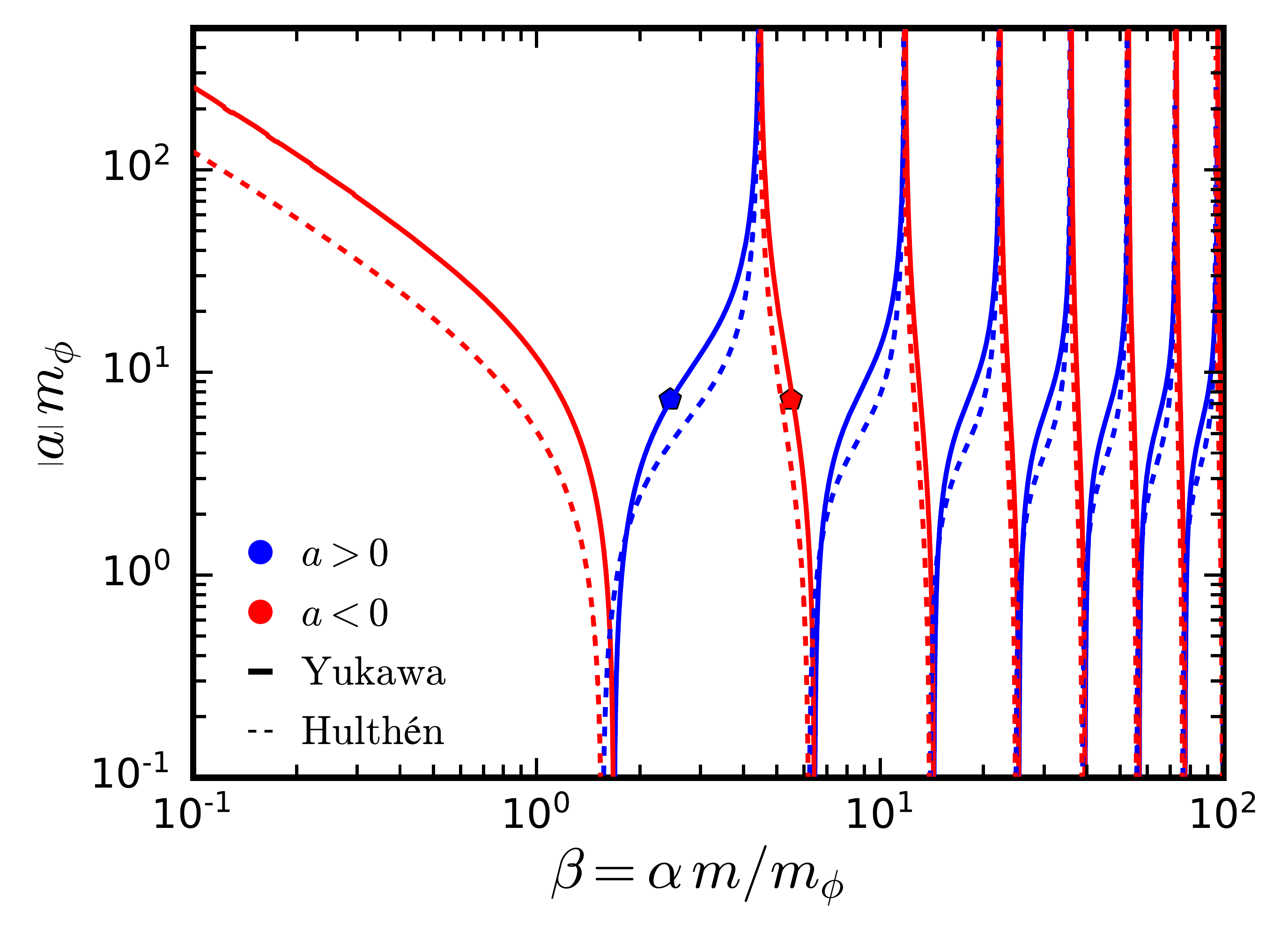}
\includegraphics[width=0.48\textwidth]{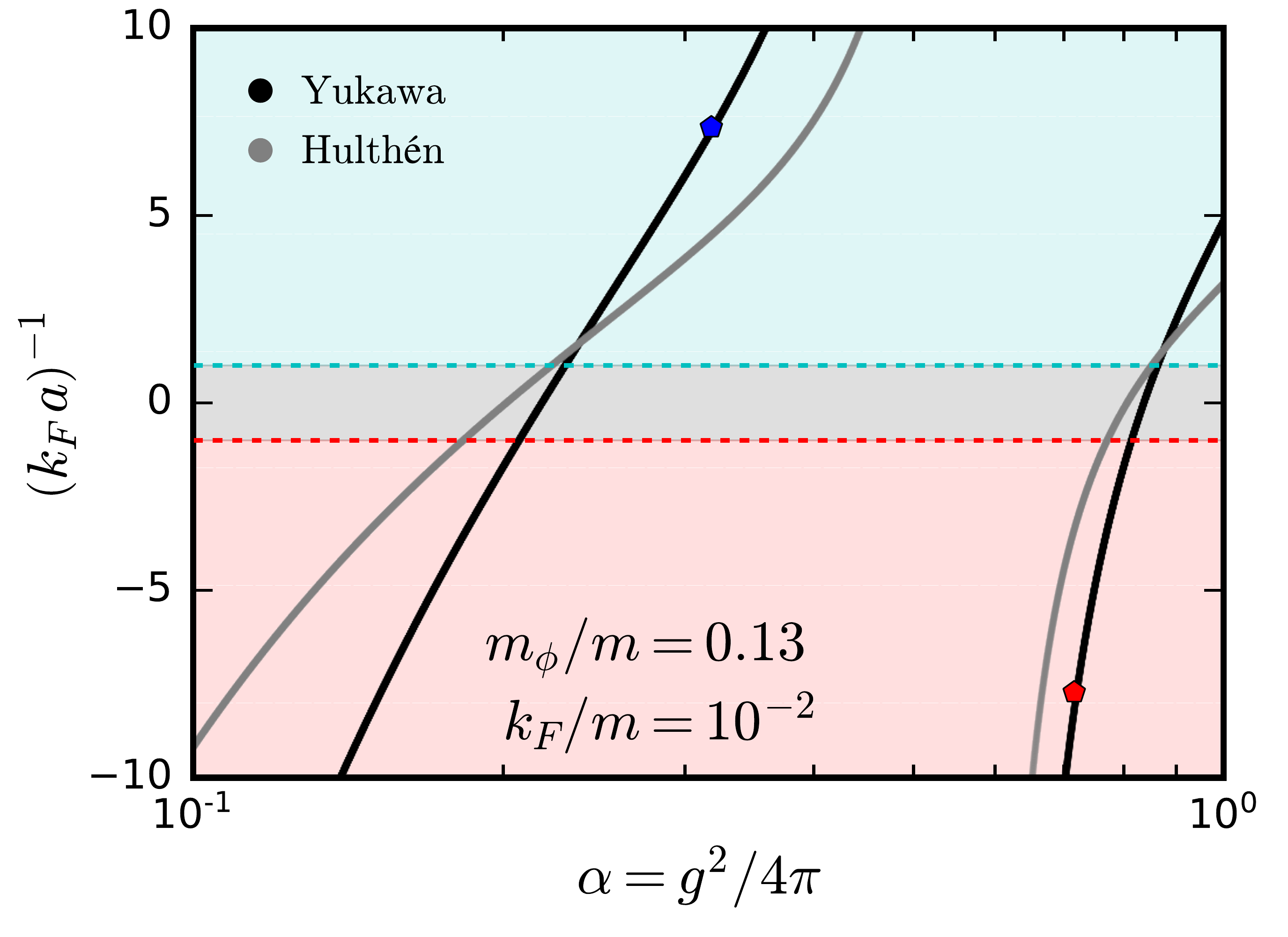}
\caption{{\it Left panel:} scattering length is shown as a function of dimensionless parameter $\beta$ for Yukawa and Huelthen potential in solid and dashed lines, respectively. {\it Right panel:} we show $(k_F\, a)^{-1}$ as a function of $\alpha$ for specific values of $k_F/m = 10^{-2}$, $m_\phi /m = 0.13$. The red regions indicate BCS phase while the blue regions indicate BEC phase. }
\label{fig:scattering_length_HultYuk}
\end{figure}

 In fig.~\ref{fig:scattering_length_HultYuk} (left) we present the scattering length in units of the mediator mass ($m_\phi$) as a function of dimensionless variable $\beta = \alpha m/m_\phi$, for Yukawa (Huelthen) potential in solid (dashed) lines. Blue (red) color indicates instances where the scattering length is positive (negative). As expected, we find that the Huelthen potential could be a good approximation for the Yukawa potential.
 In fig.~\ref{fig:scattering_length_HultYuk} (right) we show $1/(k_F \,a)$ as a function of $\alpha$, for the case $m_\phi/m = 0.13$ with $k_F/m = 10^{-2}$, for Yukawa (Huelthen) potential in black (gray) colors. We conclude that the Huelthen potential is qualitatively accurate enough to describe the unitary limit. 

\subsection{BCS-BEC crossover}

We end this appendix by recalling the well known phenomenon of BCS-BEC crossover in the limit of contact interactions. This transition is best parameterized in terms of the s-wave scattering length of the theory in the non-relativistic limit at finite density. For such a system, within the constant gap approximation, the equation for energy gap and the chemical potential for given a scattering length should be solved simultaneously~\cite{Schmitt:2014eka}. The gap equations take the form
\beq\label{eq:delmu}
-\frac{1}{k_F \,a} =  \frac{2}{\pi} \left(\frac{2}{3\, I_2\left(\frac{\mu}{\Delta}\right)} \right)^{1/3}  I_1\left(\frac{\mu}{\Delta}\right) \quad \quad{\rm and} \quad\quad \frac{\Delta}{E_F} = \left(\frac{2}{3\, I_2\left(\frac{\mu}{\Delta}\right)} \right)^{2/3}~,
\eeq
with
\beq
I_1(z) = \int^\infty_0 d x\, x^2 \left( \frac{1}{\sqrt{(x^2 -z )^2 +1}} - \frac{1}{x^2}\right)\quad \quad {\rm and} \quad \quad I_2(z) = \int^\infty_0 d x\, x^2 \left(1 - \frac{x^2 -z}{\sqrt{(x^2 -z)^2 +1}} \right)~.
\eeq

\begin{figure}[h]
	\centering
\includegraphics[width=0.48\textwidth]{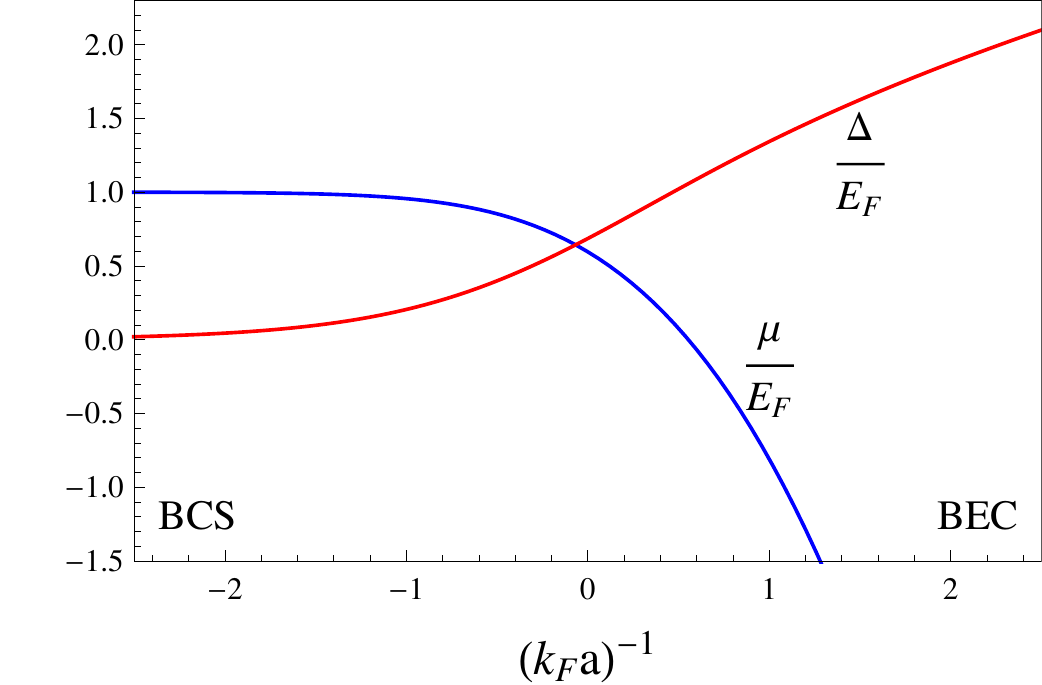}
\caption{BCS-BEC cross over for contact interactions. Reproduced from ref.~\cite{Schmitt:2014eka}. }
\label{fig:sch_cross}
\end{figure}

There two classes of asymptotic solution to eq.~\eqref{eq:delmu} depending on the sign and magnitude of the dimensionless parameter $k_F \,a$. For large and negative values of $(k_F a)^{-1}$, the chemical potential $\mu$ is close to the Fermi energy and the gap is exponentially small as could be expected in the BCS theory. For large and positive values $(k_F a)^{-1}$, the parameter $\mu$ receives large corrections making it negative, which indicates the formation of di-fermionic molecules and their condensation, often referred to as the BEC phase. While the limit $(k_F a)^{-1} \rightarrow 0$ corresponds to the unitary limit, where the phase of matter is neither in BCS nor in BEC phases, called the crossover regime. Quantitatively, $(k_F a)^{-1} < -1$ delimits well the BCS phase from the crossover. The BEC and crossover regimes are similarly delimited by the inequality $(k_F a)^{-1} > 1$~\cite{Matsuo:2005vf,Schmitt:2014eka}. The BEC-BCS phase transition is illustrated in fig.~\ref{fig:sch_cross}. In theories with light mediators the interactions can nevertheless be considered to be short-range at low enough densities. Analogous to contact interactions, the scattering length in these theories can be used to understand the phase diagram of non-relativistic systems.

\section{Condensation in the Yukawa theory}
\label{app:sec:model}

Cooper pairing according to BCS theory in metals results from an extremely small effective attractive interaction between electrons sourced by background lattice of positively charged ions. In weakly coupled theories Cooper pairing is a collective phenomena in which the Cooper pairs are spatially separated at distance scales larger than the average separation between the constituent fermions of the system. In contrast to di-fermionic molecules that are bosonic, physically Cooper pairs are thought to be quasi-particles that could be called a boson. Formation of Cooper pairs could also be sourced by fundamental attractive interaction between fermions. 
To this end, we consider a simple model where fermion ($\psi$) interacts by the exchange of scalar boson ($\phi$) through an attractive Yukawa potential, given by
\beq\label{eq:lagr}
\mathcal{L} = \bar{\psi}(i\slashed{\partial} + \gamma^0 \mu -m) \psi + \frac{1}{2}\left(\partial^\mu \phi \partial_\mu \phi -m^2_\phi \phi^2 \right) - g \,\bar{\psi} \Gamma \psi \phi\,~,
\eeq
where $\Gamma= \mathbb{I}\, (i \gamma_5)$ for scalar (pseudo scalar) couplings. The chemical potential is denoted by $\mu$ and $g\,(>0)$ is the coupling constant. The bare fermion (boson) mass is denoted by $m$ ($m_\phi$). The fermion $\psi$ and the mediator $\phi$ are singlets under the SM gauge group. However $\psi$ is charged under a dark global $U(1)$ symmetry conserving dark fermion number. The spontaneous breaking of this global symmetry results in modes that propagate with energies smaller than Fermi energy, the so-called Cooper pairs. Consequently the system exhibits superfluidity (and not superconductivity) at finite density and temperature~\cite{Pisarski:1999av}. The Dirac structure is analogous to single-flavour, single-colour limit of QCD. This simple structure already captures essential features and dynamics of superfluidity that could arise in models that have particles charged under more non-trivial groups~\cite{Schmitt:2014eka}.

The study of fermion-fermion condensate (gaps) and fermion-anti-fermion condensate (scalar condensate) is well established through the computation of partition function $Z(\mu,T,V)$, within the mean-field approximation. The partition function and the action reads
\bea\label{eq:partition_org}
Z(\mu,T,V) &=& \int \mathcal{D}\bar{\psi} \mathcal{D} \psi \mathcal{D} \phi\, e^S\,,\\
S=\int\limits_x\mathcal{L}&=&\int \limits_{x,y}\left[\bar{\psi}\left(x\right)G_0^{-1}\left(x,y\right)\psi\left(y\right)-\frac{1}{2}\phi\left(x\right)D^{-1}\left(x,y\right)\phi\left(y\right)\right]-g\int\limits_x \bar{\psi}\left(x\right)\Gamma\psi\left(x\right)\phi\left(x\right), 
\ena
with the inverse propagators defined as $D^{-1}\left(x,y\right)=\delta^4\left(x-y\right)\left(\partial_\mu \partial^\mu + m_\phi^2\right)$ and $G_0^{-1}\left(x,y\right)=\delta^4\left(x-y\right)\left(i\slashed{\partial}+\gamma^0\mu-m\right)$, for scalar mediator and fermionic particles, respectively. We follow the standard finite temperature field theory notation for space-time integral, abbreviated as $\int_x \equiv \int^{1/T}_0 d\tau \int d^3 \bold{x}$~\cite{Kapusta:2006pm}.

 The model presented here is reminiscent of the so-called $\sigma-\omega$ model of dense nuclear matter interactions~\cite{Walecka:1974qa}. Broadly speaking, protons and neutrons feel a long-range attractive force, arising from the exchange of a scalar meson ($\sigma$), and a short-range repulsive force, due to exchanges of a vector boson ($\omega$). Thus, eq.~(\ref{eq:lagr}) corresponds to the case of purely attractive interactions, $i.e$ when the $\omega$ meson plays little role. The features of our \enquote{$\sigma$ model} often shares qualitative similarities that appears in the description of formation of large bound states of nuclei which appear beyond a particular value of saturation density, and with molecular interactions via van der Waals forces.

 We proceed by first integrating out the mediator ($\phi$). Since superfluidity is the phenomenon at the surface of the Fermi-sea the most relevant scale in the problem is identified to be the Fermi-momentum ($k_F$). Before we begin, however, it is instructive to ponder whether it is justified to integrate out the mediator also when $m_\phi$ is the smallest scale in the problem (i.e. $m_\phi \ll k_F$). The answer is provided in~\cite{Polchinski:1992ed}. For completeness we reiterate the arguments here and specify by what we mean by integrating out the mediator in the context of superfluidity, and the resulting computational advantage. The partition function in eq.~\eqref{eq:partition_org} sums over all possible configurations for the fields. Shifting the scalar field arbitrarily does not change any physical quantities. For example, we can shift the scalar field the following way: $\phi(x) \rightarrow \phi(x) - i \int_y F(y) D(y,x)$ with $F(x) \equiv g \,\bar{\psi}(x)\Gamma \psi(x)$. After integrating over all the field configurations of $\mathcal{D} \phi$ we get the new effective action for the fermions

\bea\label{eq:effaction}
S^\prime&=&\int \limits_{x,y}\bigg[\bar{\psi}\left(x\right)G_0^{-1}\left(x,y\right)\psi\left(y\right) + g^2 \, \bar{\psi}\left(x\right)\,\Gamma\,\psi\left(x\right)D(x,y)\bar{\psi}\left(y\right)\,\Gamma\,\psi\left(y\right)   \bigg]. 
\ena

 The effective approximation is reasonable to describe the formation of cooper pairs since the mediator $\phi$ is never produced on-shell around the Fermi surface. In other words the momenta exchanged between fermions that would form a cooper pair is always in the t-channel. Note that s-channel diagram is absent due to the presence of a finite chemical potential. The scattering is causal and space-like ($q^{2}<0$). Thus, for extremely large $m_\phi \gg k_F, m$ the bosonic propagator in momentum space scales simply as $m^{-2}_\phi$. Whereas, in the limit $m_\phi \ll k_F, m$ the propagator scales as $t^{-1}$. The exchange momenta can never be much larger than the Fermi momentum due to degeneracy, as the fermions very close to the Fermi surface effectively participate in scattering. In this work we keep the momentum dependence of the bosonic propagator explicitly and examine in detail how the superfluid gap depends on $m_\phi$ and momentum of the gaps. The above procedure clearly brings the fermion bilinears in quadratic form, thus easing the computation of partition function. 
Before we get into details in the next section, we first examine the possible combination of bilinear that can get expectation values.

\subsection{Scalar density condensate}

The number density of $\psi$ is conserved. Therefore the mediator $\phi$ (scalar) will couple to the number density. Diagramatically this is shown in fig.~\ref{fig:feynman_cond}. In terms of Wick contraction, we have
\beq
{g^2\over 2} \wick{ \c1 {\bar \psi}(x) \Gamma \c1 \psi(x) D(x,y)  \c2  {\bar \psi}(y) \bar{\Gamma} \c2 \psi(y)} \,,
\eeq
where the contractions are taken one at a time. This in effect modifies the free fermion propagator through the correction to the bare fermion mass. Thus the in-medium mass of the fermion could be written as $m_\ast = m - \Sigma$, with
\beq
\Sigma(x) = {g^2\over 2}  D(x,y) \left<\bar{\psi}(x) \Gamma \psi(x) \right> = \Sigma(y)={g^2\over 2}  D(x,y) \left<\bar{\psi}(y) \bar{\Gamma} \psi(y) \right>   \implies \text{ in medium correction to mass}\,, \nonumber
\eeq
where the minus sign appears due to the fermion loop (see fig.~\ref{fig:feynman_cond}). The self energy $\Sigma$ is related to the scalar density through $\Sigma = g^2\, n_s/m^2_\phi$. Finally, note that if the mediator were to interact with pseudo scalar couplings to the fermions (i.e. $\Gamma = i \gamma_5$) $\Sigma$ is identically zero. Thus, we do not expect any in-medium correction to the bare mass for pseudo scalar interactions.

\subsection{Fermion-Fermion condensate: Cooper pairs}

As mentioned above, in the presence of attractive interactions (however small), the Fermi surface is unstable and generically leads to mixing between particles and holes at the edge of the surface that manifests as Cooper pairs. Diagramatically, this is shown in fig.~\ref{fig:feynman_gap}, where the blob represents the gap function. In contrast to fermion-anti-fermion condensate (scalar density condensate), Cooper pairs in a superfluid are condensates of fermion-fermion pairs. The BCS contribution to this condensate are obtained by the following Wick contraction of the 4-point function in eq.~\eqref{eq:effaction}, 

\beq
{g^2\over 2} \wick{\c2 {\bar \psi}(x) \Gamma \c1 \psi(x)  D(x,y)   \c2 {\bar \psi}(y) \bar{\Gamma} \c1 \psi(y)}\, . \nonumber 
\eeq

Thus the object that will get an expectation value is of the form $\psi(x) \psi(y)$, i.e. fermions at two different co-ordinates behave collectively in the same way. Clearly, the product $\psi \psi$ (or $\bar{\psi} \bar{\psi}$) is not well defined. With the introduction of charge conjugate spinor ($\psi_c \equiv C \bar{\psi}^T$), the Cooper pairs can now be written as $\psi \bar{\psi}_c$, where $C = i\gamma^2\gamma^0$. In the following appendix we describe the Dirac structure of the gap and derive the gap equations.

\section{The consistent set of gap equations} \label{app:sec:gaps}
We outline the derivation of the set of consistent gap equations from the free energy ($\Omega$) starting from the effective action eq.~\eqref{eq:effaction}. The four-fermion interaction part in eq.~\eqref{eq:effaction} describes all possible 2$\rightarrow$2 scattering processes involving the fermions. Due to the biquadratic nature of this term we cannot perform integrations over $\psi$ and $\bar{\psi}$ to obtain the partition function. One possible way to correctly describe the superfluid ground state of the system is to approximate the the above biquadratic term as a product of a fermion bilinear and a fermion condensate, the so-called mean field approximation. To this end, we rewrite the four-fermion interaction in eq.~\eqref{eq:effaction} in terms of the usual spinor $\psi$ and its charge conjugated counterpart $\psi_c(x) = C\bar{\psi}^T(x)$, as described in~\cite{Pisarski:1999av,Schmitt:2014eka}. In the mean field limit, the fermion bilinears $\psi_c(x) \bar{\psi}(y)$ ($\psi(y) \bar{\psi}_c(x)$) and $ \bar{\psi}(x) \psi(x)$ ($ \bar{\psi}_c(x) \psi_c(x)$) are approximated by their expectation values and small fluctuations around their expectation values, respectively. Additionally the mean fields are assumed to be static. Assuming $D(x,y) = D(y,x)$, we have

\bea
\bar{\psi}\left(x\right)\psi\left(x\right)\bar{\psi}\left(y\right)\psi\left(y\right) &=& \frac{1}{2} \left[\bar{\psi}\left(x\right)\psi\left(x\right)\bar{\psi}_c\left(y\right)\psi_c\left(y\right)  + \bar{\psi}_c\left(x\right)\psi_c\left(x\right)\bar{\psi}\left(y\right)\psi\left(y\right) \right] \nonumber \\
&&- \frac{1}{4} {\rm Tr} \left[\psi_c\left(x\right) \bar{\psi}\left(y\right) \psi\left(y\right) \bar{\psi}_c\left(x\right) + \psi\left(x\right) \bar{\psi}_c\left(y\right) \psi_c\left(y\right) \bar{\psi}\left(x\right)     \right]~.
\ena

The following fermion bilinears can get expectation values
\bea
\Sigma(x) &\sim& \left<\bar{\psi}(x) \psi(x) \right> = \left< \bar{\psi}_c(x) \psi_c(x)\right>  \implies \text{ in medium correction to mass} \nonumber \\
\Delta^+(x,y) &\sim& \left<\psi_c(x) \bar{\psi}(y) \right> \implies \text{ superfluid energy gap} \nonumber \\
\Delta^-(x,y) &=& \gamma^0 [\Delta^+(y,x)]^\dagger \gamma^0 \sim  \left<\psi(x) \bar{\psi}_c(y) \right> ~.
\ena

The parametric and function forms of $\Sigma$ and $\Delta^+$ have been computed individually in several works, most notably in refs.~\cite{Pisarski:1999av,Gresham:2018rqo}. In this work we want to consistently capture condensation in multiple channels (see below). In the mean field approximation, we capture one-particle irreducible (1PI) diagrams. The most general ansatz for the structure of $\Delta$ in the rest frame of the medium must be consistent with Grassmanian nature of the fermions, i.e. they should not vanish upon anti-symmetrization. Depending on the model at hand, parity and helicity impose restrictions on the allowed gap structure. In general, the ansatz can contain up to eight terms (expressed in the basis of linear combination of $\gamma$ matrices) that are translationally invariant~\cite{Bailin:1983bm,Pisarski:1999av}. Assuming parity is not violated and that the condensation is in $J=0^+$ channel, one possible ansatz for the fermion pair Dirac structure (that would dominate) and the scalar condensate would have the form
\bea\label{eq:gap_ansatz_app}
\Delta\equiv\langle \psi_c(x) \bar{\psi}(y)\rangle &\sim& \Delta_1\, \gamma_5 + \Delta_2 \,\bold{\gamma\cdot\hat{k}}\,\gamma_0 \gamma_5 + \Delta_3\, \gamma_0 \gamma_5 \,. 
\ena

Of the possible eight structures only four structures apply to the case of scalar mediators as they are parity conserving. As we have only one type of fermionic particle $\psi$ in the fundamental theory, only three gap structures are most significant, and contribution from $\Delta_7$ is identically zero.

As mentioned before, we include all channels of condensation within a single framework. To this end, we suitably modify and closely follow the imaginary time coherent state path integral formalism as presented in~\cite{Kleinert:1977tv,Kleinert:2011rb,Kleinert:2018yjk} and simultaneously include all possible channels of condensation. Having identified the relevant gap structures we introduce fermionic (bosonic) auxiliary fields $\Phi^\pm_{\alpha\beta}$ ($\Sigma_{\alpha\beta}$) that captures the degrees of freedom of the low energy theory via Hubbard–Stratonovich transformation (HST).

We introduce Hubbard–Stratonovich auxiliary fields $\Delta$, $\bar{\Delta}$ and $\rho (\equiv\,n_s)$.
The transformation is carried out by inserting two 'fat-identites' into the partition function. The first for the scalar condensate being
\beq
\mathds{1} \propto \int \mathcal{D}\rho \, \exp\left\{ \int_{x,y} -\frac{1}{2} (\psibar \psi -\rho) \frac{ g^2 D(x,y)}{2} (\psibar \psi -\rho) -\frac{1}{2} (\psibar_c \psi_c -\rho)  \frac{ g^2 D(x,y)}{2} (\psibar_c \psi_c -\rho) \right\}.
\eeq

For the fermionic HS fields the following Gaussian integral is introduced
\beq
\mathds{1} \propto \int \mathcal{D}\bar{\Delta}\mathcal{D}\Delta \exp\left\{ \int_{x,y}-\frac{1}{2} ( \Delta -\psi_c \psibar ) \frac{g^2 D(x,y)}{2} (\bar{\Delta} - \psi \psibar_c) -\frac{1}{2} ( \bar{\Delta} -\psi \psibar_c ) \frac{g^2 D(x,y)}{2} (\Delta - \psi_c \psibar) \right\}.
\eeq

Rewriting eq.~\eqref{eq:lagr} in terms of the auxiliary fields results in the following interaction Lagrangian in co-ordinate space  
 \bea\label{eq:hst_1}
 \mathcal{L}_{int}&=&\frac{1}{4}\bigg\{\bar{\psi}_\alpha\left(x\right)\psi_\alpha\left(x\right)+\bar{\psi}_{c,\alpha}\left(x\right)\psi_{c,\alpha}\left(x\right) \bigg\} \Sigma_{\beta \beta}\left(y\right) -\frac{1}{4}\frac{1}{g^2 D\left(x,y\right)}\Sigma_{\alpha\alpha}\left(x\right)\Sigma_{\beta\beta}\left(y\right)\\ \nonumber
&&-\frac{1}{2}\bar{\psi}_{c,\alpha}\left(x\right)\Phi^+_{\alpha \delta}\left(x,y\right) \psi_\delta\left(y\right)-\frac{1}{2}\bar{\psi}_\alpha\left(x\right)\Phi^-_{\alpha \delta}\left(x,y\right)\psi_{c,\delta}\left(y\right)-\frac{1}{4}\frac{1}{g^2D\left(x,y\right)}\Phi^+_{\alpha \delta}\left(x,y\right)\Phi^-_{\delta \alpha}\left(x,y\right)\,,
 \ena
 with
\bea\label{eq:hst_pars}
D\left(x,y\right)&=&  \left(\partial_x^2 + m^2_\phi\right)^{-1}\, ,\\
\Phi^+_{\alpha \delta}\left(x,y\right)&=&g^2\,  D\left(x,y\right)\Delta_{\alpha \delta}\left(x,y\right)\, ,\\
\Phi^-_{\alpha \delta}\left(x,y\right)&=&g^2 \,D\left(x,y\right)\bar{\Delta}_{\alpha \delta}\left(x,y\right)\, ,\\
\Sigma_{\alpha \beta}\left(x,y\right)&=&g^2\,  D\left(x,y\right) \rho_{\alpha\beta}\left(x\right)~. 
\ena

Notice that now the Lagrangian~\eqref{eq:hst_1} is quadratic in $\psi,\,\phi$ thus enabling integration over the fundamental fields $\psi,\, \phi$. In other words we can trade integration over fundamental fields for an integration over auxiliary fields. In other words, setting the auxiliary fields to their vevs ($i.e.$ the mean-field approximation) allows for computing the path integral of a theory of free particles, albeit with anomalous ($i.e.$ off-diagonal) propagators. To this end, we go to momentum space by Fourier transforming the fields that are in co-ordinate space. We now treat the action to be a functional of charge conjugated spinor $\psi_c$ as well (cf.~\cite{Schmitt:2014eka}), resulting in the following action in Nambu-Gorkov space (i.e. $8\times 8$  matrix)
\bea\label{eq:lag_nambu+gorkvo}
S &=& \sum_{k>0} \bar{\Psi} \frac{\mathcal{S}^{-1}}{T}\Psi - \frac{1}{g^2}\frac{V}{T} \sum_{k,q} \Sigma_{\alpha\alpha}(k)\, D^{-1}(k+q)\, \Sigma_{\beta\beta}(q) - \frac{1}{g^2}\frac{V}{T} \sum_{k,q} \Phi^+_{\alpha\delta}(k) \,D^{-1}(k+q)\, \Phi^-_{\delta\alpha}(q),
\ena
with the new spinors defined as
\bea\label{eq:ng_spinors}
\bar{\Psi} \equiv \left(\bar{\psi}, \bar{\psi}_c \right) \quad \text{and}\quad \Psi \equiv \begin{pmatrix} \psi\\ \psi_c\end{pmatrix}\,. 
\ena
The inverse propagator reads
\bea\label{eq:ng_invprop}
\mathcal{S}^{-1} = \begin{pmatrix}
\slashed{k}+\mu\gamma^0-m+\Sigma_{\alpha\alpha}\left(0\right) & \Phi^-\left(k\right)\\
\Phi^+\left(k\right) & \slashed{k}-\mu\gamma^0-m+\Sigma_{\beta\beta}\left(0\right)
\end{pmatrix}.
\ena

We now have all the ingredients at hand to write down the free energy of the theory through the partition function that reads
\bea\label{eq:partition}
\mathcal{Z} &=& \left({\prod_{k>0} \rm Det} \frac{\mathcal{S}^{-1}(i\omega_n,k)}{T} \right)^{1/2}
\exp\left(-\frac{1}{g^2}\frac{V}{T} \sum_{k,q} \Sigma_{\alpha\alpha}(k) D^{-1}(k+q) \Sigma_{\beta\beta}(q) - \frac{1}{g^2}\frac{V}{T} \sum_{k,q} \Phi^+_{\alpha \delta}(k) D^{-1}(k+q) \Phi^-_{\delta\alpha}(q)\right)\,.\nonumber\\
~
\ena
Inserting our ansatz for the gap structure eq.\eqref{eq:gap_ansatz_app} in the above and exploiting the fact that the determinant of a matrix is the product of its eigenvalues ($\epsilon_i(k)$), the Helmholtz free energy reduces to the form

\bea\label{eq:helm_free}
\Omega &\equiv& \Omega_\epsilon + \Omega_\Sigma + \Omega_\Phi \, \nonumber\\
\Omega = -\frac{T}{V} \log\mathcal{Z} &=& -\frac{T}{V}  \sum_k \sum_{i=\pm} \frac{1}{2} \log \frac{\omega^2_n + \epsilon^2_i(k)}{T^2} + \frac{1}{g^2} \sum_{k,q} \Sigma_{\alpha\alpha}(k) D^{-1}(k+q) \Sigma_{\beta\beta}(q) \, \nonumber\\
& & + \frac{1}{g^2}\sum_{k,q} D^{-1}(k+q) \bigg(\Delta_1(k) \Delta_1(q)  - \bold{k\cdot q} \,\Delta_2(k) \Delta_2(q) - \Delta_3(k) \Delta_3(q)\bigg)~.  
\ena
We note that at this level the difference between the contribution of the scalar condensate and the gap functions is factor four. By setting the HST fields to their expectation values a stationary phase analysis can be performed~\cite{Alford:2003fq,Alford:2004hz} and the dynamics of the low energy theory that describe condensed DM can be examined. The advantage of HST is now manifest.

The expression for the free energy of the system above includes the contribution of 1PI diagrams only. The contribution of $\Omega_\epsilon$ is UV finite upon subtraction of the free energy of fermion in the non-interacting theory. The scalar density condensate contribution $\Omega_\Sigma$ is also finite owing to their momentum dependence. However, the contribution from the gaps $\Omega_\Phi$ are UV divergent which implies divergent pressure. To remedy this problem one should consistently include 2PI contributions~\cite{Alford:2007xm}, which is the CJT formalism. Upon inclusion of these diagrams the free energy contribution of the gaps reads
\beq
\Omega_\Phi =-\sum_\eta \int \frac{d^3 k}{(2\,\pi)^3} \frac{1}{2\, \epsilon_\eta(k)} \left(\Delta^2_1(k) + \Delta^2_2(k) +\Delta^2_3(k) \right) ~.
\eeq

\subsection{Dispersion Relation}\label{app:subsec:dr}
The roots of the determinant of the inverse propagator are four fold degenerate. The dispersion relations take the form
\beq\label{eq:disp1}
\epsilon_\pm^2=\mu^2+\omega^2+\Delta_1^2+\Delta_2^2+\Delta_3^2
\pm 2\bigg(\mu^2\omega^2+ 2\mu k \Delta_1 \Delta_2+m^2_\ast\Delta_2^2+\Delta_1^2\Delta_2^2+2m_\ast\mu \Delta_1\Delta_3-2m_\ast k\Delta_2\Delta_3+k^2\Delta_3^2+\Delta_1^2\Delta_3^2\bigg)^{1/2}.
\eeq
The effective mass is denoted by $m_\ast = m - \Sigma(0)$ and the energy is $\omega^2 = m_\ast^2 +|\bold{k}|^2 $. The dispersion relation for particles (anti-particles) corresponds to $\epsilon_-$ ($\epsilon_+$). In the above expression all the gap functions are momentum dependent. Admittedly not much physics can be extracted from the above expression. However, we know that the gap functions are not larger than the chemical potential upon including the momentum dependence in the energy gaps~\cite{Pisarski:1999av}. We make use of this fact and expand the dispersion relation in $\Delta_i/\sqrt{\mu \omega}$ and a-posteriori check that they are consistent. The approximated dispersion is of the form
\beq\label{eq:disp_approx1}
\epsilon_\pm^2\approx \left(\omega\pm \mu \right)^2+\left(\Delta_1\pm\left(\frac{k}{\omega}\Delta_2+\frac{m_\ast}{\omega}\Delta_3\right)\right)^2+\left(\frac{m_\ast}{\omega}\Delta_2-\frac{k}{\omega}\Delta_3\right)^2\pm\frac{\left(m_\ast\Delta_2-k\Delta_3\right)^2}{\mu \omega}.
\eeq
We further note that the last two terms approach zero as $m_\ast\rightarrow 0$ or $|k|\rightarrow 0$. We believe neglecting the last two terms is nevertheless a good approximation as all the relevant dynamics are encoded in the first few terms. This results in the following much simpler dispersion relation
\beq\label{eq:disp_approx_app}
\epsilon_\pm^2\approx \left(\omega\pm \mu \right)^2+\left(\Delta_1\pm\left(\frac{k}{\omega}\Delta_2+\frac{m_\ast}{\omega}\Delta_3\right)\right)^2.
\eeq
For convenience we introduce the parameters
\bea\label{eq:dtilde_app}
\dtild_\pm &=& \Delta_1\pm\left(\frac{k}{\omega}\Delta_2+\frac{m_\ast}{\omega}\Delta_3\right), \\
\ktild &=& \frac{m_\ast}{\omega}\Delta_2-\frac{k}{\omega}\Delta_3\,.
\ena

As stated in the main text, when $m_\ast \rightarrow 0$, the gap $\tilde{\Delta}_\pm \approx \Delta_1 \pm \Delta_2$ and $|\tilde{\kappa}| \approx  -\Delta_3$. In the non-relativistic limit as $k \rightarrow 0$,  the gap $\tilde{\Delta}_\pm \approx \Delta_1 \pm \Delta_3$ and $|\tilde{\kappa}| \approx  \Delta_2$. The gap parameter $\tilde{\Delta}_-$ is the same as the order parameter $d$ obtained in Bailin and Love, see eq. (3.54a), evaluated at the Fermi momentum $k_F$.

\subsection{Gap equations and scalar condensate}\label{app:sub:gesc}
We use a variational approach to derive the gap equations. Upon setting the auxiliary fields to their expectation values we minimize the free energy eq.~\eqref{eq:helm_free} with respect to the gap parameters, i.e. non-trivial energy gaps would be the solution to
\beq\label{eq:minimize_app}
\frac{\partial \Omega}{\partial \Delta_1} =0\,, \quad\quad \frac{\partial \Omega}{\partial \Delta_2} =0\,,\quad\quad \frac{\partial \Omega}{\partial \Delta_3} =0\,, \quad\quad \frac{\partial \Omega}{\partial \Sigma}=0\,.
\eeq
The gap equations take the form
\bea\label{eq:pre_gaps1}
\Delta_1(p_0,p)& =& g^2\,\frac{T}{V} \sum_k \sum_{\eta=\pm}  D(p-k) \frac{\dtild_\eta(k)}{k^2_0- \epsilon^2_\eta(k)}\,,\\ 
\Delta_2(p_0,p)& =& -g^2\,\frac{T}{V} \sum_k \sum_{\eta=\pm}\eta\, D(p-k)\bold{k\cdot p}\frac{k}{\omega_k} \frac{\dtild_\eta(k)}{k^2_0- \epsilon^2_\eta(k)}\,,\\ 
\Delta_3(p_0,p)& =& g^2\,\frac{T}{V} \sum_k \sum_{\eta=\pm}\eta\, D(p-k) \frac{m_\ast}{\omega_k} \frac{\dtild_\eta(k)}{k^2_0- \epsilon^2_\eta(k)}\,,\\ 
\Sigma(p_0,p)& =& g^2\,\frac{T}{V} \sum_k \sum_{\eta=\pm}2 \frac{D(0)}{{k^2_0- \epsilon^2_\eta(k)}} \bigg(\frac{m_\ast}{\omega_k} (\mu +\eta \omega_k)  -\eta \frac{k}{\omega_k} \frac{\ktild(k)}{\omega_k} \dtild_\eta(k)  \bigg)\,.
\ena

Few comments are in order: We need to  perform the Matsubara sum in $k_0 = i (2 n+1) \pi T$. As we are interested in stationary solutions, we suppose that $\Sigma$ and $\Delta$ depend only on three momentum $\bold{k}$. Furthermore, as in Pisarski and Rischke~\cite{Pisarski:1999av}, we assume that the exchanged boson has zero energy, i.e. $p_0 = k_0=0$. This removes the $k_0$ dependence of the gaps. This approximation is justified, as $k\sim\mu$. For large energy exchange of bosons, the fermions lie far away from the Fermi surface (which implies that it is too expensive to have an energy gap). Note that, these considerations are reminiscent of Eliashberg theory of superconductivity where the energy dependence of the gap is retained while the momentum dependence is neglected~\cite{Son:1998uk}. The final set of gap equations in the $T \rightarrow 0$ limit reads
\bea\label{eq:gap_penultimate_app}
\Sigma\left(0\right)&=&\frac{-g^2}{m_\phi^2}\sum\limits_\eta\int \frac{d^3k}{\left(2\pi\right)^3}\left\{\frac{m_*}{\omega_k}\left(\frac{\omega_k + \eta \mu}{\epsilon_\eta\left(k\right)} -1 \right)-\eta\frac{k}{\omega_k}\frac{\ktild(k)}{\omega_k}\frac{\dtild_\eta(k)}{\epsilon_\eta(k)}  \right\}\,,\\
\dtild_\pm(p)&=& \frac{g^2}{32\pi^2} \sum_\eta\int^\infty_0 d k \frac{k}{p} \left\{\log\frac{m_\phi^2+(p+k)^2}{m_\phi^2+(p-k)^2} \mp \eta \frac{k p}{\omega_p \omega_k}\left(-2 + \frac{m^2_\phi +k^2 +p^2}{2 k p} \log\frac{m_\phi^2+(p+k)^2}{m_\phi^2+(p-k)^2} \right)\right. \\ \nonumber
& & \left. \quad \quad \quad \quad \quad \quad \quad \quad\pm \eta \frac{m^2_\ast }{\omega_p \omega_k} \log\frac{m_\phi^2+(p+k)^2}{m_\phi^2+(p-k)^2}\right\} \frac{\dtild_\eta(k)}{\epsilon_\eta(k)}\, ,\\
\ktild(p)&=& \frac{g^2}{32\pi^2} \sum_\eta\int^\infty_0 d k \frac{k}{p} \left\{-\eta \frac{m_\ast k}{\omega_p \omega_k}\left(-2 + \frac{m^2_\phi +k^2 +p^2}{2 k p} \log\frac{m_\phi^2+(p+k)^2}{m_\phi^2+(p-k)^2}\right) \right. \\ \nonumber
& & \left. \quad \quad \quad \quad \quad \quad \quad \quad - \eta \frac{m_\ast p}{\omega_p \omega_k} \log\frac{m_\phi^2+(p+k)^2}{m_\phi^2+(p-k)^2}  \right\} \frac{\dtild_\eta(k)}{\epsilon_\eta(k)} .
\ena
Note we have renormalized the expression for the scalar density condensate by subtracting the vacuum part (hence the $-1$). We find that the condensate contribution is momentum independent. The advantage of having a consistent set of gap equations is that we can integrate the above equations all the way to infinity without having the need to introduce spurious cut-off dependence. 
\subsubsection{The BCS limit of the gap equations}\label{app:sub:BCSlimit}
We now consider the limit $m_\phi \gg k_F, m$, $i.e.$ when the mediator mass is the largest scale. We aim to recover the well-known BCS equation from the above set of equations. In the relativistic limit ($m_\ast\rightarrow 0$), we find that
\begin{align}\label{eqapp:bcs_rel}
\Delta_1 = \frac{g^2}{2m_\phi^2}\int \frac{d^3k}{\left(2\pi\right)^3}\frac{\Delta_1}{2\epsilon_-\left(k\right)}
\end{align}
the gap channels $\Delta_2$ and $\Delta_3$ being identically zero in the heavy mediator, relativistic limit. Within the BCS approximation, the gaps are assumed to be momentum independent. Such an assumption signifies possible UV divergence for the gap equation.

In the non-relativistic limit ($\omega \rightarrow m$), we find that
\begin{align}\label{eq:bcs_non_rel}
\tilde{\Delta}_- &=\frac{g^2}{2m_\phi^2}\int \frac{d^3k}{\left(2\pi\right)^3} \frac{\tilde{\Delta}_-}{\epsilon_-\left(k\right)}\\
\tilde{\kappa}\left(p\right) &=\frac{p}{m} \frac{g^2}{2m_\phi^2}\int \frac{d^3k}{\left(2\pi\right)^3}\frac{\tilde{\Delta}_-}{2\epsilon_-\left(k\right)}
\end{align}
with $\tilde{\Delta}_+$ being identically zero. In this limit, two pairing channels $\Delta_1$ and $\Delta_3$ dominantly contributes to $\tilde{\Delta}_-$. The gap equation for $\tilde{\Delta}_-$ is, up to a factor $2$, the BCS gap equation. Note that $\tilde{\kappa}$ retains an overall momentum dependence $\tilde{\kappa}\propto p/m$, hence it is suppresed compared to $\Delta_-$. 

\subsubsection{$T\ne 0$ Case }

The above equations were evaluated in the limit $T \rightarrow 0$ corresponding to temperatures well below the critical temperature of phase transition. More generally, the gap equations including the dependence of finite non-zero temperature takes the form

\bea\label{eq:gap_penultimatetne0}
\Sigma\left(0,T\right)&=&\frac{-g^2}{m_\phi^2}\sum\limits_\eta\int \frac{d^3k}{\left(2\pi\right)^3}\left\{\frac{m_*}{\omega_k}\left(\frac{\omega_k + \eta \mu}{\epsilon_\eta\left(k\right)}\, \tanh \left(\frac{\epsilon_\eta(k)}{2 T}\right) -1 \right)-\eta\frac{k}{\omega_k}\frac{\ktild(k)}{\omega_k}\frac{\dtild_\eta(k)}{\epsilon_\eta(k)} \, \tanh \left(\frac{\epsilon_\eta(k)}{2 T}\right)  \right\}\,,\\
\dtild_\pm(p,T)&=& \frac{g^2}{32\pi^2} \sum_\eta\int^\infty_0 d k \frac{k}{p} \left\{\log\frac{m_\phi^2+(p+k)^2}{m_\phi^2+(p-k)^2} \mp \eta \frac{k p}{\omega_p \omega_k}\left(-2 + \frac{m^2_\phi +k^2 +p^2}{2 k p} \log\frac{m_\phi^2+(p+k)^2}{m_\phi^2+(p-k)^2} \right)\right. \\ \nonumber
& & \left. \quad \quad \quad \quad \quad \quad \quad \quad\pm \eta \frac{m^2_\ast }{\omega_p \omega_k} \log\frac{m_\phi^2+(p+k)^2}{m_\phi^2+(p-k)^2}\right\} \frac{\dtild_\eta(k)}{\epsilon_\eta(k)}\, \tanh \left(\frac{\epsilon_\eta(k)}{2 T}\right)\, ,\\
\ktild(p,T)&=& \frac{g^2}{32\pi^2} \sum_\eta\int^\infty_0 d k \frac{k}{p} \left\{-\eta \frac{m_\ast k}{\omega_p \omega_k}\left(-2 + \frac{m^2_\phi +k^2 +p^2}{2 k p} \log\frac{m_\phi^2+(p+k)^2}{m_\phi^2+(p-k)^2}\right) \right. \\ \nonumber
& & \left. \quad \quad \quad \quad \quad \quad \quad \quad - \eta \frac{m_\ast p}{\omega_p \omega_k} \log\frac{m_\phi^2+(p+k)^2}{m_\phi^2+(p-k)^2}  \right\} \frac{\dtild_\eta(k)}{\epsilon_\eta(k)} \, \tanh \left(\frac{\epsilon_\eta(k)}{2 T}\right)\, .
\ena

\section{Solution to the gap equation}\label{sec:solgaps}

In this appendix we present the numerical results we have obtained. Before we discuss the numerical results it is instructive to examine the analytically solutions in the BCS theory (i.e. for contact interactions) for non-relativistic systems. We work in the constant gap approximation, i.e. we introduce a cut-off scale in order to understand the dependence of the solution on the parameters of the theory. However, in the numerical results we present no approximation schemes are used.  

\subsection{Non-relativistic ansatz in the BCS limit }
The gap equation in the non-relativistic heavy mediator limit is given by eq.~\eqref{eq:bcs_non_rel},
with the upper limit of the integral being some cut-off associated with the system~\cite{Schmitt:2014eka} (for electronic superconductors the cut-off would be the Debye frequency). However our setup is more general than the case of electronic superconductor and hence we need to know what the cut-off associated with our system is. To this end we follow the approach of renormalization for short range interactions from ultra-cold physics as discussed in ref.~\cite{Bulgac:2001ei,Alexander:2016glq} (reminiscent of renormalization scheme is Pisarski and Rischke~\cite{Pisarski:1999av} but in the non-relativistic limit). The renormalized gap equation (and their solution) takes the form
\beq\label{eq:bcs_nr_ren}
\frac{m^2_\phi}{g^2} =  \int^{k_c}_0 \frac{\dd k}{2\,\pi^2} \frac{k^2}{2\,\epsilon_-} - \frac{m k_c}{2\, \pi^2} \bigg[1 - \frac{k_F}{2\,k_c} \log \frac{k_c + k_F}{k_c - k_F}  \bigg] = -\frac{m\, k_F}{2\, \pi^2} \log \frac{m\, \Delta}{k^2_F} + \frac{m k_c}{2\, \pi^2} \bigg[1 - \frac{k_F}{2\,k_c} \log \frac{k_c + k_F}{k_c - k_F}  \bigg]. 
\eeq
In the above we have introduced a cut-off momentum $k^2_c = 2 m (E_c + \mu) $. This way the rhs of the renormalized gap equation converges as $k_c \rightarrow \infty$ and $E_c \gg \mu$.

In obtaining the above we have also assumed $m \Delta \ll k^2_F$ and $m \Delta \ll \epsilon k_F \ll k^2_F$, with $\epsilon \ll k_F$. Thus, the solution to eq.~\eqref{eq:bcs_nr_ren} is
\bea
\frac{m^2_\phi}{g^2} &\approx& - \frac{m\, k_F}{2\, \pi^2} \log \frac{m\, \Delta}{k^2_F}\\
\Delta &\approx&  \frac{k^2_F}{m}\,e^{- \frac{2\,\pi^2\,m^2_\phi}{ g^2\,m\,k_F}}.
\ena
Adding the condensate contribution does not bring about further complications.
The Fermi momentum is defined as $k^2_F/(2 m) =  \mu + g^2/m^2_\phi \,n$, with fermion number density $n$.  For this case, one finds a parametrically similar solution as above when we proceed following eq.~\eqref{eq:delmu}. \\

The solution in the relativistic limit is similar and reads
\beq
\Delta \approx \mu e^{- \frac{4\,\pi^2\,m^2_\phi}{ g^2\,\mu^2}}~.
\eeq

These ansatz are exactly applicable only in the case of strictly contact interactions. For all other cases the full set of gap equations should be solved numerically.

\subsection{Structure of momentum dependent gap equations and their numerics}
\label{app:numerics}
Momentum dependent gap equations are of the form

\bea\label{eq:typical-gap}
\Delta(p) = - \int K(p,\kp) \frac{\Delta(\kp)}{\eps(\kp)}  \frac{\dd^3 \kp}{(2 \pi)^3},
\ena
with the kernel function denoted by $K(p,\kp)$ and the dispersion relation  written as $\eps(\kp)$. The task is to solve for $\Delta(p)$. \\

In mathematics, the above equation goes by the name non-linear Fredholm equations of the first kind (NLFEFK). It is well known that a direct brute-force solution to the above equation is very sensitive to the initial guess value and consequently unreliable.

In the context of superfluidity in high density neutron matter where such energy gap equations are often encountered,  new techniques have been proposed to overcome the numerical difficulty imposed by eq.~\eqref{eq:typical-gap}. One such technique is discussed in refs.~\cite{KHODEL1996390,PhysRevLett.81.3828,KHODEL2001827}, where eq.~\eqref{eq:typical-gap} is rewritten in the form of non-linear Fredholm equations of the second kind (NLFESK), which is more tractable and offers numerical stability in all regimes without making severe physical assumptions. 

 NLFESK take the form
\bea\label{eq:second-ex}
y(x) = F(x) + \int G(x,t) H(y(t)) \dd t,
\ena
where $G(x,t)$ is the kernel function and the non-linearity is hidden in the function $H(y(t))$. In the context of gap equation the first step is to  recast eq.~\eqref{eq:typical-gap} in the form eq.~\eqref{eq:second-ex}, as suggested by ref.~\cite{KHODEL1996390}. The idea is that the kernel function $K_{p,\kp} \equiv K(p,\kp)$ can be written as a sum of variable separable parts close to the Fermi surface (i.e. at  $p=k=k_F$), and a remainder part that is non-zero only away from the Fermi surface (i.e. the excitations/deformations of the sphere). More strictly $K_{p,\kp} \ge 0$ or $K_{p,\kp} \le 0$ $\forall\, p,\kp \in [0,\infty]$ (as they are our integration limits). Furthermore, $F(x), y(x): [0,\infty] \mapsto \mathbb{R}$. These conditions seem to be trivially satisfied by the gap equations we have. The ansatz can be written as
\bea\label{eq:ansatz}
K_{p,\kp} = \underbrace{K_{(k_F,k_F)}}_\text{propagator func at Fermi surface}\times  \overbrace{\phi(p) \phi(\kp)}^\text{corrections close to surface} + \underbrace{W_{p,\kp}}_\text{remainder function $\equiv$ all deformations away from Fermi surface},
\ena
with
\bea\label{eq:ansatz_c1}
W_{(p,k_F)} &=& W_{(k_F,\kp)} \overset{!}{=}0\,,\\
\phi(k_F) = 1\,\quad \quad &\text{and}& \quad \quad \phi(p) = \left(\frac{K_{p,k_F}}{K_{k_F,\kp}}\phi(\kp) \right)_{\kp=k_F}\,\, , \\
\phi(p) &=& \frac{K_{p,k_F}}{K_{k_F,k_F}}
\ena
where the separable function is $\phi$, the remainder function is $W_{p,\kp}$ that vanishes on the  surface, and  $K_{(k_F,k_F)} = K(p=k_F,\kp=k_F)$. Using the above two conditions in eq.~\eqref{eq:ansatz}, we obtain the following
\bea\label{eq:remainder}
W_{k,\kp} &=& K_{k,\kp} - \frac{1}{K_{(k_F,k_F)}} K_{(p,k_F)} K_{(k_F,\kp)}\\
\ena

Thus the gap equation eq.~\eqref{eq:typical-gap} becomes
\bea
\Delta(p) &=& - K_{(k_F,k_F)} \phi(p) \int  \phi(\kp) \frac{\Delta(\kp)}{\eps(\kp)}  \frac{\dd^3 \kp}{(2 \pi)^3}- \int W_{p,\kp}  \frac{\Delta(\kp)}{\eps(\kp)}  \frac{\dd^3 \kp}{(2 \pi)^3}, \\
\Delta(k_F) &=& - K_{(k_F,k_F)} \int  \phi(\kp) \frac{\Delta(\kp)}{\eps(\kp)}  \frac{\dd^3 \kp}{(2 \pi)^3},
\ena
then we have
\bea\label{eq:solve_gaps}
\Delta(p) &=&   \phi(p)\Delta(k_F)  - \int W_{p,\kp}  \frac{\Delta(\kp)}{\eps(\kp)}  \frac{\dd^3 \kp}{(2 \pi)^3}, \\
\Delta(k_F) &=& - K_{(k_F,k_F)} \int  \phi(\kp) \frac{\Delta(\kp)}{\eps(\kp)}  \frac{\dd^3 \kp}{(2 \pi)^3}.
\ena
To obtain the last equation (constraint equation) we have set $p=k_F$ in the reformulated gap equation above. As suggested by refs~\cite{KHODEL1996390,PhysRevLett.81.3828,KHODEL2001827}, it is numerically convenient to divide both sides by the normalization of the gap at Fermi momentum and recast the equations in terms of dimensionless variables, by introducing the so-called shape function $y(p)= \Delta(p)/\Delta(k_F)$,
\bea
y(p) &=&   \phi(p)  - \int W_{p,\kp}  \frac{y(\kp)}{\eps(\kp,y(\kp))}  \frac{\dd^3 \kp}{(2 \pi)^3}, \\
1 &=&- K_{(k_F,k_F)} \int  \phi(\kp) \frac{y(\kp)}{\eps(\kp,y(\kp))}  \frac{\dd^3 \kp}{(2 \pi)^3}. 
\ena

Notice that the above equations are finally of the form eq.~\eqref{eq:second-ex}, with $F\equiv \phi$, $W_{k,\kp} \equiv G$ and $y(\kp)/\eps(\kp,y(\kp)) \equiv H(y)$.

\begin{figure}[H]
	\centering
\includegraphics[width=0.48\textwidth]{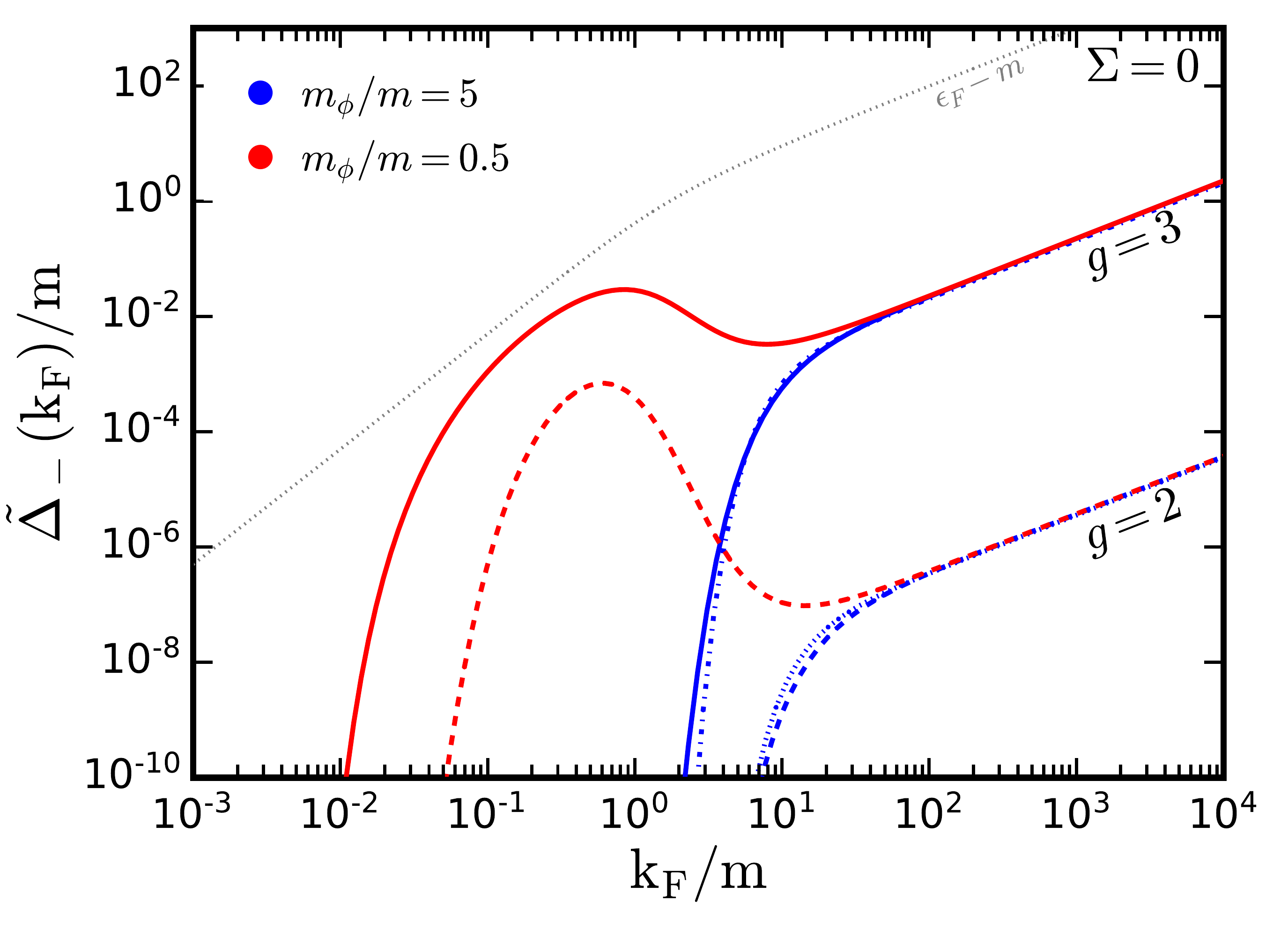}
\includegraphics[width=0.48\textwidth]{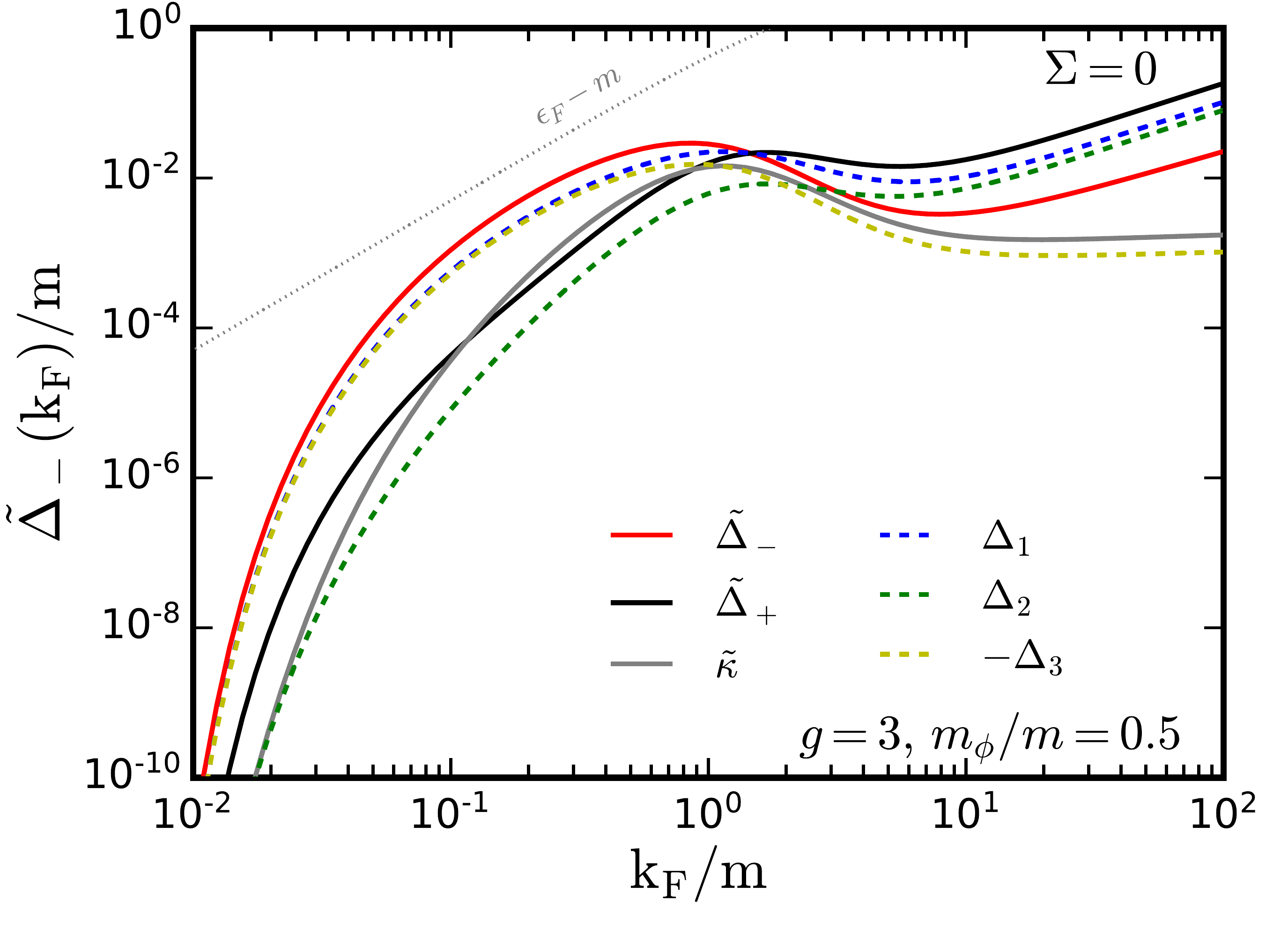} \\ 
\includegraphics[width=0.48\textwidth]{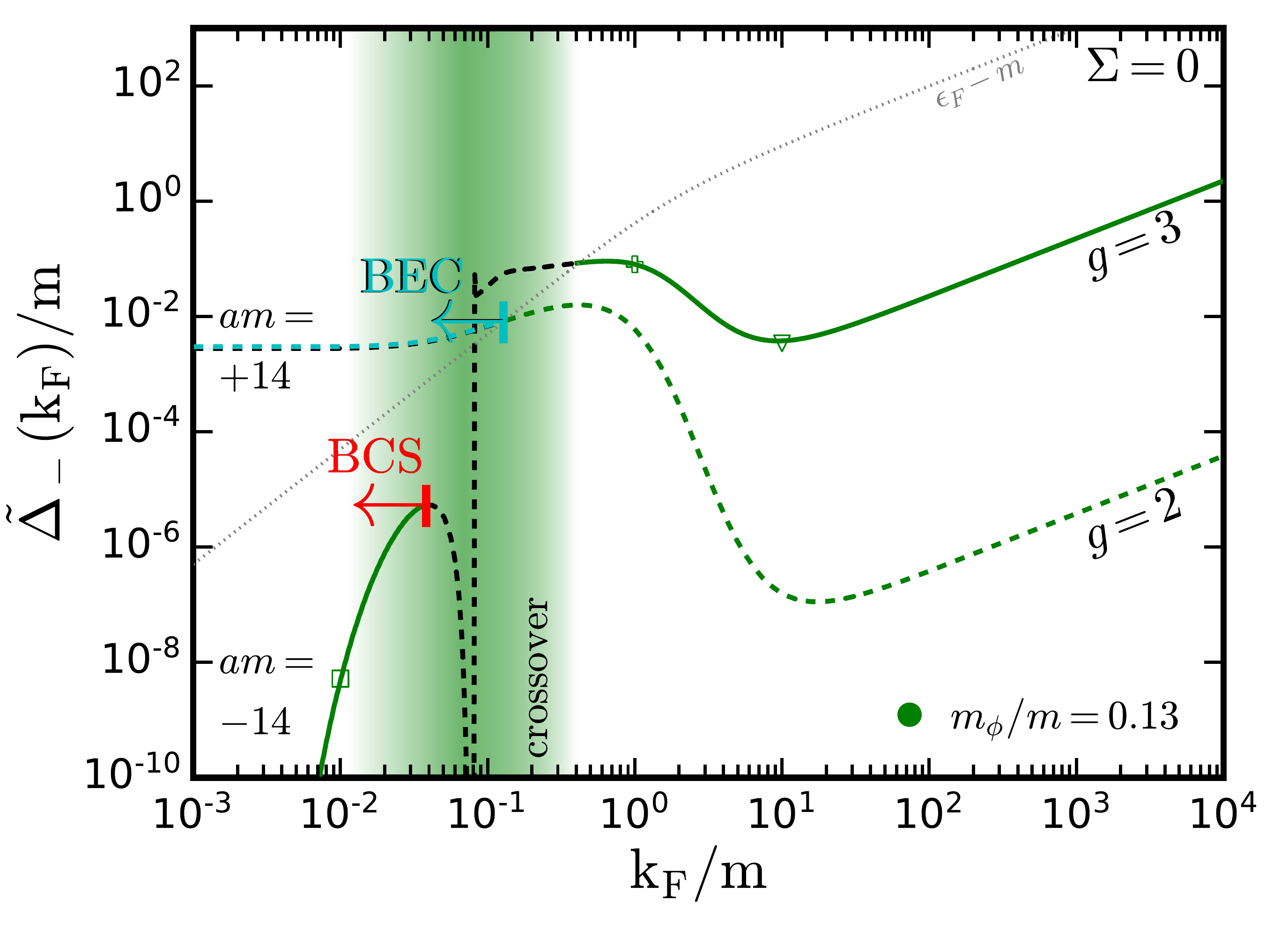}
\includegraphics[width=0.48\textwidth]{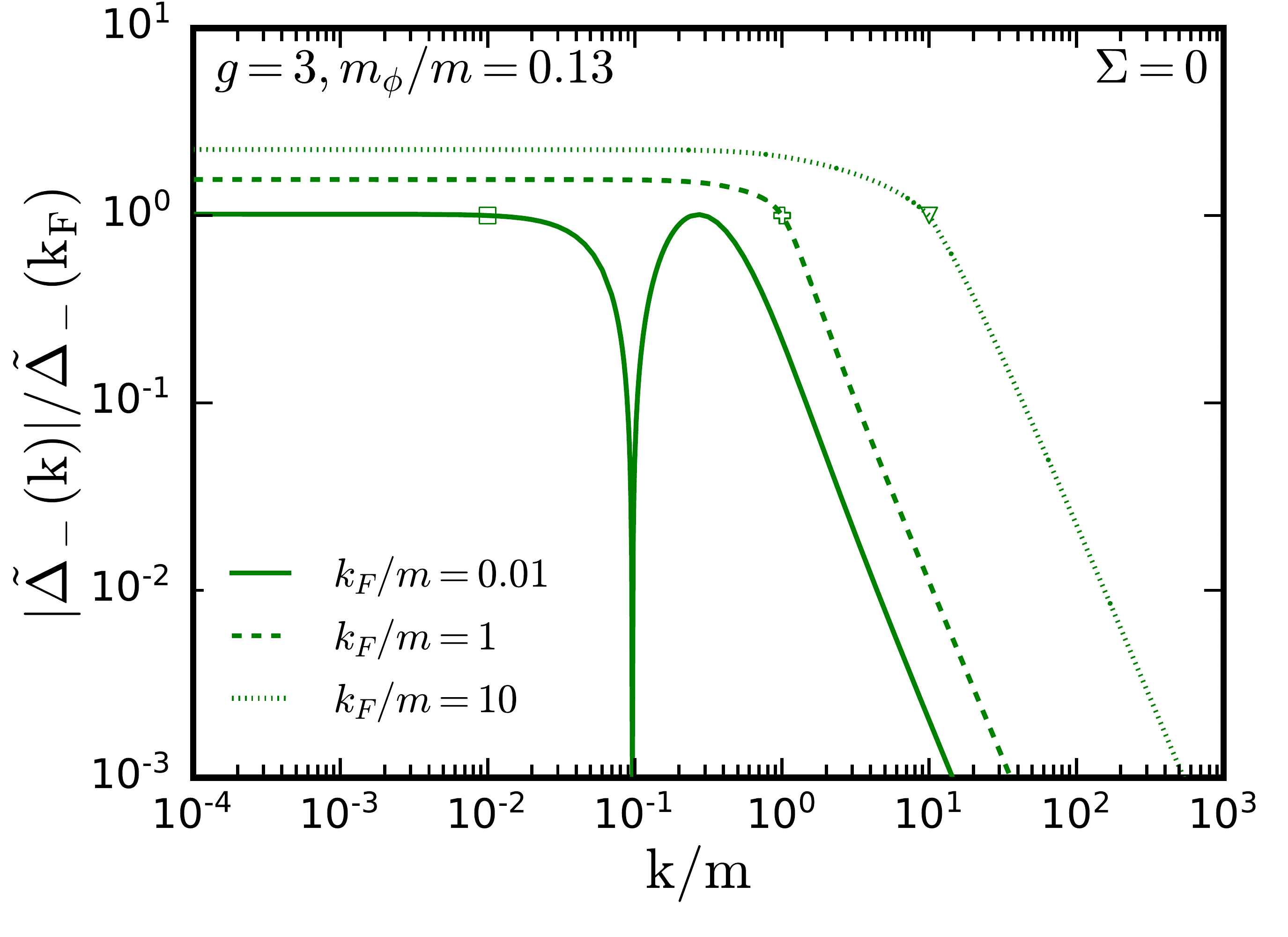}
\caption{The solution to the gap equation excluding the effect of scalar density condensate is shown as function of dimensionless variable $k_F/m$. In the upper left panel we show the results for values of $g=3$ (solid curves) and $g=2$ (dashed curves). Blue (red) colored curves correspond to mediator mass $m_\phi= 5\, m $ ($m_\phi= 0.5\, m$). In the upper right panel we show the individual pairing channels $\Delta_1, \Delta_2, \Delta_3$ (equivalently $\tilde{\Delta}_-,\tilde{\Delta}_+,\tilde{\kappa}$) as a function of $k_F/m$. In the lower left panel the results for values of $g=3$ (solid curves) and $g=2$ (dashed curves) are presented corresponding to $m_\phi = 0.13\,m$. The green shaded region represents the possible crossover regime. In the lower right panel the shape functions are presented for model parameters $g=3$ and $m_\phi = 0.13 \,m$, at various values of $k_F/m$.}
	\label{fig:res_nocond_scalar}
\end{figure}

\subsection{Results}\label{app:subsec:resgap}
As mentioned previously, we solve the full gap equation keeping the momentum dependence using techniques described in ref.~\cite{KHODEL1996390,PhysRevLett.81.3828} (see above). We systematically present results for the full set of gaps below. We begin with the discussion of the solution of gap equations in the absence of scalar density condensate. These results are shown in fig.~\ref{fig:res_nocond_scalar}. All but the right panels are the same as in fig.~\ref{fig:results} (see gray dotted curves) but where the effect of the scalar density condensate is neglected, i.e. $\Sigma=0$. In the top left panel we show in this scenario the numerical results for heavy mediator masses $m_\phi= 5\, m$ (blue curves) and moderately heavy mediator masses $m_\phi=0.5 \,m$ (red curve), for $g=3\, \left(2\right)$ in solid (dashed) respectively, neglecting the effect of the scalar density condensate. In blue dash-dotted (dotted) is given, for $g=3\, (2)$, the following analytical ansatz
\begin{align}\label{geq:bcs_rel_rp}
\tilde{\Delta}_-\left(k_F\right)=\frac{4}{3}k_F\exp\left(-\frac{8\pi^2}{g^2}\right)\exp\left(-\frac{4\pi^2 m_\phi^2}{g^2 k_F^2}\right)~,
\end{align}
for which good agreement is found with the numerical results in the heavy mediator mass regime. The first exponential factor was derived in ref.~\cite{Pisarski:1999av} under the assumption of massless fermions, the second exponential factor is the well-known BCS ansatz. In the top right panel, we show the solutions for the gaps for each individual pairing channel, as well as along the two orthogonal directions, $\tilde{\Delta}$ and $\kappa$. As discussed in the main text (section~\ref{sec:gaps}), we find that not all the pairing channels contribute at all densities. At small densities $\tilde{\Delta}_-$ is the largest gap and it is dominated by the combination of pairing channels $\Delta_1 - \Delta_3$. At large densities however  $\tilde{\Delta}_+$ dominates through the combination of pairing channels  $\Delta_1 + \Delta_2$. These results are consistent with expectations, even though the transition from one regime to the other had not been worked out  so far. Also, as anticipated, at all densities $\kappa$ is subdominant, $\Delta_3$ ($\Delta_2$) being small in the relativistic (resp. non-relativistic) limit.  Finally, we note that the dominant gap combination, $\tilde{\Delta}$, smoothly evolves as the system changes from the non-relativistic regime to the relativistic regime. The bottom panels focus on the light mediator case with $m_\phi/m=0.13$. In the left panel we present the gaps evaluated at $k=k_F$ for $g=3\, (2)$ in solid (dashed). In the right panel we show the momentum dependence of the gap $\tilde{\Delta}_-\left(k\right)$, the so-called shape function, at representative values of densities, $k_F=0.01, \, 1,\, 10$, represented by markers. At large momentum ($k\gg k_F$), the momentum dependence of the gap is found to be $\sim k^{-1.85}$, ensuring self-consistent UV convergence of the gap equations. At low densities, the gap can exhibit one (or multiple) change of signs, and this is exemplified by the dip of the shape function for the lowest shown density, $k_F=0.01$. For such choice of parameters, the gap is negative at large momentum, but nevertheless goes smoothly to $0$ as $k\rightarrow \infty$.\\

\begin{figure}[H]
	\centering
\includegraphics[width=0.48\textwidth]{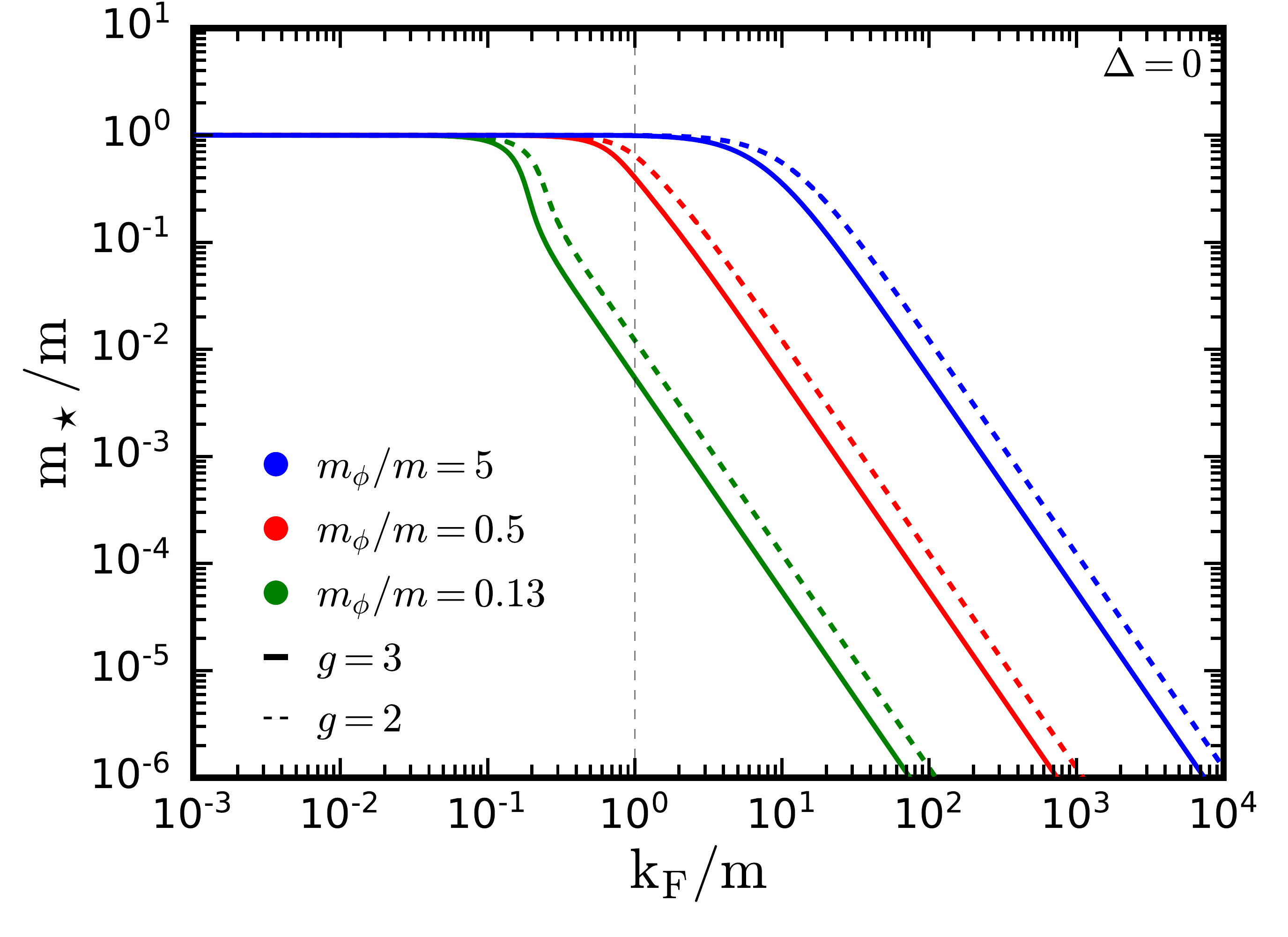} 
		\caption{The effective fermion mass as a function of dimensionless Fermi momentum $k_F/m$, evaluated assuming there are no Cooper pairs.}
	\label{fig:res_cond_nogaps}
\end{figure}

The effective mass $m_\ast$ as a function of density is shown in fig.~\ref{fig:res_cond_nogaps}, for the same model parameters considered in the above figures and in the main text. We follow the same color scheme.  The non-relativistic regime $k_F/m<1$ and relativistic regime $k_F/m>1$ are delimited by the vertical gray dashed line. For the lowest effective coupling $C_\phi^2$, $i.e.$ comparatively large $m_\phi/m$ and low $g$, the effective mass $m_\ast$ deviates from the bare mass $m$ in the relativistic regime. The effect of the scalar density condensate on the gaps is thus expected to be small, see the blue curves along their gray dotted companion in the right panel of fig.~\ref{fig:results}. At low densities, as $m_\phi/m$ becomes smaller than unity, for the studied choice of the coupling, large impact on the gaps due to the scalar density condensate are expected. This is in particular shown by the differences between the red curves along their gray dotted companion in the left panel of fig.~\ref{fig:results}. Similar effect is also seen in the right panel of fig.~\ref{fig:results}.

\begin{figure}[H]
	\centering
\includegraphics[width=0.48\textwidth]{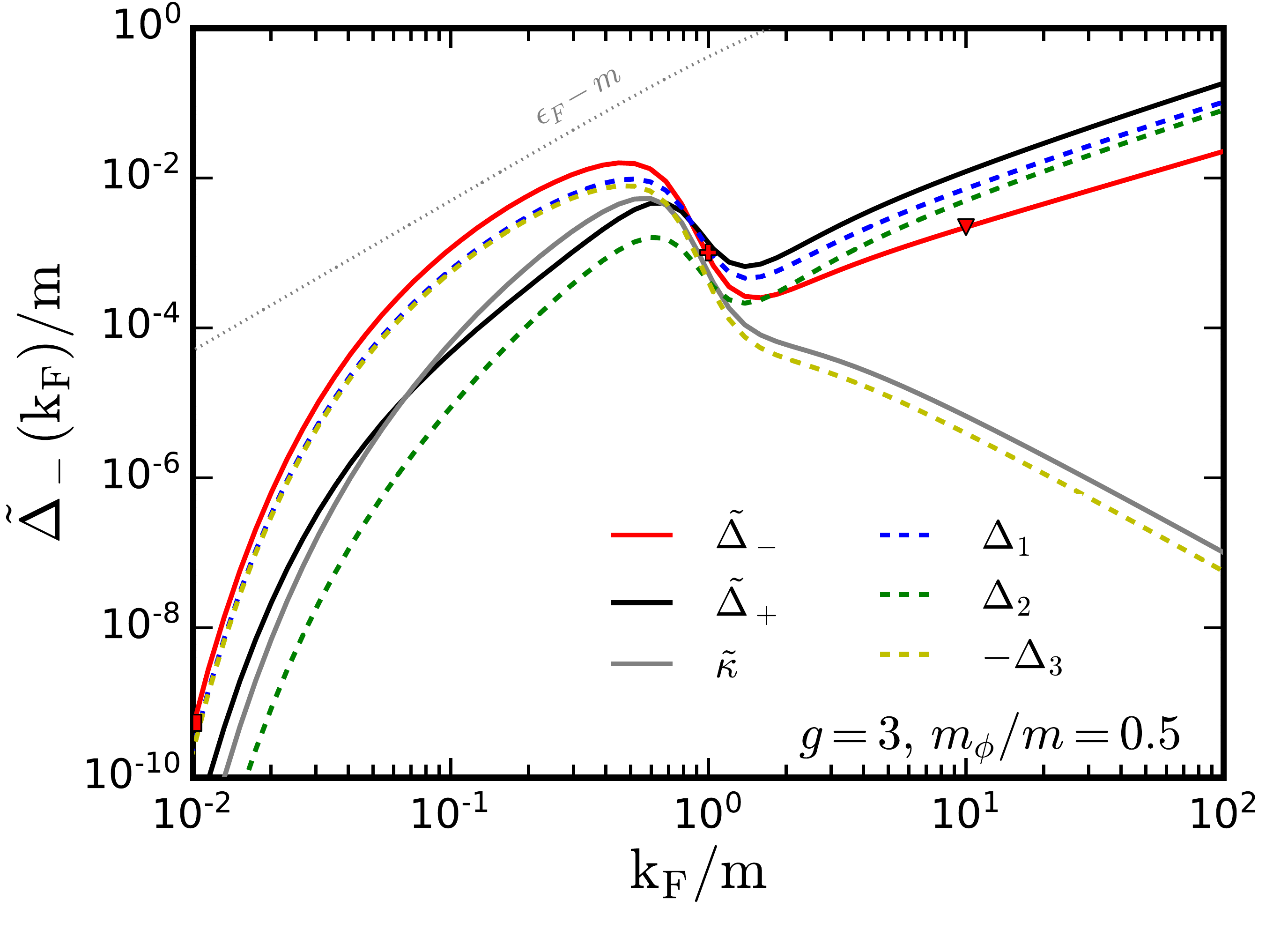} \includegraphics[width=0.48\textwidth]{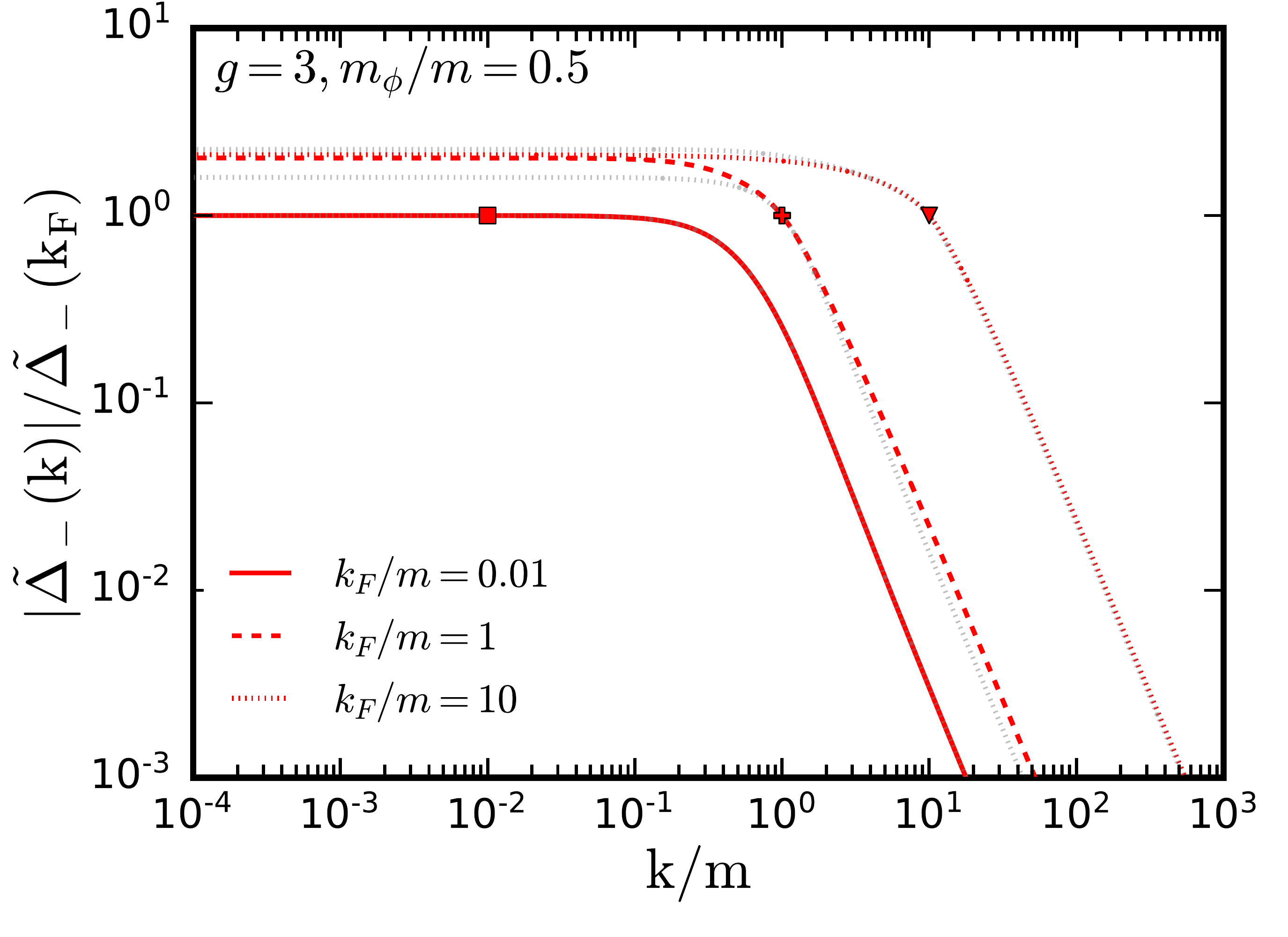} \\
	\includegraphics[width=0.48\textwidth]{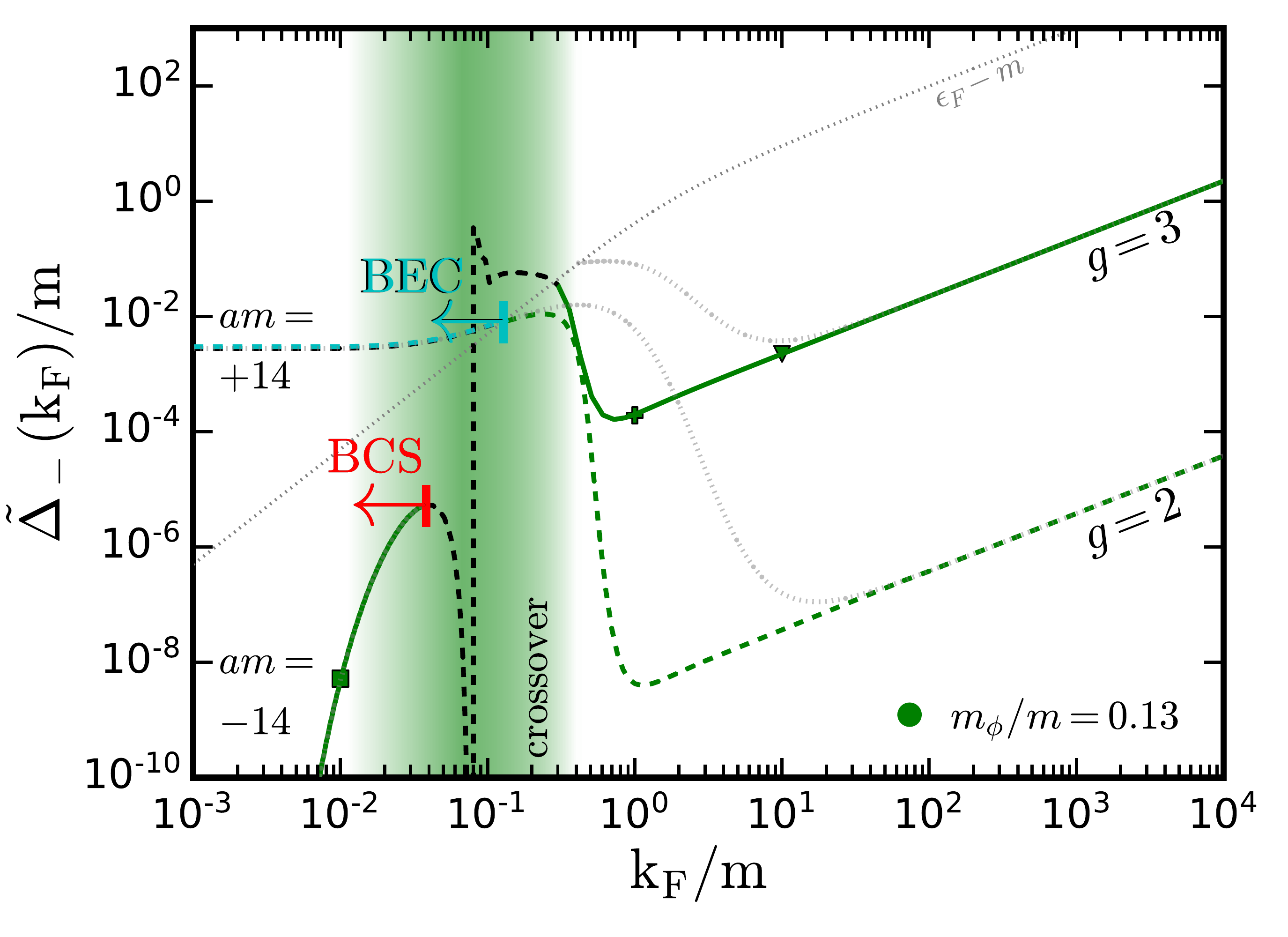} 
\includegraphics[width=0.48\textwidth]{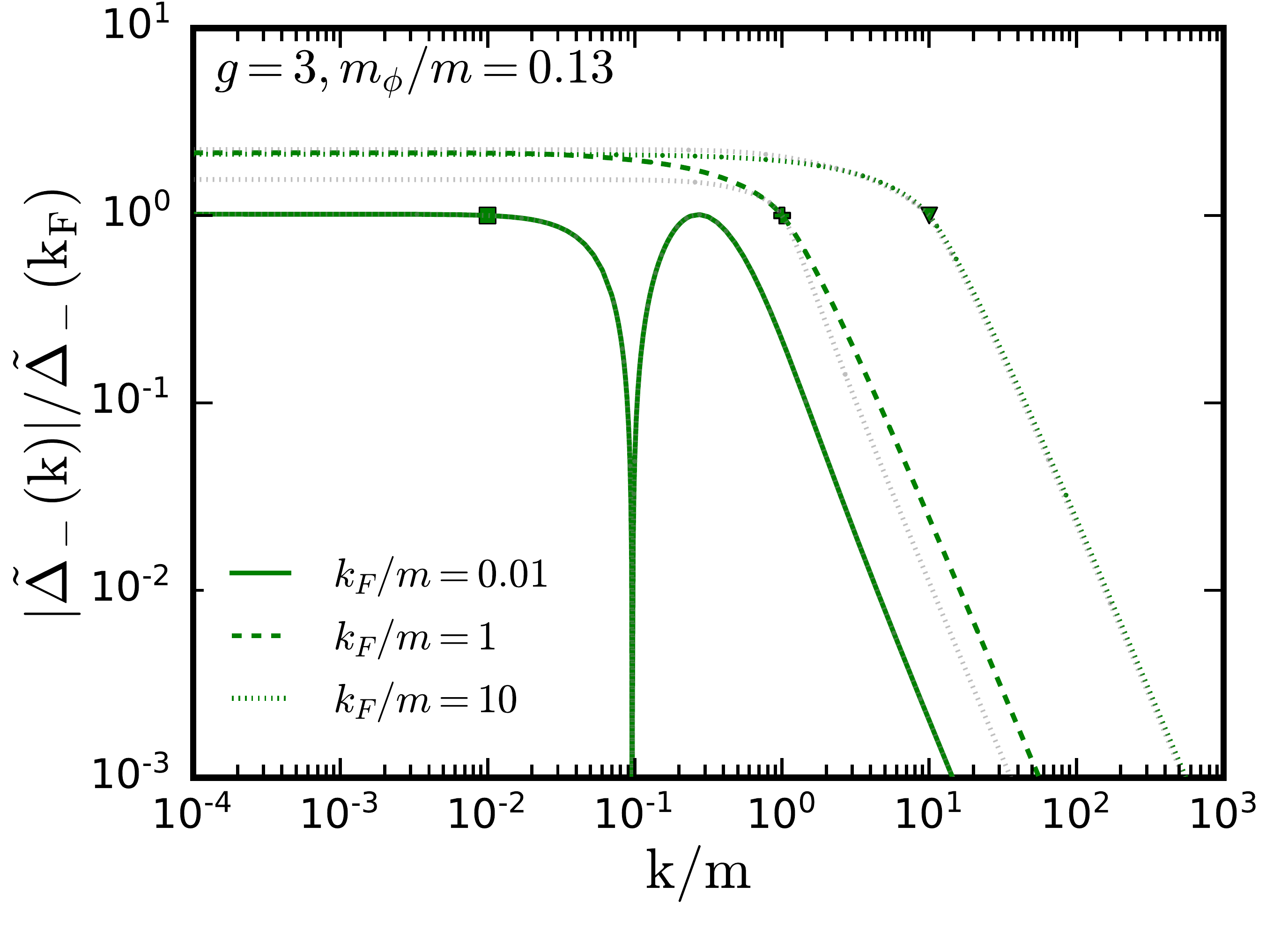}

	\caption{Complimentary to fig.~\ref{fig:results}, the results for the gaps including the effect of scalar condensate are shown. In the top left panel, the dashed curves show the individual pairing channels $\Delta_1, \Delta_2, \Delta_3$, while the solid curves correspond to $\tilde{\Delta}_-, \tilde{\Delta}_+, \kappa$, for $g=3$ and $m_\phi = 0.5\, m$. In the top right panel we show a few shape functions corresponding to marked points in the left panel. In the lower left panel the results for values of $g=3$ (solid curves) and $g=2$ (dashed curves) are presented corresponding to $m_\phi = 0.13\, m$. The green shaded region represents the possible crossover regime. In the lower right panel the shape functions are presented for model parameters $g=3$ and $m_\phi = 0.13 m$, at various values of $k_F/m$ as indicated by the corresponding markers.}
	\label{fig:ex}
\end{figure}

In fig.~\ref{fig:ex}, the focus is put on the dependence of the shape functions on the scalar density condensate for light mediator masses. The bottom left panel is the same as the right panel of fig.~\ref{fig:results}, the markers represent the densities at which the shape function is given on the bottom right panel. As before, the gray dotted lines correspond to the case of no scalar density condensate, $\Sigma =0$. For $g=3$ and $m_\phi/m=0.13$, due to the change in the effective mass, at moderate density $k_F/m=1$, the system is however relativistic as $k_F/m_\ast\gg 1$. In turn, the shape function for $k_F/m=1$ closely resembles that for $k_F/m=10$, albeit shifted to lower $k_F$. Note that this is dissimilar to the case of $\Sigma=0$. The top left panel differs from the top right panel of fig.~\ref{fig:res_nocond_scalar} by the fact that in the limit $k_F\rightarrow \infty$, $\Delta_3, \tilde{\kappa} \rightarrow 0$, as both gaps are proportional to the effective mass $m_\ast$ instead of the bare mass $m$. This vindicates the simplification we proposed in the expression for the dispersion relation, see also section~\ref{sec:disp}.

\section{Thermodynamics and Phase Diagram}\label{sec:phase}

Building upon the solutions obtained for the gaps and condensate, we can now determine the free energy of the system of fermions and study its  phase diagram. We evaluate them in the limit of zero temperature. As the interaction between the fermions is attractive, a gas to liquid transition is expected at large fermion densities~\cite{Walecka:1974qa}. The free energy of the system when superfluid gaps are only a small correction is given by~\cite{Kapusta:2006pm}
\begin{align}
\Omega&= -2T\int\frac{d^3p}{\left(2\pi\right)^3}\left[\log \left(1+e^{-\beta\left(\omega_*-\mu\right)}\right)+\log \left(1+e^{-\beta\left(\omega_*+\mu\right)}\right)\right] +\frac{1}{2}m_\phi^2\phi_0^2~, \label{eq::freenergy}
\end{align}
where $\omega_*=\sqrt{m_*^2+p^2}$. The pressure $P$ is simply minus the free energy and takes the following form at zero temperature \cite{Gresham:2018rqo}
\begin{align}
P=-\Omega= \frac{m^4}{3\pi^2}\left(-\frac{\varphi^2}{2C_\phi^2}+\int\limits_0^{k_F/m}\frac{x^4}{\sqrt{x^2+\left(1-\varphi\right)^2}}dx\right)~,
\end{align}
where we have denoted $\varphi=\Sigma/m$ and $C^2_\phi = 4 \,\alpha m^2_\chi/(3\pi m^2_\phi)$. The adimensional condensate renormalizes the mass as $m_\ast=m\left(1-\varphi\right)$ and $C_\phi^2$ is the effective coupling strength. The energy density $\epsilon$ (not to be confused with the dispersion relation $\epsilon_\pm$) at zero temperature is then found through
\beq
\epsilon =\mu n - P=\frac{m^4}{3\pi^2}\left(\frac{\varphi^2}{2C_\phi^2}+3\int\limits_0^{k_F/m}x^2\sqrt{x^2+\left(1-\varphi\right)^2}dx\right)~,
\eeq
where the number density of particles is given by $n=k_F^3/3\pi^2$.
 Alternatively, the energy density of the fermions is simply given by the integral over all momentum of the Fermi-Dirac distribution weighted by the energy of each mode. The chemical potential is then related to $k_F$ by $\mu^2=m_\ast^2+k_F^2$ where $m_*$ is the effective mass of the fermions in the medium. Subsequently, $\varphi$ is determined through the following equation
\begin{align}
\frac{\varphi}{C_\phi^2}&=3\int \limits_0^{k_F/m}x^2\frac{1-\varphi}{\sqrt{x^2+\left(1-\varphi\right)^2}}dx~.\label{eq::Zurekcond}
\end{align}
This equation is the same as that of eq.~(\ref{eq:gap_penultimate_app}) when the gaps are zero. Furthermore, it is easy to accommodate for repulsive interactions by introducing a vector boson. Such simplified system has been treated extensively in the nuclear physics literature (the so-called $\sigma-\omega$ model) in order to understand the properties of nuclear matter at very high densities. Neutrons, protons interact attractively by exchange of the  $\sigma$ meson and repulsively by exchange of the vector meson $\omega$. Condensation of $\sigma$ renormalizes the nucleon mass, condensation of $\omega$ (more precisely, the $0^{\rm th}$ component of $\left\langle \omega^\mu\right \rangle$ by virtue of rotational symmetry) renormalizes the chemical potential. By fitting the couplings $g_{\sigma, \omega}$ to measured quantities,  some characteristics of nuclear matter can be reproduced. Interestingly, this system is similar in spirit to a Lennard-Jones-type potential and some features are expected to be shared between the nuclear matter and more mundane matter. For the scenario that is treated here, i.e. DM, the same parallel can be made, with the attractive interaction rising from the exchange of the scalar mediator and the degenerate Fermi pressure at relativistic Fermi momentum balancing it.

Thermodynamic quantities of the system are obtained for a given $m$ and $C_\phi^2$ by solving eq.~(\ref{eq::Zurekcond}) for $\varphi$ at every $k_F$ and then computing $P\left(k_F\right)$ and $\epsilon\left(k_F\right)$. The pressure and energy density of a free gas (no interactions) is recovered by taking the limit $C_\phi^2\rightarrow 0$ which, in turn, sends $\varphi\rightarrow 0$ and the in-medium mass becomes the bare mass, $m_*\rightarrow m$.\\

\begin{figure}[H]
\centering
\includegraphics[width=0.625\linewidth]{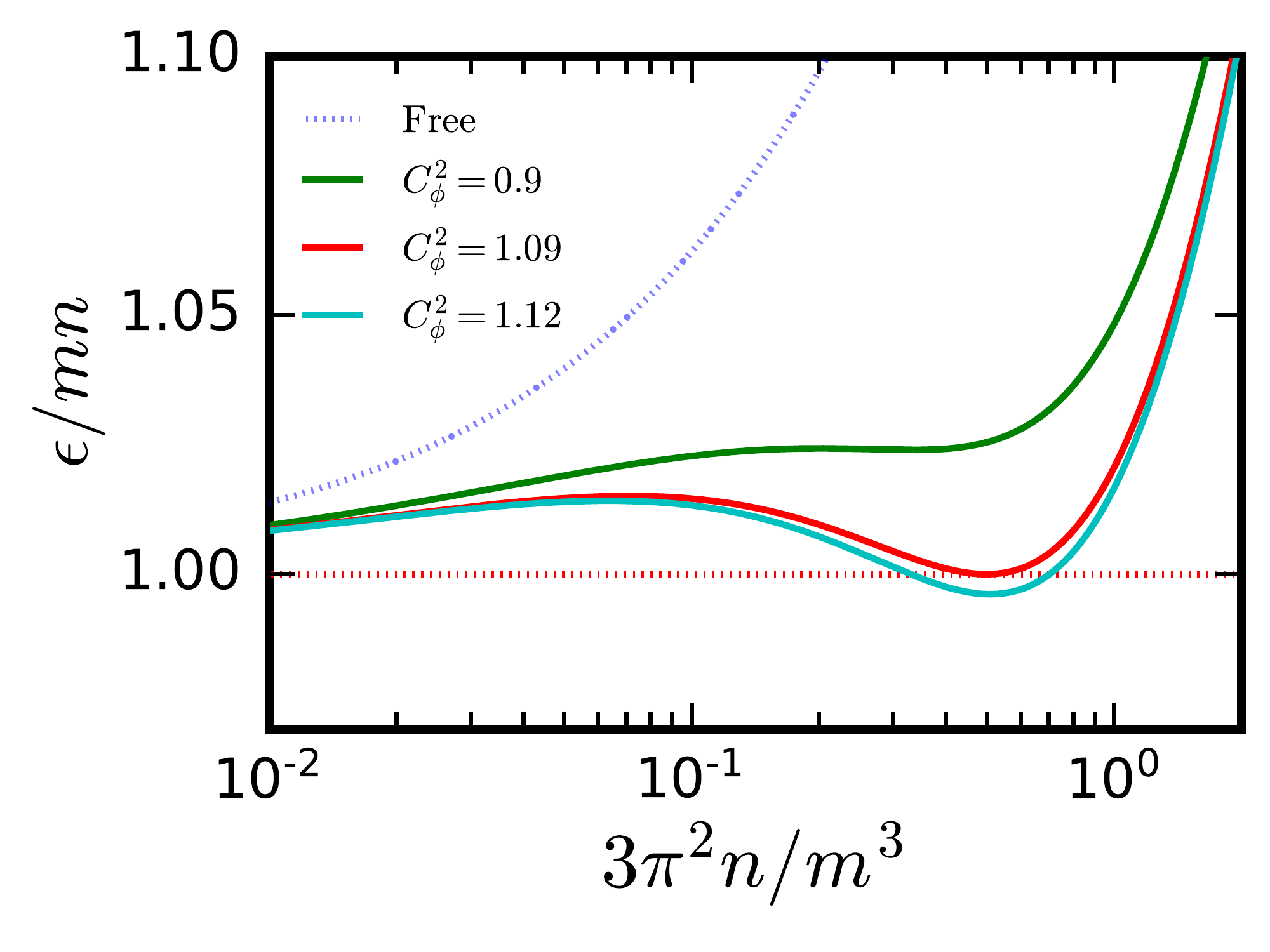}
\caption[Binding energy per particle as a function of density for different coupling $C_\phi^2$.]{Binding energy per particle as a function of density for different coupling $C_\phi^2$. The blue dotted line corresponds to a free gas. Below $\epsilon/mn=1$ (dashed green), matter can clump into nuggets.} \label{fig::bindingenergy}
\end{figure}

It is interesting to first consider the binding energy per fermion as a function of density. This broadly paints the behavior of the system. In fig.~\ref{fig::bindingenergy} we show $\epsilon/mn$ as a function of the adimensional density $\left(k_F/m\right)^3 $ for a few illustrative values of the effective coupling. For $C_\phi^2\ll 1$ the non-interacting case is recovered. For $C_\phi^2>1.09$, the binding energy becomes bigger than the rest mass of the fermions and therefore a global minima develops at high density. This signals that the matter in the system can clump into stable bound states of high densities, the so-called \textit{nuggets} \cite{Gresham:2017cvl, Gresham:2017zqi}, which are identified as the liquid phase.
\begin{figure}[H]
\centering
\includegraphics[width=0.495\linewidth]{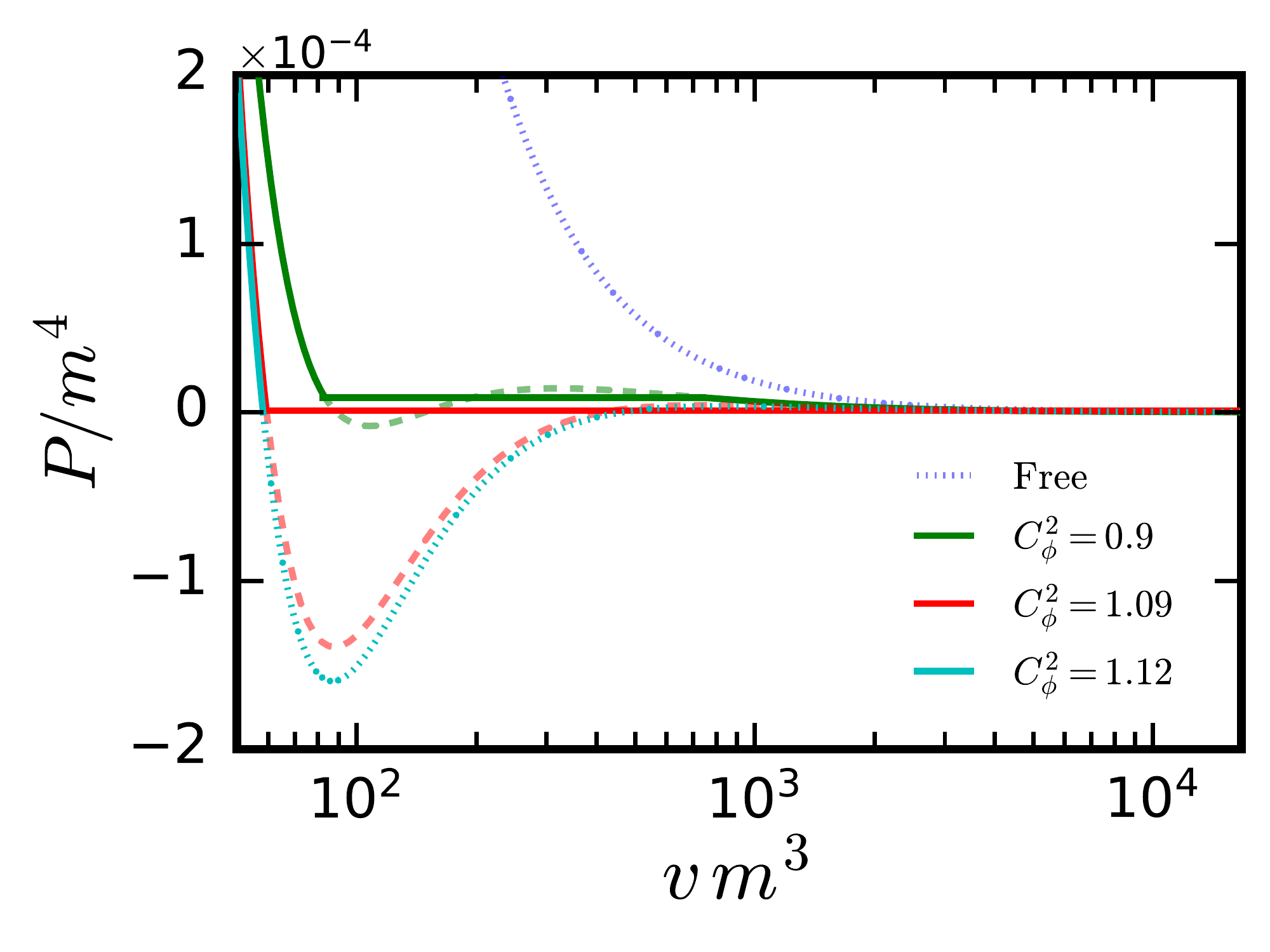}
\includegraphics[width=0.495\linewidth]{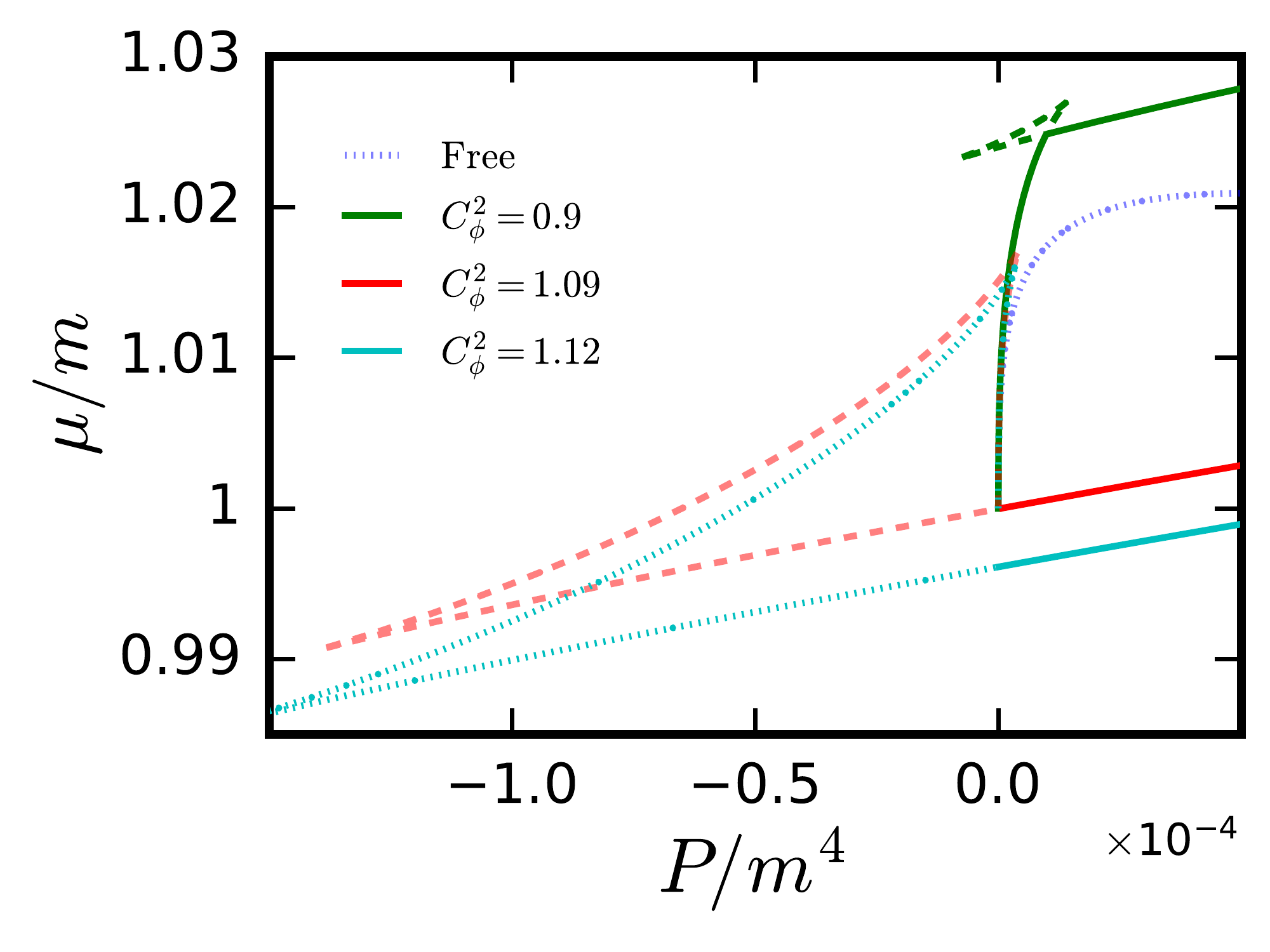}
\caption{Maxwell construction of the physical equation of state. {\it Left panel:} Pressure is shown as a function of volume. {\it Right panel:} the chemical potential is shown as a function of pressure. The dashed and red dotted states do not correspond to equilibrium configurations.}\label{fig::isothermal}
\end{figure}

To understand better the behaviour of the system at different DM densities and coupling, the \textit{isothermal curves} of the pressure $P$ as a function of volume $v=1/n$ are shown in fig.~\ref{fig::isothermal}; such figures are reminiscent of textbook treatment of the van der Waals equation of state, see for example ~\cite{Callen:450289}. The inverse coupling $\tau=1/C_\phi^2$ plays a role analogous to the temperature. Indeed, at high $\tau$ (small $C_\phi^2$) the system can only be in the gas phase whereas at small $\tau$ (high $C_\phi^2$) the system can be put in a high density phase. The conjugate variable to $\tau$ is identified to be the square of the scalar density $\varphi^2$, as seen from the expression of free energy above. For $C_\phi^2>0.841$, the pressure $P\left(v\right)$ start developing a local minima for small volumes, and for $C_\phi^2>0.885$ the minima becomes global with the pressure dropping below $0$. Therefore for strong enough interactions, some parts of the $P\left(v\right)$ curves have $dP/dv>0$, meaning that the compressibility of the system is negative and that these regions are unstable under any small perturbations. If prepared in such states, the system is expected to collapse to a mixture of low and high density matter at the same pressure.

Constructing the physical EoS is a matter of finding the equilibrium states and of identifying the metastable and unstable configurations. Let us summarize how the physical EoS is built from the standard Maxwell constructions \cite{Callen:450289}. The Gibbs-Duhem relation $d\mu=-s dT+v dP$ provides, at constant temperature (in our case, it would be constant $\tau$), a relation between the chemical potentials between two states $A$ and $B$ taken to be, say, at low and high density respectively and the area under the curve $v\left(P\right)$ between those two states
\begin{align}\label{eq::mubmua}
\mu_B-\mu_A=\int\limits_A^B v\left(P\right)dP~,
\end{align}
which, of course, depend on the coupling. For appropriately chosen states $A$ and $B$, it possible for the chemical potential of, say, $B$, to be bigger, lower or equal to the chemical potential of $A$. As can be seen on the right of fig.~\ref{fig::isothermal}, many different state share the same equal pressure $P$ but different volumes $v$. The essence of the Maxwell construction is, for an EoS $P\left(v\right)$, to identify the two states $A$ and $B$ such that they are at equal pressure and at the same chemical potential. Using eq.~(\ref{eq::mubmua}), this amounts to looking at two points on the $P\left(v\right)$ curve  such that the areas between the curve and a constant $P$ line (which is the pressure of both the state $A$ and $B$) is the same below and above. In fig.~\ref{fig::isothermal}, the green curve ($C_\phi^2=0.9$) showcases this construction. 

In the right panel of fig.~\ref{fig::isothermal},  we show $\mu\left(P\right)$ for the free case, and for $C_\phi^2=0.9$, $1.09$, $1.12$, respectively. Following the Maxwell construction of the physical EoS, the physical iso-coupling for $C_\phi^2=0.9\, (1.09)$ is explicitly shown and are presented as the solid green (red) line respectively, while the unphysical iso-coupling in shown as dashed lines. As the coupling grows, the gas to liquid phase transition happens at smaller and smaller densities. Interestingly, for $C_\phi^2>1.09$ the  Maxwell construction cannot be made as there are  no pairs of states with different densities but with the same $\mu$ and $P$. Technically, for large enough coupling, the area between the $P<0$ region of the curve and $P=0$ horizontal axis is bigger than the area under the curve of the region where $P>0$, even if integrated up to infinite volume. Amusingly, a similar feature is encountered in a fluid analogy to black hole thermodynamics~\cite{Kubiznak:2014zwa}.
 
\begin{figure}[H]
\centering
\includegraphics[width=0.48\linewidth]{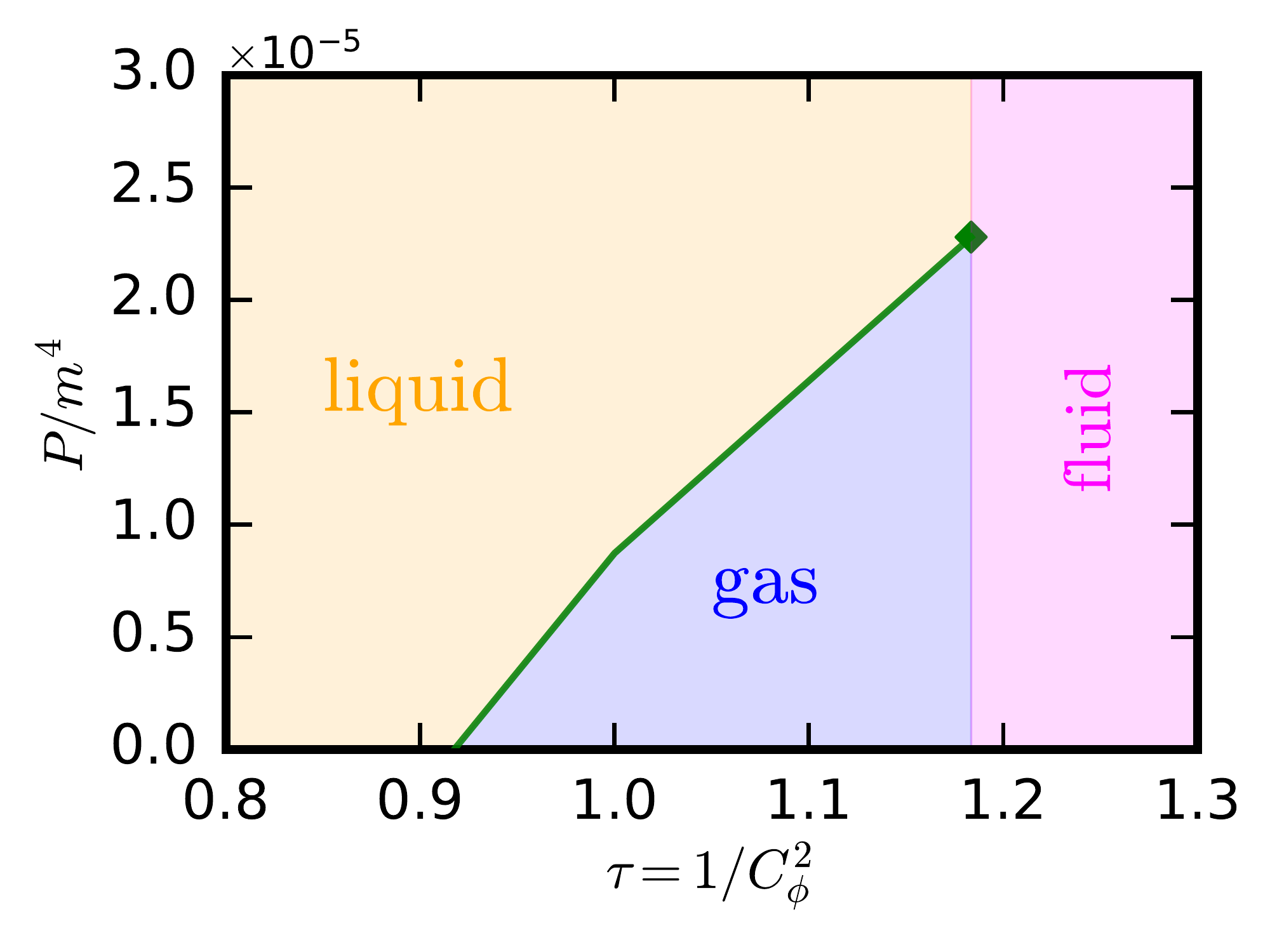} \includegraphics[width=0.48\textwidth]{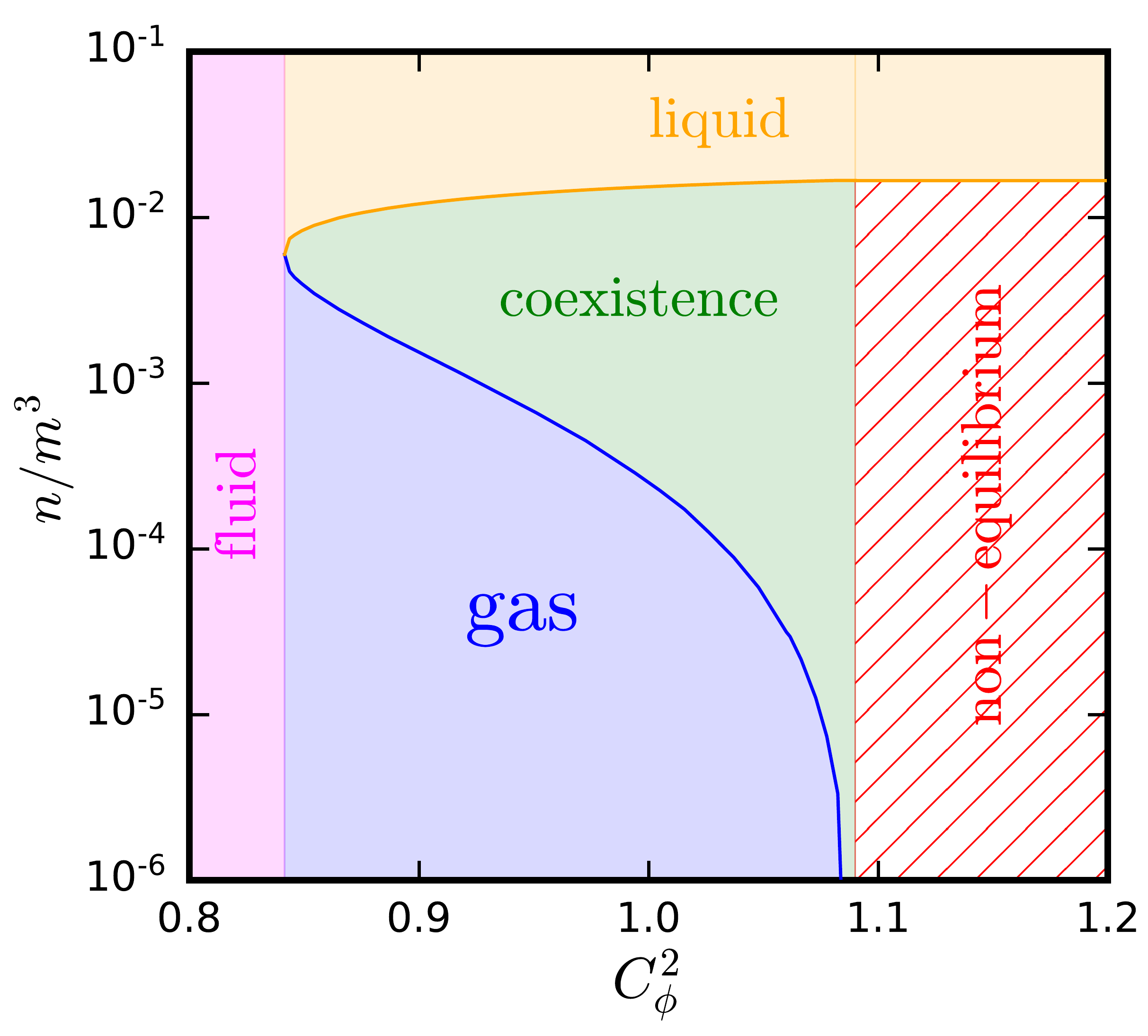}
\caption{{\it Left panel:} Phase diagrams in the dimensionless pressure$-$inverse effective coupling plane for the state of matter described by eq.~(\ref{eq::freenergy}). {\it Right panel:} Phase diagram in the dimensionless number density$-$ effective coupling plane in the Yukawa theory, with $C^2_\phi = 4 \,\alpha m^2_\chi/(3\pi m^2_\phi)$.}\label{fig::Binodal}
\end{figure}

As mentioned before, the EoS in this model closely resembles that of the Lennard-Jones potential and so a similarity between both phase diagrams is expected. This is mostly conveniently seen in $P-\tau$ plane which is shown in fig.~\ref{fig::Binodal} (left panel) using the  effective coupling $C_\phi^2 = 4 \,\alpha m^2_\chi/(3\pi m^2_\phi)$ and the number density  $n/m^3$.
We now interpret the phase diagram presented on the right panel of fig.~\ref{fig::Binodal}. Given a coupling $C_\phi^2$, a system of DM particles at density $n/m^3$ and  in an equilibrium configuration could behave as a 'gas' (blue regions) or as a 'liquid' (orange regions) or coexist as a mixture of the two (green regions) or in the  so-called fluid phase (purple regions), respectively. It is also possible for the system to have no equilibrium configuration strictly speaking but to be instead in metastable (very long lived) or unstable states; the region where such behavior occurs is show in hatched red. In the context of our DM model, say captured at the core of neutron stars, we can assume that such a situation could not be realized.

At this point, it may be instructive to distinguish what we mean by fluid, gas and liquid phases. As is evident from the phase diagram, the gas phase corresponds to low density configuration of the system at moderate $C^2_\phi \approx 1$. Therefore, the expected behavior is very similar to a non-interacting degenerate Fermi gas. For similar and larger values of the coupling, the liquid phase is characterized by very large densities.  Near the saturation density, indicated by a solid orange line, the EoS of the liquid phase closely resembles that of an incompressible fluid~\cite{Gresham:2018rqo}.  The intermediate coexistence region is reached as we increase the density for similar values of coupling. This transition from the gas phase happens to be of first order~\cite{Kapusta:2006pm}. Furthermore, in the so-called fluid phase there are no phase transitions, and consequently the system smoothly goes from the limit of non-interacting Fermi gas to a relativistic one.   

Finally, we briefly address the possible impact of the superfluid phase on this phase diagram. We find that the presence of superfluid phase only gives as small correction to the pressure (as the gaps are parametrically smaller than the kinetic energy at the Fermi surface). More precisely, the gaps contributes positively to the total pressure $\propto \mu^2 \Delta^2(k_F)$. Qualitatively, the correction to picture in fig.~\ref{fig::Binodal} is modest.

\end{document}